\documentclass[12pt,prd,eqsecnum,tightenlines,nofootinbib]{revtex4-2}
\usepackage[letterpaper,margin=1.0in]{geometry}
\usepackage{amsmath}
\usepackage{amsfonts}
\usepackage{mathtools}
\usepackage{bm}
\usepackage{hhline}
\usepackage{setspace}
\usepackage{afterpage}
\usepackage{graphicx}
\usepackage{hyperref}
\graphicspath{{plots/}}

\global\arraycolsep=2pt	% Fix space around '=' in eqnarray's

\def\tr{{\rm tr} \,}

\begin {document}
\count\footins = 1000

\title
    {%
    Resurrecting the coherent state variational
    \texorpdfstring{\\}{}
    algorithm for large \texorpdfstring{$\bm N$}{\it N} gauge theories
    }%
\author	{Laurence G.~Yaffe}
\email	{yaffe@phys.washington.edu}
\affiliation
    {%
    Department of Physics,
    University of Washington,
    Seattle, Washington 98195--1560
    }%

\begin{abstract}
    The feasibility of studying, numerically, properties of infinite volume
    QCD-like theories in the large $N$ limit
    using coherent state variational methods is reassessed.
    An entirely new implementation of this approach is described,
    applicable to SU($N$) lattice gauge theories,
    with or without fundamental representation fermions,
    on cubic lattices of up to four dimensions.
    In addition to various test cases, 
    initial results are presented for 
    Hamiltonian Yang-Mills theory on an infinite two-dimensional
    spatial lattice.
\end{abstract}

\maketitle
\bgroup
\advance\parskip -2pt
\tableofcontents
\egroup
\setstretch{1.1}

\newpage 
\section {Introduction}

For more than forty years, it has been understood that the
large $N$ limit of SU($N$) (or U($N$)) gauge theories
is a type of classical limit \cite{YaffeRMP}.
Suitably defined coherent states provide an overcomplete basis
for the gauge-invariant Hilbert space.
Off-diagonal coherent state overlaps,
as well as off-diagonal coherent state matrix elements of
``reasonable'' operators,%
\footnote
    {%
    Including single trace operators of bounded length,
    or finite order products of such operators.
    }
vanish exponentially as the gauge group rank $N \to \infty$.
As a result, one may show that
the dynamics of the quantum field theory, in the large $N$ limit,
is reproduced by classical dynamics on a phase space essentially
isomorphic to the space of coherent states, with a classical
Hamiltonian given by the large $N$ limit of the coherent
state expectation value of the quantum Hamiltonian
(rescaled by $N^{-2}$).
This structure precisely parallels the usual $\hbar \to 0$ limit
of point particle quantum mechanics, with $1/N^2$ playing the role
of $\hbar$.
For SU($N$) gauge theories containing
fundamental representation fermions
there is a nested structure to the large $N$ limit,
with the large $N$ coherent states of the pure Yang-Mills
theory leading to the gauge sector $N = \infty$ classical
dynamics,
while coherent states of the fermionic degrees of freedom
(generated by exponentials of fermion bilinears)
lead to an analogous phase space structure and classical dynamics
which reproduce the
subleading $O(N)$ quantum dynamics of the full theory
\cite{CSVAI,CSVAII}.%
\footnote
    {%
    This assumes that the number of fermion flavors is held
    fixed as $N \to \infty$.
    }

\advance\parskip 3.0pt plus 1.0pt minus 2.0pt	% needs some stretchability

The classical nature of the large $N$ limit implies that
``solving'' the quantum field theory
 --- meaning accurate computation of ground state and 
physically relevant low energy properties ---
reduces, when $N \to \infty$, to a
\emph{classical} minimization problem:
finding the minimum of the classical Hamiltonian
which reproduces the large $N$ quantum dynamics.
In semiclassical point particle quantum mechanics,
one expands around the minimum of the classical potential
to determine quantum level spacings and anharmonic corrections.
Similarly,
expanding to quadratic order about the minimum of the
large $N$ classical Hamiltonian enables one to compute
the frequencies of small oscillation normal modes which,
in an SU($N$) or QCD-like gauge theory,
amounts to determining the low-lying glueball or meson mass spectrum.
Cubic terms in the Taylor expansion about the minimum determine
the leading large-$N$ behavior of two particle decay amplitudes
of mesons or glueballs,
while quartic terms determine the leading behavior of
two-to-two particle meson or glueball scattering amplitudes.%
\footnote
    {%
    Baryons are solitons in the large $N$ limit \cite{Witten:1979kh},
    with masses scaling as $O(N)$.
    Computation of large $N$ baryonic properties will not
    be considered in this paper.
    }

The above summary applies directly to lattice regulated
gauge theories in a Hamiltonian formulation
(i.e., spatial lattice with continuous time).
A parallel formulation is applicable to Euclidean gauge theories
on a space-time lattice, where the natural language is that
of statistical mechanics.
Instead of minimizing the expectation value of a quantum Hamiltonian,
the goal is to minimize the free energy, viewed as a functional
of an arbitrary statistical density matrix $\rho$,%
\footnote
    {%
    The factor of temperature which would conventionally
    multiply the entropy in the free energy expression (\ref{eq:F})
    is omitted.
    One may view this as a choice of units, or equivalently
    regard the energy $A$ and free energy $F$ used here
    as $\beta \equiv 1/T$ times the conventional
    energy and free energy.
    }
\begin{equation}
    F[\rho] \equiv A[\rho] - S[\rho] \,,
\label{eq:F}
\end{equation}
where the energy (or action) $A[\rho]$ is a linear functional of the density
matrix, while the entropy is given by the von Neumann definition,
\begin{equation}
    S[\rho] \equiv - \tr (\rho \ln \rho) \,.
\label{eq:S}
\end{equation}
The Boltzmann distribution,
$
    \rho_{\rm B} \equiv Z^{-1} \, e^{-A}
$,
minimizes the free energy (\ref{eq:F}) but, in a non-trivial gauge theory,
computing properties of this ensemble is very challenging.
As in the Hamiltonian approach, one may define a manifold
of ``coherent state'' statistical ensembles which provide an
overcomplete basis, meaning that any statistical density matrix
may be expressed as a positively weighted mixture of
coherent state density matrices.%
\footnote
    {%
    In the gauge sector of a $D$-dimensional Euclidean theory,
    these coherent state density matrices are just the 
    absolute squares of coherent state wavefunctionals
    for a $D{+}1$ dimensional Hamiltonian theory.
    }
For large $N$, the entropy and free energy of a
coherent state density matrix are $O(N^2)$, while the
entropy of mixing in a linear combination of such coherent states
remains $O(1)$.
As $N \to \infty$, each coherent state density matrix
acts like an extremal thermodynamic ensemble in which
``reasonable'' observables satisfy large $N$ factorization.
The coherent state ensemble of minimal free energy is
indistinguishable, via measurements of any such
reasonable operator, from the exact Boltzmann ensemble.
So solving the Euclidean theory in the large $N$ limit 
reduces to the minimization of the
free energy of individual coherent state ensembles,
followed by computing physically relevant
properties of that minimizing coherent state ensemble.

This coherent state approach for Euclidean theories is,
of course, of practical interest
only if minimizing the coherent state free energy
and computing physically relevant properties is
less demanding then performing the stochastic simulations
needed to accurately estimate physically
interesting properties of the Boltzmann ensemble.
The key point is that the coherent state formulation
allows one to work directly at $N = \infty$,
exploit large $N$ factorization,
and entirely avoid both finite volume effects and
statistical sampling variance.

The outline of this paper is as follows.
Possible variational strategies for
both Hamiltonian and Euclidean formulations are
discussed in Section \ref{sec:strategy},
emphasizing the choices leading to the
specific form of the coherent state variational
algorithm first presented in Ref.~\cite{CSVAII}.
Section \ref{sec:gordion} then briefly
describes the recent (re)implementation
of this approach in the form of a unified program
named \emph{Gordion}.
Section \ref{sec:results} presents results
from one-plaquette model test cases
as well as for both 2D Euclidean and 2+1 dimensional
Hamiltonian
theories on an infinite two-dimensional cubic lattice,
using the simple observable truncation scheme described below.
The following section \ref{sec:approx} presents results
from an initial effort to reduce truncation errors by using
a loop-factorization based approximation for expectation values
of non-retained observables.
Features, lessons and implications of these various results are
discussed in Section \ref{sec:discussion}, while
a final section \ref{sec:conclusion} offers
concluding discussion and remarks.

All results presented in this paper are limited to
pure Yang-Mills theories using the simplest
Wilson action \cite{Wilson:1974sk} or
Kogut-Susskind Hamiltonian \cite{Kogut:1974ag}.
(A subsequent publication will examine mesonic properties
in QCD-like theories with fermions.)
All computations presented in this work were performed on
a desktop computer,%
\footnote
    {%
    A 2022 Apple Mac Studio with 20-core M1 Ultra cpu,
    128 Gb of memory, and 2 Tb of solid state disk.
    }
not on any large cluster or supercomputer.
So the presented results should not 
be regarded as fully exploring the potential of this approach.

There are, of course, other possible approaches for studying
large-$N$ gauge theories.
Highly developed numerical simulation methods for
Euclidean lattice gauge theories,
involving Monte-Carlo sampling of gauge field configurations,
have been applied to SU($N$) Yang-Mills theories
in three and four dimensions, 
for values of $N$ ranging from 2 up to 12 (4D) or 16 (3D)
\cite{Meyer:2004hv,
Teper:2008yi,
Bursa:2012ab,
Athenodorou:2015nba,
Athenodorou:2016ebg,
Athenodorou:2021qvs}.
These works have studied the glueball spectrum,
$k$-string tensions, and deconfinement temperatures,
with a notable finding of remarkably
weak dependence of these physical quantities on $N$.
There are also recent results from large $N$ lattice
simulations using twisted Eguchi-Kawai reduction
\cite{Gonzalez-Arroyo:1982hyq}.
This approach employs twisted boundary conditions to suppress
unwanted small-volume center symmetry breaking phase transitions,
enabling simulations which take advantage of the
volume independence of large $N$ gauge theories
\cite{Kovtun:2007py}.
Recent results from this approach include calculations of
large $N$ meson masses, the QCD chiral condensate, and more
\cite{Bonanno:2025hzr,Bonanno:2023ypf}.

There has also been interesting recent work 
applying so-called ``bootstrap'' methods
\cite{Anderson:2016rcw,
Kazakov:2022xuh}
to the $N \,{=}\, \infty$ loop equations
of Euclidean lattice Yang-Mills theories
\cite{Makeenko:1979pb,Wadia:1980rb}.
This approach combines (subsets of) the
lattice loop equations with positivity constraints
to derive rigorous inequalities on the range of possible
expectation values of selected sets of Wilson loops.
Related work has applied similar ideas to 
finite $N$ Yang-Mills theory
\cite{Kazakov:2024ool,
Guo:2025fii},
as well as various matrix models
\cite{Lin:2020mme}.

The direct applicability of the coherent state approach
to Hamiltonian lattice gauge theories is one major contrast
to these alternative approaches based on Euclidean lattice formulations.%
\footnote
    {%
    Closely related to the approach of this paper is recent
    work on Hermitian multi-matrix models
    using the collective field Hamiltonian
    \cite {Koch:2021yeb,
    Mathaba:2023non,
    Rodrigues:2025sbu}.
    See footnote \ref{fn:Joao} for further comments on
    connections to this work.
    }
The Hamiltonian formulation
allows far more direct access to the spectrum
of glueballs or mesons, without having to extract
masses from the long-distance fall-off of correlation functions.
Although not yet fully realized, of even greater significance is
the potential to obtain decay widths and two-particle scattering amplitudes
from cubic and quartic terms in the Taylor expansion of the
large $N$ classical Hamiltonian about its minimum.
While there has been significant recent progress in the
development of methods to extract mesonic scattering amplitudes
from Euclidean lattice simulations
\cite{Dawid:2025zxc,
Dawid:2025doq},
this remains a exceptionally difficulty endeavor.

Finally, although not yet applicable to non-trivial Yang-Mills theories,
there are also interesting recent efforts applying
Hamiltonian truncation schemes to model field theories
\cite{Fitzpatrick:2022dwq}.
To date, this approach is very far from reaching the goal of
accurate calculations in non-Abelian gauge theories in two or more
space dimensions.
While the computational challenge involved in applying
the coherent state variational method to large $N$
Yang-Mills theories in 2+1 or 3+1 dimensions is very substantial,
it seems clear that this classical minimization problem must 
nevertheless be far more tractable than any direct attack
on finite $N$ Yang-Mills theory via Hamiltonian truncation
capable of yielding physically interesting results.

\section {Variational Strategy}
\label{sec:strategy}

\subsection {State representation}

Designing any variational minimization 
begins with a choice of coordinates, or more generally
a choice for how to represent information characterizing
properties of individual states in the
minimization domain.
For large-$N$ lattice Yang-Mills theories, the conceptually
simplest, most natural answer is the set of Wilson loop
expectation values for all closed paths,
\begin{equation}
    \left\{
    W_\Gamma \equiv 
    \lim_{N\to\infty} N^{-1} \big\langle \tr U_\Gamma \big\rangle
    \right\} .
\label{eq:Loop}
\end{equation}
Here $\Gamma$ denotes an arbitrary closed path on the lattice,
with $U_\Gamma$ the ordered product of link matrices (or holonomy) around the path $\Gamma$.

In Hamiltonian lattice Yang-Mills theory,
the loop traces $\{ \tr U_\Gamma \}$ are all
commuting operators defined at equal time on the spatial lattice.
The Wilson loop expectation values (\ref{eq:Loop}) 
may be viewed as coordinates on the large-$N$ classical configuration space,
i.e., the time-reversal invariant subspace of the large-$N$ phase space.%
\footnote
    {%
    Throughout this work, all theories under consideration 
    possess both time reversal and charge conjugation symmetry.
    }
The associated classical momenta are time-derivatives of these coordinates,
which amount to expectations of Wilson loop
operators with one electric field insertion,
\begin{equation}
    \left\{
    W_{\ell,\,\Gamma} \equiv
    \lim_{N\to\infty} N^{-1} \big\langle \tr E_\ell \, U_\Gamma \big\rangle 
    \right\} .
\label{eq:Eloop}
\end{equation}
Here, $\ell$ labels links on the lattice,
$E_\ell$ is the U($N$) electric field operator on link $\ell$,
and the path $\Gamma$ starts with link $\ell$
or ends with the conjugate link $\bar\ell$.%
\footnote
    {%
    The difference between U($N$) and SU($N$) gauge theories is
    subleading in the large-$N$ limit, and is irrelevant to this
    discussion.
    The U($N$) electric field operators satisfy the lattice gauge theory
    commutation relations,
    \begin{equation}
	\left[ (E_\ell)_{ij},\, (U_{\ell'})_{kl} \right]
	=
	\tfrac {1}{N} \> \delta_{\ell \ell'} \, \delta_{kj} \, (U_\ell)_{il} \,,
    \quad
	\left[ (E_\ell)_{ij},\, (E_{\ell'})_{kl} \right]
	=
	\tfrac {1}{N} \> \delta_{\ell \ell'}
	\left(
	    \delta_{kj} \, (E_\ell)_{il} - \delta_{il} \, (U_\ell)_{kj}
	\right),
    \label{eq:YMcomms}
    \end{equation}
    with $i,j,k,l$ denoting U($N$) gauge indices.
    }

In theories with fundamental representation fermions, the set of all fermion bilinear
expectation values,
\begin{equation}
    \left\{
    G_{\Gamma_{xy}} \equiv
    \lim_{N \to \infty}
    \big\langle \bar\psi_x \, U_{\Gamma_{xy}} \, \psi_y \big\rangle 
    \right\},
\label{eq:bilinear}
\end{equation}
can serve as coordinates on the fermionic large-$N$ phase space.
Here $x,y$ label lattice sites, $\Gamma_{xy}$ is some lattice path
from site $x$ to site $y$,
and the fermion fields $\{ \bar\psi_x , \psi_y \}$ satisfy
conventional anticommutation relations.
These fermion bilinear expectation values may also be partitioned
into time-reversal even ``coordinates'' and time-reversal odd ``momenta''
on the fermionic large-$N$ phase space.

These sets of Wilson loop and fermion bilinear expectations, 
plus (in Hamiltonian theories) Wilson loop time derivative expectations,
encode the information in any large-$N$ coherent state.
All such coherent states satisfy large-$N$ factorization, with
different coherent states distinguished by differing
values in their sets of expectation values.

On a translationally invariant lattice
(without spontaneous breaking of translation symmetry, which is
assumed throughout),
one may identify expectation values of operators
which merely differ by a lattice translation.
Likewise for observables related by other lattice symmetries
(rotations or reflections).
The resulting set of
of all closed loops, or fermion bilinears, modulo lattice symmetries
is denumerable but remains infinite,
even on a finite lattice of just one plaquette.
So the challenge is to formulate an effective variational strategy 
for this infinite dimensional minimization problem.

Any practical numerical calculation will necessarily involve some truncation
of the variational domain to a finite dimensional subdomain.
One simple approach is to select, in some manner,
a finite set of Wilson loop expectations (and fermion bilinears)
and simply neglect --- approximate by zero --- all other expectations
not in the retained set.
This will be referred to as a ``loop list'' truncation scheme.

In the infinite coupling limit, expectation values of all Wilson loops
vanish identically, $W_\Gamma = 0$, except for the trivial identity ``loop''
which goes nowhere and has unit holonomy.
For large but finite values of the lattice gauge coupling $\lambda \equiv g^2 N$,
Wilson loop expectation values are non-zero but small, with a hierarchy
of sizes determined by the order in a strong coupling expansion
(i.e., a Taylor series in $1/\lambda$)
at which a given loop first acquires a non-zero expectation.
So, at least for sufficiently strong coupling, a truncation scheme which
neglects expectation values of all observables outside some set
of selected observables can work well, especially if the selection of 
retained observables is directly based on the strong coupling order at
which different loops first acquire expectation values.
How to perform such a selection is described below.

The downside of using a state representation based on the neglect
of expectation values of observables outside some 
finite (perhaps large) set of retained Wilson loops is that
the truncation error produced by neglect of
non-retained loop expectations necessarily grows as the lattice
gauge coupling decreases, since the correct expectation value of any
fixed Wilson loop tends to unity in the weak coupling limit.
So this type of state representation will have a domain of utility
which extends downward from strong coupling but terminates at
some non-zero value of the coupling when observable truncation errors 
become too large to ignore.
How this domain of utility depends on the size of the truncation set,
and whether it can extend into the weak coupling regime of the theory
for sufficiently large truncations, can only be answered by
doing the requisite calculations and examining results.

An alternative approach for representing, and truncating,
the information defining a particular gauge theory coherent state
is provided by large-$N$ ``master fields''.
In the $N \to \infty$ limit, one can argue that a single
gauge field configuration can reproduce all Wilson loop
expectation values 
\cite{Witten:1979pi,
Coleman:1980nk,
Coleman:1985rnk}.
On a translationally invariant lattice,
any master field realization must also be translationally invariant
up to a gauge transformation.
Without loss of generality,
one may choose a realization in which the link matrices are
translation invariant without any additional gauge transformation.
For a $d$-dimensional simple cubic lattice, this means that 
a set of just $d$ unitary matrices, $\{ u_i \}$, $i = 1, ...,d$,
in the limit in which the size of these matrices tends to infinity,
can lead to holonomies whose traces exactly reproduce all large-$N$
Wilson loop expectation values.
In effect, one is using unboundly large matrices to
encode an unboundly large amount of information.
By introducing additional matrices
to represent electric field insertions or fermions,
this master field approach can be extended to handle
the gauge field conjugate momenta (\ref{eq:Eloop})
and fermion bilinears (\ref{eq:bilinear}) \cite{CSVAI}.

The master field formulation suggests an obvious alternative truncation
strategy for approximate numerical calculations:
simply restrict master field matrices to some large but finite size.
Then use ordinary matrix multiplication to evaluate the
(truncated approximation) to any Wilson loop or fermion bilinear expectation value.
However, no finite-dimensional set of unitary matrices
can reproduce the correct large-$N$
expectation values of all Wilson loops in the infinite
coupling limit.

%Ref.~\cite{CSVAI} explored both these approaches:
%a truncated ``loop list'' or a finite-dimensional
%master field approximation, in the context
%of one-plaquette models.
%The results of that wark generally favored the truncated loop list approach,
%but there are pros and cons associated with either choice.

Choosing the gauge field master field approximations
$\{ u_i \}$ to be independent random $K \times K$ unitary matrices
provides, in the limit that $K \to \infty$,
a valid master field realization for the
ground state at infinite gauge coupling.
But for finite $K$, any particular realization
of $d$ such random U($K$) matrices will generically lead to
non-vanishing values for the trace (divided by $K$)
of the product of these matrices around any closed loop,
with both the mean and variance of the resulting normalized random matrix
loop trace vanishing with increasing matrix rank only as $O(1/K^2)$.
The typical error in the approximation to the correct infinite
coupling answer of zero only decreases
as the square root of this variance, namely
linearly with $1/K$.
This is awfully slow convergence.
One can engineer alternative sequences of finite dimensional approximations
to an infinite coupling unitary master field which force 
traces of certain select classes of loops to exactly vanish.
But, even in a lattice theory with only a single plaquette,
no set of finite rank unitary master field approximations can
correctly reproduce all Wilson loop expectations at infinite coupling.%
\footnote
    {%
    Formulating some reasonable notion of optimality
    for a finite-dimensional master field approximation,
    just at infinite gauge coupling,
    or constructing specific approximations which are,
    in some meaningful sense, superior to random unitaries,
    are interesting problems which have received little attention.
    }

Hence, at least in the strong coupling regime,
a state representation based on finite size master field approximations
will have greater truncation error than a state representation using a
comparably sized loop-list truncation.
But a state representation using some
approximate master field provides, by construction, a valid unitary
gauge field configuration.
Such an approximation will automatically satisfy all
(rank-independent) positivity constraints, such as the
trivial bound $|W_\Gamma| \le 1$.
As will be discussed further below,
respecting such positivity constraints is non-trivial when using
a truncated loop-list state representation.

Clearly, different truncation schemes for the state representation
will have differing pros and cons.
Properly assessing the utility of either of the above approaches
requires fully implementing a variational procedure using a given
truncation scheme, applying this procedure with different sized
truncations to various theories, and examining results.
This paper presents work using state representations
involving a finite truncated list of physical expectation values.
A future work will examine the alternative master field based approach.

Within the loop-list class of truncation schemes, a key question is how
to select those observables whose expectation values will be retained.
Classifying loops based on their length is a particularly simple
scheme, but in the strong coupling domain of a lattice gauge theory
the most effective truncation scheme --- essentially by definition ---
involves classifying Wilson loops according to their
\textit{strong-coupling order}.
This is defined as the order in a
strong coupling expansion of the ground state energy (or
Euclidean free energy) at which an error is first made if the
particular loop's expectation value is omitted from an otherwise
correct expansion.
As shown in Ref.~\cite{CSVAII}, 
the strong-coupling order of a Wilson loop is the sum of its
\textit{creation order} plus \textit{expectation order}.
The creation order is defined as twice the number of nested commutators
involving a single plaquette with one electric field insertion
acting on an initial single plaquette
which are needed to generate a given loop,
while the expectation order is twice the number of such single plaquette
commutations required to return the loop in question to the identity.
These two orders can differ due to the unitary nature of
the gauge field and consequent automatic cancellation of
backtracking links which can occur in the latter process.
This approach can be extended to
Wilson loops with electric field insertions, as well as
fermion bilinears;
see Ref.~\cite{CSVAII} for details.%
\footnote
    {%
    When fermions are present in the theory,
    the classification is based on a simultaneous
    strong coupling and large mass expansion,
    and the creation order is 
    defined as the minimal sum of the number of commutations with single-link
    fermion hopping terms plus twice the number of commutations with single
    plaquette generators needed to produce the given observable.
    The expectation order is equal to the minimal sum of the number of commutations
    with single-link fermion hopping terms plus twice the number of commutations with single
    plaquette generators needed to yield the identity operator, or a single-site fermion bilinear,
    starting from the given observable.
    If there are compactified directions,
    see Ref.~\cite{CSVAII} for precise definitions.
    These definitions of observable creation, expectation, and strong-coupling orders
    are all twice those used in Ref.~\cite{CSVAII}, so as to avoid half-integral creation or
    expectation orders for fermion bilinears.
    The counting of plaquette operations with double the weight of fermion hopping operations
    reflects the structure of the standard Kogut-Susskind Hamiltonian \cite{Kogut:1974ag}.
    In $H/\lambda$, the gauge kinetic energy is $O(1)$ while the fermion kinetic energy
    (or hopping term) is $O(1/\lambda)$ and the gauge potential energy (or plaquette term)
    is $O(1/\lambda^2)$.
    }

So a state representation based on storing a finite
subset of Wilson loop
expectation values,
with a selection criterion based on an observable's
strong-coupling order,
provides a highly effective representation of the properties of
a large-$N$ coherent state in the large gauge coupling
(and large fermion mass) regime of the theory.
There is, of course, no guarantee that the same observable truncation
scheme will continue to provide an effective approximate state representation
in the weak coupling regime.
But, simply put, no clearly superior computationally useful
representation is known today, and the only way to determine the
limit of utility of this approach is to put it into practice
and examine results.

\subsection {Variational parameters}

The next essential ingredient in any variational minimization is the
choice of variational parameters;
what gets varied to move around in the minimization domain?
The most typical (and obvious) choice is to identify one's
variational parameters with whatever is providing the state representation.
In other words, to use Wilson loop (and fermion bilinear)
expectation values themselves as variational parameters.
%\footnote
%    {%
%    Or, if using a finite dimensional master field approximation,
%    to regard individual components of the master field link
%    matrices as variational parameters.
%    }
For large-$N$ gauge theories, this is a bad choice, for at least
two independent reasons:

First, traces of holonomies
(i.e., Wilson loops) necessarily satisfy an intricate set of inequalities.
The most basic is just $|W_\Gamma| \le 1$, but there are infinitely many
more inequalities relating different traces which follow
just from unitarity.%
\footnote
    {%
    For example \cite{Anderson:2016rcw},
    $
	|W_\Gamma - W_{\Gamma'} W_{\Gamma''}|^2
	\le
	(1-|W_{\Gamma'}|^2)
	(1-|W_{\Gamma''}|^2)
    $
    for any the self-intersecting loop $\Gamma$
    with sub-loops $\Gamma'$ and $\Gamma''$,
    such that $\Gamma = \Gamma'\Gamma''$.
    }
As a result, if Wilson loop expectation values are viewed as 
coordinates on the large-$N$ classical phase space, then the physical domain
has a highly non-trivial boundary arising at finite values of
these coordinates.
Moreover, as the lattice gauge coupling is varied (and decreased), 
the minimum of the large-$N$ Hamiltonian can move from the interior of the
physical domain toward the boundary, hit the boundary at some critical value
of gauge coupling, and thereafter move within the physical domain boundary.
This is one way to understand the origin of the third-order large-$N$ phase 
transition in one-plaquette models
\cite{Gross:1980he,Wadia:2012fr,Wadia:1980cp,Friedan:1980tu}.
It is quite awkward to formulate an effective numerical minimization algorithm which
can handle, correctly, the desired minimum reaching and then moving along this non-trivial
boundary surface.

The second problem is that the functions one wants to minimize, namely
the classical Hamiltonian (defined as the coherent state expectation of the
quantum Hamiltonian, rescaled by $N^{-2}$) or, for the Euclidean formulation,
the free energy (\ref{eq:F}) of a coherent state ensemble (also rescaled by
$N^{-2}$),
do not have simple expressions in terms of Wilson loop expectation values.
The standard Kogut-Susskind lattice Hamiltonian,
\begin{equation}
	H_{\rm gauge}/N = \tfrac 14 \lambda \sum_\ell \> \tr (E_\ell^2)
	    + \lambda^{-1} \sum_p \> \tr (2 - U_{\partial p} - U_{\partial p}^\dagger) \,,
\label{eq:YM-ham}
\end{equation}
is a sum of kinetic energy, proportional to $\sum_\ell \tr E_\ell^2$, and potential energy,
depending on the sum of single plaquettes, $\sum_p \tr U_{\partial p}$
(with $U_{\partial p}$ denoting the holonomy around the boundary of plaquette $p$).
The kinetic energy expectation value, in time-reversal invariant states,
can be formally expressed in terms of Wilson loop expectation values,
but the result is far from computationally convenient.
One may show that \cite{YaffeRMP}
\begin{equation}
    \lim_{N \to \infty} \frac 1{N} \,
    \sum_\ell \> \left\langle \tr (E_\ell^2) \right\rangle
    =
    \sum_{\Gamma,\Gamma'} \>
    \omega_\Gamma \, (\Omega^{-1})_{\Gamma \Gamma'} \, \omega_{\Gamma'} \,,
\label{eq:EE}
\end{equation}
where the double sum runs over the complete set of all possible closed loops.%
\footnote
    {%
    In time reversal non-invariant coherent states, there is an additional
    term involving a quadratic form in which the canonical conjugates of Wilson loops
    are doubly contracted with the loop-joining matrix $\Omega$.
    }
The ``loop-joining'' matrix $\Omega$ has components
\begin{equation}
    \Omega_{\Gamma\Gamma'}
    \equiv
    \lim_{N\to\infty} N \,
    \sum_\ell
    \big\langle
	\left[\, \tr U_\Gamma, (E_\ell)_{ij} \right] 
	\left[(E_\ell)_{ji}, \tr U_{\Gamma'} \right]
    \big\rangle,
\end{equation}
while the ``loop-splitting'' vector $\omega$ has components
\begin{equation}
    \omega_\Gamma \equiv
    \lim_{N\to\infty} \,
    \sum_\ell
    \big\langle
	\left[ (E_\ell)_{ij}, \,
	\left[ (E_\ell)_{ji}, \, \tr U_\Gamma \right] \right]
    \big\rangle .
\end{equation}
For any pair of loops $\Gamma$ and $\Gamma'$,
the loop-joining matrix element $\Omega_{\Gamma\Gamma'}$ is a 
linear combination of Wilson loops which result when
$\Gamma$ and $\Gamma'$ are sewn together at some commonly-traversed link $\ell$.
Similarly, each component $\omega_\Gamma$ of the loop-splitting vector is
a linear combination of the original loop $\Gamma$ and
quadratic products of subloops which result when the loop $\Gamma$ is
split apart at some multiply-traversed link $\ell$.

A key point is that the quadratic form (\ref{eq:EE}) involves components of
the inverse loop-joining matrix, $\Omega^{-1}$,
i.e., the inverse of an infinite dimensional matrix.
One finds in simple models, and can argue more generally,
that the loop-joining matrix $\Omega$ is insufficiently diagonally dominant
for a truncation to a finite set of loops of its inverse to be well approximated
by the inverse of its truncation to that finite set.
(I.e., the inversion of a truncation provides
a poor approximation to the corresponding truncation of the true inverse.)
So, despite the validity of the formal result (\ref{eq:EE}), expressing
the gauge field kinetic energy in terms of Wilson loop expectations in some
computationally useful form is problematic.%
\footnote
    {%
    However,
    recent work in Hermitian multi-matrix models has obtained promising results
    from an approach in which a truncated ``loop-joining'' matrix
    is computed (and then inverted) using
    a master field state representation which automatically
    ensures positivity of the truncated loop-joining matrix
    \cite{Koch:2021yeb,Mathaba:2023non,Rodrigues:2025sbu}.
    \label{fn:Joao}
    }

An entirely analogous issue arises in the Euclidean formulation,
where the free energy (\ref{eq:F}) depends on the entropy of a statistical ensemble
of interest.
In the large $N$ limit, that entropy is, in principle, expressible in terms
of Wilson loop expectations.
But, just as with the gauge kinetic energy in the Hamiltonian formulation,
there is no explicit computationally useful formula.%
\footnote
    {%
    The Migdal-Makeenko loop equations
    \cite{Makeenko:1979pb,Wadia:1980rb},
    an infinite set of polynomial
    equations satisfied by large-$N$ Euclidean Wilson loop expectations,
    are a formulation of the conditions defining a saddle-point of the
    large-$N$ free energy.
    For computational purposes, it is vastly preferable to have a variational
    formulation with a computable free energy, bounded below, which one is
    minimizing.
    The loop equations themselves can have multiple solutions
    (even within the physical domain)
    corresponding to multiple saddle points of the free energy.
    Knowledge of the
    free energy is required to identify the correct physical solution.
    }

For both the above reasons, Wilson loop expectations fail to serve
as good variational parameters.
The alternative adopted in the coherent state variational algorithm of
Refs.~\cite{CSVAI,CSVAII} takes advantage of the intrinsic geometry of the
large $N$ phase space (or manifold of coherent states) which follows from
the underlying structure of the infinite dimensional \textit{coherence group}
$\mathcal G$ which generates gauge theory coherent states.

The coherence group $\mathcal G$ consists of unitary operators
exponentiating elements (or generators) of the \textit{coherence Lie algebra} $\mathbf g$,
\begin{equation}
    \mathcal {G} \equiv \{ e^\Lambda \>|\> \Lambda \in \mathbf g \} \,.
\end{equation}
For QCD-like theories, elements of this infinite dimensional Lie algebra
are anti-Hermitian linear combinations of arbitrary Wilson loops, loops with one
electric field insertion, and fermion bilinears,%
\footnote
    {%
    This is the Hamiltonian theory description. 
    See Ref.~\cite{CSVAII} for the appropriate normal-ordering specifications,
    and for discussion of the parallel Euclidean formulation.
    }
\begin{equation}
    \mathbf {g}
    \equiv
    \bigl\{ \Lambda(a,b,c) \bigr\} \,,
\end{equation}
with
\begin{equation}
    \Lambda(a,b,c)
    \equiv
	\sum_\Gamma \>
	    N \, a_\Gamma \, \tr (U_\Gamma)
	+
	\sum_{\ell,\Gamma} \>
	    N \, b_{\ell,\Gamma} \, \tr (E_\ell \, U_\Gamma)
	+
	\sum_{\Gamma_{xy}} \>
	c_{\Gamma_{xy}} \, \bar\psi_x \, U_{\Gamma_{xy}} \, \psi_y \,.
\label{eq:Lambda}
\end{equation}
Elements of $\mathcal G$ acting on a base state (which may be taken to be
the infinite gauge coupling, infinite fermion mass ground state)
generate the manifold of coherent states which, collectively, form
a \textit{coadjoint orbit} of $\mathcal G$ \cite{YaffeRMP,CSVAII}.

As in any Lie group, one parameter subgroups formed by exponentiating
some given Lie algebra element,
$
    e^{s \Lambda(a,b,c)}
$
for
$\Lambda(a,b,c) \in \mathbf g$ and
$s \in \mathbb R$,
are geodesics on the group manifold.
Elements of $\mathcal G$
which lie in a neighborhood of the identity, when acting on any given
coherent state $|u\rangle$, generate coherent states lying in a
neighborhood of $|u\rangle$.
Consequently, when a coherence group element
$
    e^{\Lambda(a,b,c)}
$
acts on a coherent state $|u\rangle$,
the Lie algebra coefficients
$\{ a_\Gamma, b_{\ell,\Gamma}, c_{\Gamma_{xy}} \}$
are precisely \textit{Riemann normal coordinates},
and parameterize coherent states in the
neighborhood of $|u\rangle$.
Most importantly, such Riemann normal coordinates
may serve as variational parameters
in an iterative minimization scheme.

If, at some point in the iterative minimization, one wishes to move from a
coherent state $|u\rangle$ to some nearby state
$
    |u'\rangle \equiv e^{\Lambda(a,b,c)} \, |u\rangle
$,
then the change in the expectation value of any observable $\mathcal O$
may be computed by integrating the \emph{geodesic equation},
\begin{equation}
    \frac d{ds} \langle \mathcal O \rangle_s
    =
    \langle \left[ \mathcal O, \Lambda(a,b,c) \right] \rangle_s
\label{eq:geo}
\end{equation}
from $s = 0$ to $s = 1$.
Here, $\langle \cdots \rangle_s$ denotes an expectation value in
the intermediate states $\{ |u(s)\rangle \equiv e^{s \Lambda(a,b,c)} |u\rangle \}$
which comprise the geodesic connecting $|u\rangle$ to $|u'\rangle$.
The key point is that, for observables of interest,
the commutator defining this geodesic equation can be evaluated
analytically using the underlying lattice gauge theory commutation relations.
In particular, because coherence group generators contain at most one
electric field insertion, the derivative along a geodesic
of any Wilson loop expectation is some linear combination of 
other Wilson loop expectations.
Similarly, the derivative of any fermion bilinear is a linear combination
of other fermion bilinears.

Actually evaluating, explicitly, the commutators defining the geodesic
equations (\ref{eq:geo}) for a large (but finite) set of Wilson loops
and/or fermion bilinears and a large set of coherence group generators
can be a major undertaking, but this symbolic computation task, 
for a given set of observables and generators, need only be performed once.

This approach of using Riemann normal coordinates as variational parameters
bypasses the problem of inequality violations
which arises when Wilson loop expectations are used directly as variational parameters.
Deformations in a coherent state produced by the action of a coherence group element
necessarily leave the state within the physical domain.%
\footnote
    {%
    This, of course, presumes that errors induced by a truncated
    state representation are under control.
    With a loop-list state representation, when integrating the (truncated)
    geodesic equations, the resulting expectations may eventually violate
    positivity bounds when the truncation error in the
    state representation becomes significant.
    This can serve as a useful diagnostic for the limit of utility
    of a given truncation.
    }

The second problem discussed above, namely the computability of the
gauge kinetic energy or entropy, can also be dealt with in this approach
by augmenting the set of observables used in the state representation.
In the Euclidean formulation this is simple, one must merely add the
entropy to the set of retained Wilson loops.
The geodesic equations for the entropy
are easily expressed in terms of Wilson loop and/or fermion bilinear
expectations \cite{CSVAII}.
For example, the entropy variation induced by
the gauge generator $\Lambda(0,b,0)$ is given by
\begin{equation}
    \delta (S/N^2) 
    =
    -\frac 1N \, \sum_{\ell,\Gamma} \> b_{\ell,\Gamma} \,
    \bigl\langle \tr \big[ E_\ell, \, U_\Gamma \big] \bigr\rangle .
\end{equation}
Evaluating this single trace ``internal'' commutator leads to a quadratic polynomial
in Wilson loop expectations.

In the Hamiltonian formulation, a commutator of the gauge kinetic energy
with a coherence group generator (\ref{eq:Lambda}) leads to linear combinations
of expectations of Wilson loops with one electric field insertion,
Wilson loops with two electric field insertions,
and fermion bilinears with one electric field insertion.%
\footnote
    {%
    As explained in Ref.~\cite{CSVAII},
    in the commutator of a single-$E$ generator with a double-$E$ Wilson loop,
    the normal-ordering prescriptions also lead to terms cubic in Wilson loops.
    }
Therefore, to make it possible to integrate the variation of 
the kinetic energy along a geodesic, it is necessary
to include Wilson loops with up to two electric field insertions
and fermion bilinears with one electric field insertion
in the set of retained observables.
As noted earlier,
the classification of observables according to their strong-coupling order
can be extended in a straightforward manner to loops or bilinears containing
electric field insertions.

These additional observables with electric field insertions are,
in time-reversal invariant states,
redundant variables which could, in principle but not in practice,
be expressed in terms of Wilson loop and fermion bilinear expectation values.
By adding them explicitly to the set of retained observables, all geodesic
variations become computable polynomials in expectation values of those observables.

\subsection {Coherent state variational algorithm}

The above choices of state representation and variational parameters
lead directly to the coherent state variational algorithm
as formulated in Ref.~\cite{CSVAII}.
The basic steps are:
\begin{enumerate}\advance\itemsep -4pt
\item
    Construction of a list of observables $\{ \mathcal O_i \}$ of the chosen theory
    to be retained in the finite truncation set,
    and whose expectation values will be computed.
    In Euclidean theories, ``observables'' mean Wilson loops and fermion bilinears,
    plus the entropy, while in Hamiltonian theories observables
    are Wilson loops with up to two electric field insertions
    and fermion bilinears possibly with one electric field insertion.
\item
    Selection of the finite set $\{ e_\alpha \in \mathbf g \}$ of coherence group generators,
    for a given theory,
    which will be used to generate deformations of coherent states,
    and whose coefficients will serve as Riemann normal coordinates
    in the neighborhood of any given point
    in the large $N$ phase space.
    These are anti-Hermitian combinations of Wilson loops (in Hamiltonian theories),
    loops with one electric field insertion, and fermion bilinears.
\item
    Symbolic evaluation of the commutators of selected observables
    and generators.  These define the geodesic equations
    on the large $N$ phase space,
    encoding how observables vary as one deforms the state,
    \begin{equation}
	\frac d{ds} \,
	\langle \mathcal O_i \rangle = 
	\sum_\alpha \>
	c^\alpha \,
	\langle [\mathcal O_i, e_\alpha] \rangle \,.
    \end{equation}
\item
    Symbolic evaluation of the commutators defining the first and second
    variations of the Hamiltonian (or Euclidean free energy) with respect
    to Riemann normal coordinates,
    \begin{equation}
	(dH)_{\alpha} \equiv
	\langle [ H ,\, e_\alpha ] \rangle \,,
    \qquad
	(d^2H)_{\alpha\beta} \equiv
	\langle [ [ H,\,  e_\alpha] ,\, e_\beta ] \rangle \,.
    \label{eq:ddH}
    \end{equation}
    And for Hamiltonian theories,
    evaluation of commutators of generators whose resulting expectation values
    define the Lagrange bracket (or inverse Poisson bracket)
    of the large $N$ phase space,
    \begin{equation}
	L_{\alpha\beta} \equiv
	\langle [ e_\alpha, e_\beta ] \rangle \,.
    \label{eq:L}
    \end{equation}.
\item
    Numerical minimization of the Hamiltonian, or Euclidean
    free energy, using Newton iteration.
    Each iterative step
    involves numerical evaluation of the symbolic expressions
    for the gradient and curvature of the Hamiltonian (or free energy),
    followed by
    solution of the linear equations predicting the location of the minimum,
    \begin{equation}
	c \equiv - (d^2H)^{-1} \cdot (dH) \,.
    \label{eq:predict}
    \end{equation}
    and numerical integration of the geodesic equations describing
    the change in expectation values of all observables as one moves
    to the newly predicted minimum.
\item
    In Hamiltonian theories,
    numerical evaluation of the symbolic expressions for the curvature of the
    Hamiltonian and the Lagrange bracket, in any chosen symmetry channel,
    and evaluation of the resulting
    small oscillation frequencies around the minimum
    yielding a determination of the
    low-lying glueball or meson spectrum.
    This requires solving the generalized eigensystem
    \begin{equation}
	(d^2H) \cdot \delta = i \omega \, L \cdot \delta \,.
    \label{eq:spectra}
    \end{equation}
\end{enumerate}
Each of these steps is carried out first for the $O(N^2)$ pure gauge dynamics,
and then again for the sub-leading $O(N)$ fundamental representation
fermion dynamics.

\subsection {Why bother?}

Before describing the recent (re)implementation of the coherent state variational method,
and presenting initial results from this implementation, it may be worthwhile to address
a very basic question:
given all the advances in stochastic simulations of Euclidean lattice gauge theories over
the past four decades, including work exploring $N$ dependence by running simulations
with progressively larger values of the gauge group rank
\cite{Meyer:2004hv,
Teper:2008yi,
Bursa:2012ab,
Athenodorou:2015nba,
Athenodorou:2016ebg,
Athenodorou:2021qvs},
is an effort to solve, numerically, for properties of QCD-like gauge theories
directly at $N = \infty$ worth the trouble?
Although only classical minimization is involved, with no stochastic sampling,
it is evident that minimizing the Hamiltonian of the large-$N$ classical dynamics
(or free energy of the large-$N$ statistical mechanics) is a hard problem, 
surely exponentially hard in terms of dependence on correlation length.

From the outset, it is clear that coherent state studies of QCD at $N = \infty$
will not compete with evaluations of experimentally observable quantities
(such as light hadron masses and selected weak matrix elements) for which
it is already feasible to achieve fully-controlled percent-level accuracy in
Euclidean lattice simulations of real ($N = 3$) QCD.
If the only potential output from numerical work directly studying
the $N = \infty$ limit of lattice Yang-Mills theory
were, say, the ground state energy density and expectation values of a few small
Wilson loops, it would be hard to justify the effort.
But that is too narrow a perspective.
Reasons for pursuing numerical studies of
large $N$ Yang-Mills and QCD, despite the inherent difficulty,
include the following:
\begin{itemize}
\item
    The ability to study lattice gauge theories in a Hamiltonian formulation.
    The Euclidean formulation of the coherent state variational method
    is a desirable variant which can enable direct comparison with
    Euclidean lattice simulations, but it is the Hamiltonian formulation
    which is of most interest.

\item
    The ability to work directly in infinite volume is also a key feature
    which distinguishes the large $N$ coherent state approach from
    standard Euclidean lattice simulations.

\item
    The ability to easily study theories with dynamical fermions, without any of the
    complications needed for dealing with the Dirac determinant in lattice simulations.

\item
    Extracting from lattice simulations the spectrum of hadronic resonances
    (especially glueballs) which decay via strong interactions is
    challenging, whereas in the Hamiltonian formulation of large $N$
    dynamics this merely requires the evaluation of small oscillation
    frequencies about the minimum of the large $N$ classical Hamiltonian.

\item
    While there has been notable recent progress on the extraction of
    scattering amplitudes from Euclidean lattice
    simulations
    \cite{Dawid:2025zxc, Dawid:2025doq},
    doing so is extremely challenging and currently
    only feasible in limited cases.
    The potential to extract decay widths and 
    scattering amplitudes from third and fourth derivatives
    of the large $N$ classical Hamiltonian
    (with no finite size effects or momentum quantization constraints
    to contend with) is a novel feature.
    This capability has not yet been implemented,
    as it requires evaluation of commutators of operators with
    non-zero momentum, but is a feasible extension of current work
    and a significant motivation for the approach.

\item
    The large $N$ limits of Yang-Mills and QCD-like theories are intrinsically
    interesting in their own right.
    Performing classical minimization of the large $N$ Hamiltonian,
    despite the difficulty, is surely far more achievable than using any
    foreseeably existing quantum computer to perform real time evolution
    and extraction of, say, the glueball spectrum in infinite volume 2+1 or 3+1
    dimensional Yang-Mills theory.

\end{itemize}

\section {The program ``Gordion''}
\label{sec:gordion}

Early work in the 1980's 
\cite{CSVAI,CSVAII,
Lindqwister:1988xc,
Dickens:1987ih,
Somsky:1989}
found that the coherent state variational approach works well in
simple test cases.
In non-trivial higher dimensional lattice gauge theories
the approach, by design, works well at sufficiently strong coupling.
But as the gauge coupling decreases and the correlation length grows,
it is inevitable that at some point any given truncation will cease to provide
a good approximation.
With the computing resources available in the 1980's, it was not really feasible
to reach values of the lattice gauge coupling where one would begin to see
weak-coupling behavior.

Computing capabilities have, of course, vastly increased since the 1980's.
Might it now be possible to obtain decent results at interestingly small values
of the gauge coupling in non-trivial large-$N$ lattice gauge theories?
The only way to find out is to try.
This motivated the decision to create a unified program,
efficiently written,
which can carry out all the steps in the
coherent state variational algorithm in a variety of lattice gauge theories of
interest.

The program name \textit{Gordion} is an allusion to the Greek legend
associated with Alexander the Great, who is reputed to have sliced
through a horrendously complicated knot, instead of carefully untying it,
in Gordion (Latin: Gordium, Phrygian: Gordum),
the capital city of ancient Phrygia.
Dealing with arbitrarily complicated loops is, of course, at the
core of the coherent state variational method as applied
to lattice gauge theories.

The design goal was creation of a program capable of handling
U($N$) gauge theories on translationally invariant cubic lattices
in dimensions ranging from 1 up to 4, in either Hamiltonian or
Euclidean formulations, with or without fundamental representation fermions.
More specifically:
\begin{itemize}\advance\itemsep -5pt
\item
    The lattice dimension may be 1, 2, 3 or 4, with each dimension
    either infinite, or periodically compactified.
\item
    Zero, one or two fermion flavors are allowed, with
    each fermion flavor defined as having one conjugate pair of
    fermion operators per site per flavor.
\item
    All theories under consideration are invariant under
    cubic lattice symmetries (translations, permutations, and reflections)
    together with charge conjugation (C) and time reversal.%
    \footnote
        {%
        When fermions are present,
        the size of any compactified dimension must be
        an even number of lattice spacings and, in Hamiltonian theories,
        only translations by an even number of lattice spacings are
        symmetries of the theory.
        }
\item
    None of these symmetries are spontaneously broken, and
    observables of interest are invariant under all symmetries.
\end{itemize}

The current program includes evaluation of the curvature of the
Hamiltonian and extraction of small oscillation frequencies
in any chosen point group representation, but only for
translationally invariant (zero momentum) excitations.
The program design also allows testing a factorization-based
observable approximation scheme as discussed below in Sec.~\ref{sec:approx}.

Extending the code to handle spectrum calculations at
non-zero momentum is a potential future addition,
as are calculations of decay widths or scattering amplitudes
requiring evaluation of higher derivatives about the minimum
of the large-$N$ classical Hamiltonian.

The implementation language is C++.
This choice, instead of some higher level programming language,
allows a substantially more efficient implementation.
The program effectively uses multi-core processors and parallelizes
the major time-consuming steps in the approach including
observable generation, commutator evaluation for geodesic equations,
and numerical evaluation of geodesic equations.
Although compute capabilities have vastly increased
since the 1980's,
the computational complexity of numerically
minimizing the large $N$ Hamiltonian (or Euclidean free energy)
to a given accuracy inevitably grows exponentially with the
correlation length of the theory, since the number of Wilson loops
grows exponentially with any reasonable measure of their size.
Because the continuum limit entails diverging
correlation length,
for any given level of computational resources
a more efficient implementation will allow one to
reach larger correlation lengths.

As described above, the first step in applying the
coherent state variational method, using a loop-list
truncation scheme,
involves the generation of all relevant observables
with strong coupling orders below some specified limit.
Using the new program,
Table \ref{tab:counts} shows the resulting number of 
``canonicalized'' gauge observables
(i.e., counting as a single observable all those related by
any symmetry)
with 0, 1 or 2 electric field insertions
as a function of their strong-coupling order
and lattice dimension.%
\footnote
    {%
    More detailed counts of observables with specified
    creation and expectation orders may be found in
    the implementation notes \cite{Gordion}.
    For pure loop observables, the counts shown in
    table \ref{tab:counts} agree with those shown
    in tables 1--3 of Ref.~\cite{CSVAII} for strong-coupling
    orders below 24 (labeled as order 12 in Ref.~\cite{CSVAII}),
    but there are discrepancies at order 24,
    the highest order reported in Ref.~\cite{CSVAII},
    with undercounts in the older reference.
    The reason for this discrepancy with old results
    is not known.
    For observables with electric field insertions, there
    are more extensive discrepancies with the old results.
    This, it is believed, reflects a subtlety involving
    the expectation order determination for such observables;
    see footnotes 10 and 11 in the notes \cite{Gordion}.
    }
These observable sets extend to strong coupling orders
well beyond what was possible
to generate in earlier work \cite{CSVAII}.

Much more information about the 
program \textit{Gordion} may be found in the design and implementation notes
\cite{Gordion}.
These notes, together with the program code itself, are available on the
\href{https://github.com/lgyaffe/Gordion}{Github repository}.
Interested readers are encouraged to download the program, read the
program notes, and give it a try.

\begin{table}[t]
\footnotesize
\setlength{\tabcolsep}{5pt}
\vspace*{-12pt}\centering
\hspace*{-20pt}\begin{tabular}{|c|r|r|r|r|r|r|r|}
\hline
\multicolumn{1}{|c|}{strong-coupling}
&\multicolumn{3}{c|}{2D lattice}
&\multicolumn{3}{c|}{3D lattice}
&\multicolumn{1}{c|}{4D lattice}
\\
\hhline{|~|-|-|-|-|-|-|-|}
order
& Loop
& $E$-loop
& $EE$-loop
& Loop
& $E$-loop
& $EE$-loop
& Loop
\\[2pt]
\hline
$0$ & 1 & - & - & 1 & - & - & 1
\\
\hline
$4$ & 1 & 2 & 1 & 1 & 2 & 1 & 1
\\
\hline
$8$ & 3 & 7 & 20 & 5 & 13 & 48 & 5
\\
\hline
$12$ & 17 & 102 & 506 & 63 & 431 & 1,939 & 78
\\
\hline
$16$ & 196	& 2,524 & 19,597 & 2,134 & 28,871 & 213,759 & 3,509
\\
\hline
$20$ & 3,989	& 75,952 & 888,217 & 142,093 & 2,519,588 & 28,394,853 & 349,740
\\
\hline
$24$ & 109,454 & 2,542,659 & 40,684,963 & 11,992,955 & 250,454,123 & $> 2^{32}$ & 45,996,638
\\
\hline
$28$ & 3,380,056
\\
\hhline{|-|-|}
$32$ & 111,958,945
\\
\hhline{|-|-|}
\end{tabular}
\caption
    {%
    Counts of canonicalized gauge observables of the indicated types
    and specified strong-coupling order,
    on cubic lattices of dimension 2, 3 or 4.
    \label{tab:counts}
    }
\end{table}

% The rest of this paper will focus on presenting initial results
% obtained from this new implementation of the coherent state
% variational algorithm.

\section {Results}
\label{sec:results}

Models with exactly soluble large-$N$ limits, in both Euclidean and
Hamiltonian formulations, provide instructive testing grounds
for any approach to large-$N$ dynamics which may
have more general applicability.
The performance of the coherent state variational algorithm
for one-plaquette models was examined in Refs.~\cite{CSVAI,CSVAII}.
That examination is briefly repeated here, both to show
a check on the correctness of the new implementation of the
method and to illustrate the impact of different truncations
of coherence group generators.
Following this, results are presented for two-dimensional
Euclidean Yang-Mills theory on an infinite lattice, without
using the non-local change of variables which 
allows one to reduce the theory to a product
of decoupled single-plaquette models.
This is a far more demanding test case which allows one to
study effects of observable and generator truncations 
with the full complexity of a multi-dimensional lattice.
Finally, initial results are presented for 2+1 dimensional
Hamiltonian Yang-Mills theory.

\subsection{Euclidean one-plaquette model}
\label{sec:ym1e}

The Euclidean one-plaquette model is defined by the
probability measure
\begin{equation}
    d\mu[U] \equiv Z^{-1} \, e^{-A[U]} \, dU  \,,
\end{equation}
where $dU$ denotes Haar measure on U($N$) and
the action
\begin{equation}
    A[U] \equiv \frac N \lambda \,
    \tr (2 - U - U^\dagger) \,.
\end{equation}
The partition function normalizing the measure is
defined as usual,
$
    Z \equiv \int dU \> e^{-A[U]}
$.
The free energy 
$
    F \equiv -\ln Z
$,
and the entropy
$
    S \equiv \langle A \rangle - F
$,
with $\langle A \rangle \equiv \int d\mu[U] \, A[U]$.

The large-$N$ limit may be solved by diagonalizing the matrix $U$ and then
performing a saddle-point expansion of the resulting integral over eigenvalues
\cite{Gross:1980he}.
One finds  a density of eigenvalues
\begin{equation}
    \rho_0(\theta)
    =
    \begin{cases}
	1 + \frac 2\lambda \, \cos\theta \,, & \lambda \ge 2 \,;
	\\
	\frac 4\lambda \cos \frac\theta 2 \, 
	(\frac\lambda 2 - \sin^2 \frac\theta 2 )^{1/2} \;
	\Theta(\frac\lambda 2 - \sin^2 \frac\theta 2) \,, & \lambda \le 2 \,,
    \end{cases}
\end{equation}
where $\Theta(z)$ is the unit step function.
The large-$N$ expectation value of the
unit winding Wilson loop is given by
\begin{equation}
    w_1 \equiv \lim_{N\to\infty} \int d\mu[U] \, \tfrac 1N \tr U
    = \int \frac{d\theta}{2\pi} \, \rho_0(\theta) \, \cos\theta
    =
    \begin{cases}
	\frac 1\lambda \,, & \lambda \ge 2 \,;
	\\
	1- \frac \lambda 4 \,, & \lambda \le 2 \,,
    \end{cases}
\end{equation}
while higher winding-number $k > 1$ loops have expectation values
\begin{align}
    w_k \equiv \lim_{N\to\infty} \int d\mu[U] \, \tfrac 1N \tr U^k
    &= \int \frac{d\theta}{2\pi} \, \rho_0(\theta) \, \cos (k\theta)
\nonumber\\ &{} 
    =
    \begin{cases}
	0 \,, & \lambda \ge 2 \,;
	\\
	(1 - \frac \lambda 2) 
	\left[
	    \frac {P'_k(1-\lambda)}{k(k+1)} +
	    \frac {P'_{k-1}(1-\lambda)}{k(k-1)}
	\right]
	\,, & \lambda \le 2 \,,
    \end{cases}
\end{align}
with $P'_m(z)$ the derivative of the Legendre polynomial of order $m$.
The resulting large-$N$ free energy is given by
\begin{equation}
    f \equiv
    \lim_{N\to\infty} F/N^2
    =
    \begin{cases}
	    \frac 2\lambda - 1/\lambda^2 \,, & \lambda \ge 2 \,;
	    \\
	    \frac 3 4 - \frac 12 \ln \frac \lambda 2 \,, & \lambda \le 2 \,,
    \end{cases}
\end{equation}
and possesses a third-order large-$N$ phase transition at $\lambda = 2$.
Finally,
the large-$N$ entropy
\begin{equation}
    s \equiv \lim_{N \to\infty} S/N^2 = \tfrac 2 \lambda \,  (1-w_1) - f \,.
\end{equation}

Applying the coherent state variational algorithm to this model
is straightforward.
Within the framework of the program \textit{Gordion},
this theory is viewed as living on a one-dimensional
lattice periodically compactified to a single lattice spacing.
The link variable $U$ is the Polyakov loop around this compact direction.
The strong-coupling order of a winding-$k$ loop, as defined earlier, is
just four times the winding number.

\begin{figure}[tp]
\hspace*{-1cm}%
\includegraphics[scale=0.35]{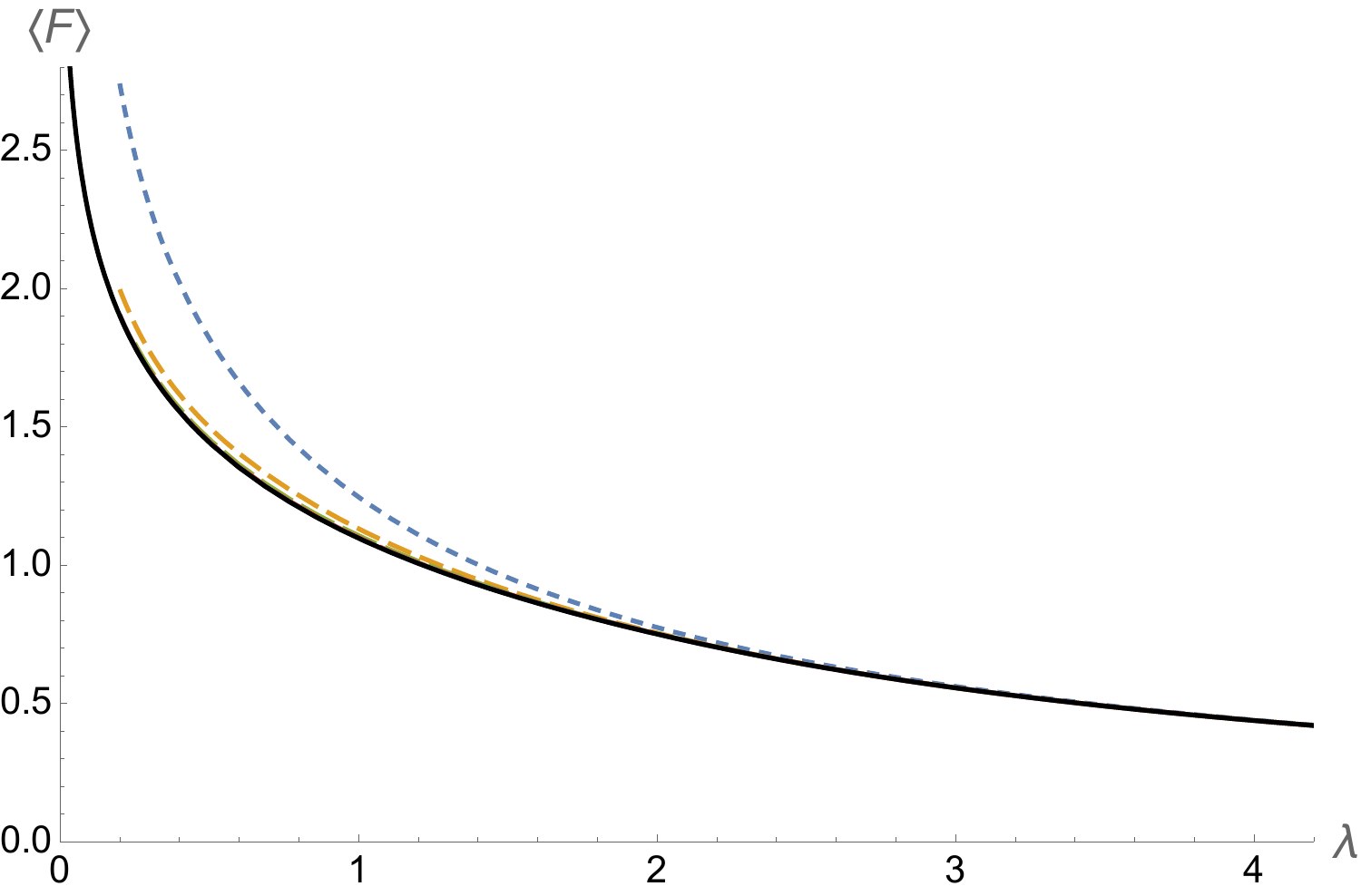}
\includegraphics[scale=0.35]{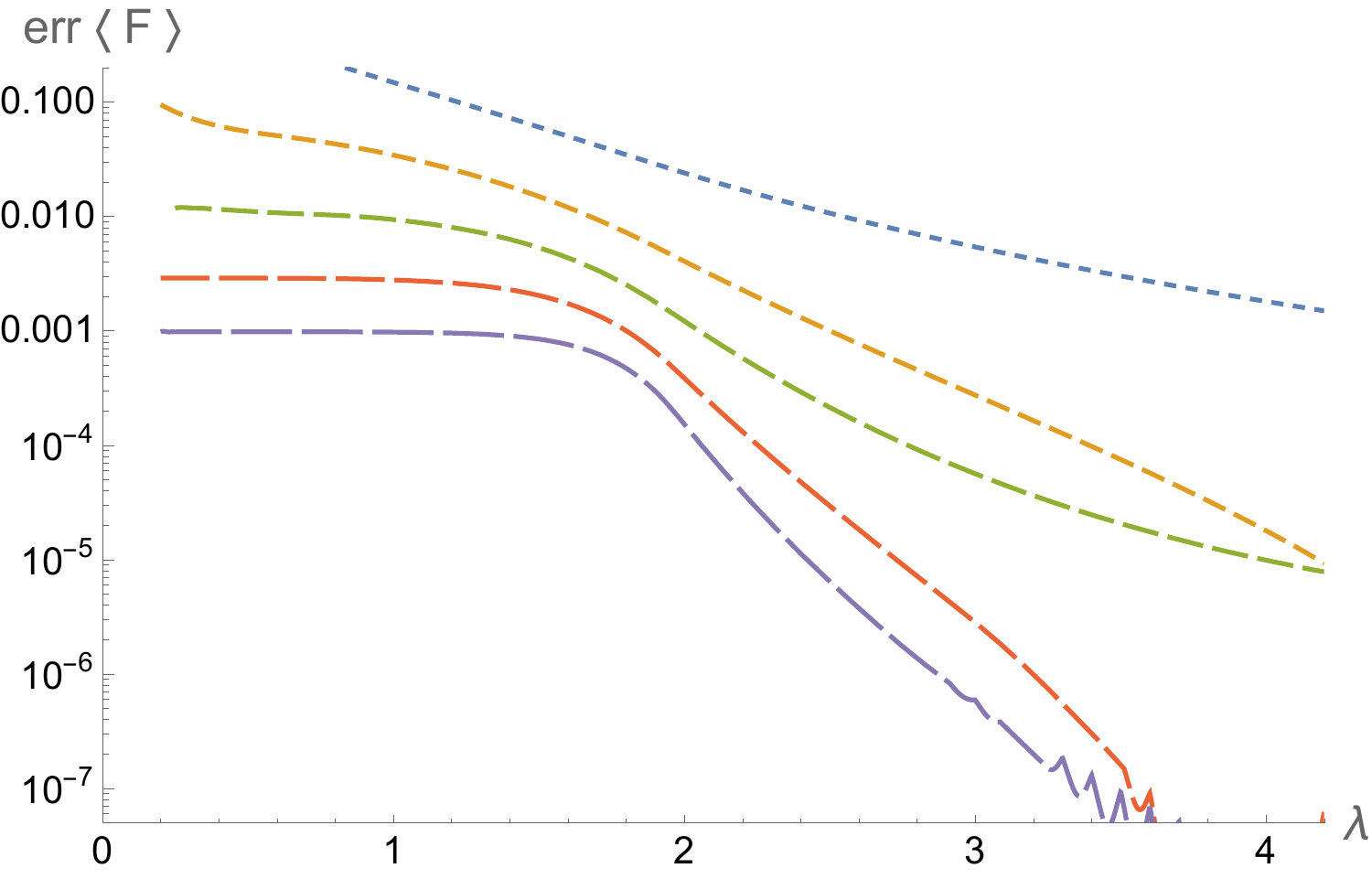}%
\hspace*{-1cm}
\caption
    {%
    Left: Free energy of the Euclidean one-plaquette model.
    Shown is the exact result (solid black line) and
    results from variational calculations using 1 through 5 generators.
    Results with three or more generators are not visually distinguishable
    from the exact curve.
    Right: Semi-log plot of the absolute error magnitude between exact and variational results
    with 1 (blue), 2 (orange), 3 (green), 4 (red), or 5 (purple) generators;
    curves with progressively longer dashes have an increasing number of generators.
    \label{fig:ym1e-f}
    }
\end{figure}

\begin{figure}[tp]
\hspace*{-1cm}%
\includegraphics[scale=0.35]{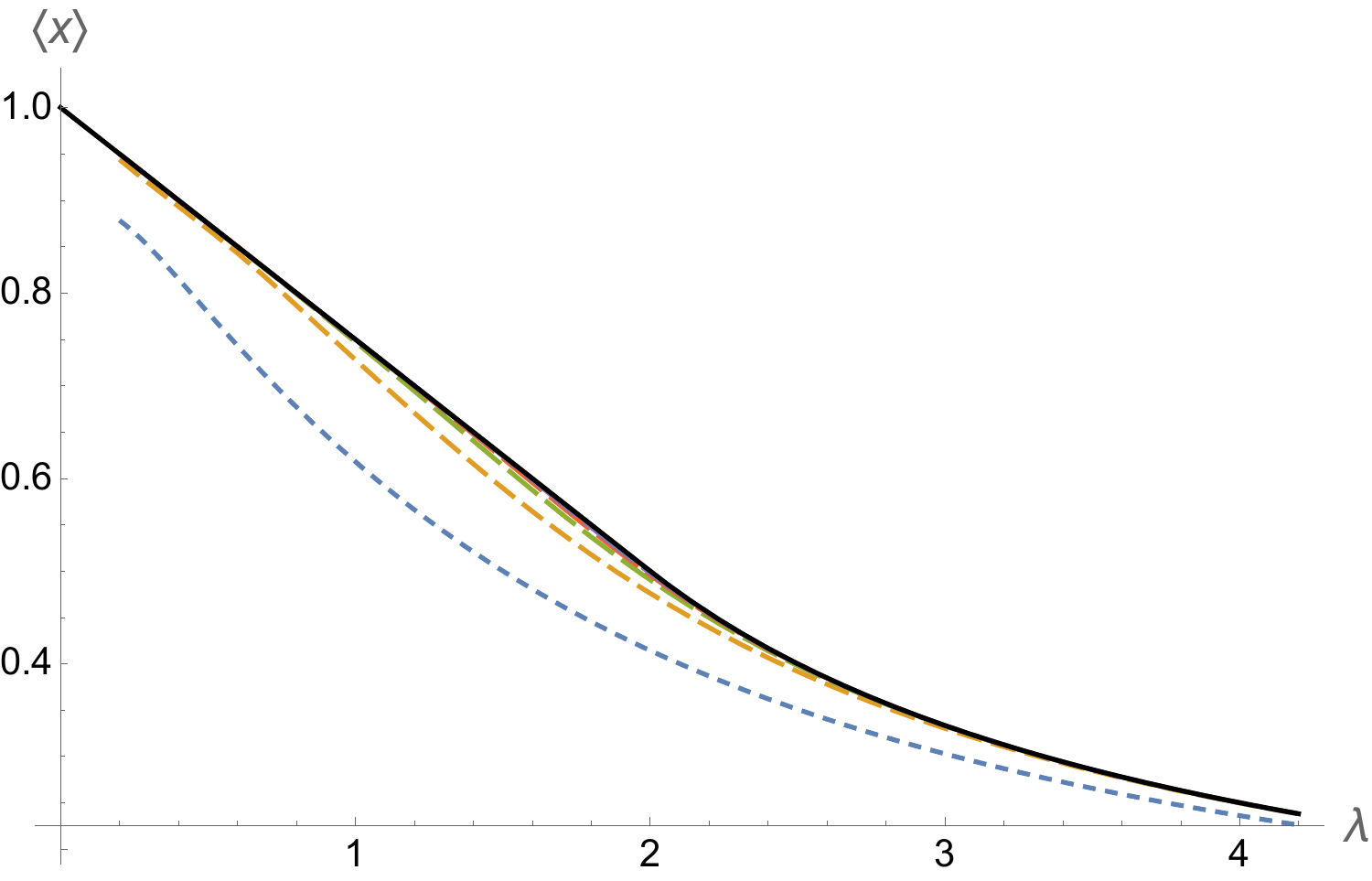}
\includegraphics[scale=0.35]{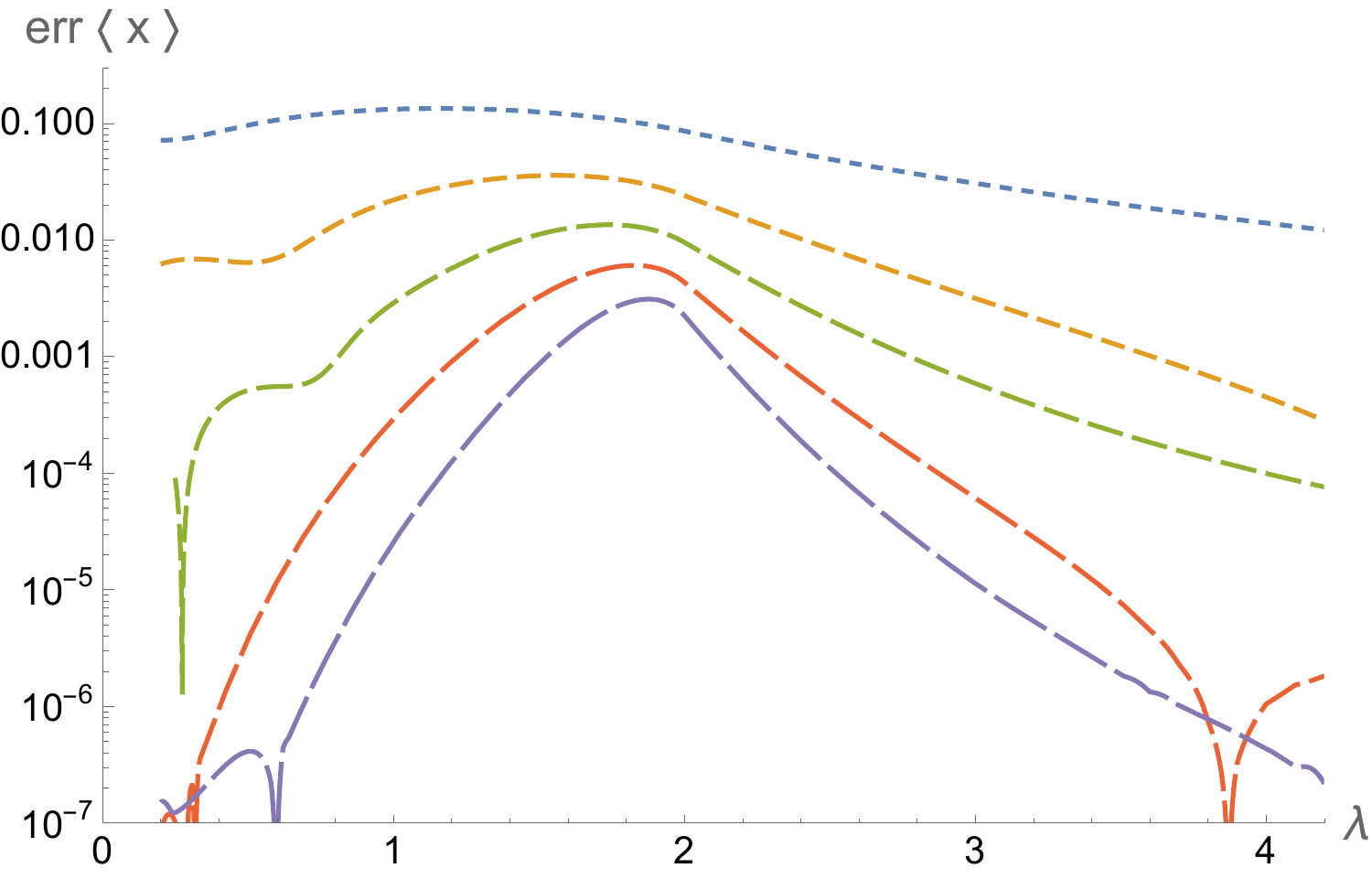}%
\hspace*{-1cm}
\vspace*{-8pt}
\caption
    {%
    Left: Winding one loop expectation value $w_1 \equiv \langle x \rangle$ in
    the Euclidean one-plaquette model.
    Shown is the exact result (solid black line) and
    results from variational calculations using 1 through 5 generators.
    Results with three generators are barely visually distinguishable
    from the exact curve.
    Right: Semi-log plot of the absolute error magnitude between exact and variational results
    with the number of generators ranging from 1 to 5.
    (The same line styles as in Fig.~\ref{fig:ym1e-f} are used.)
    \label{fig:ym1e-w1}
    }
\end{figure}

\begin{figure}[tp]
\hspace*{-1cm}%
\includegraphics[scale=0.35]{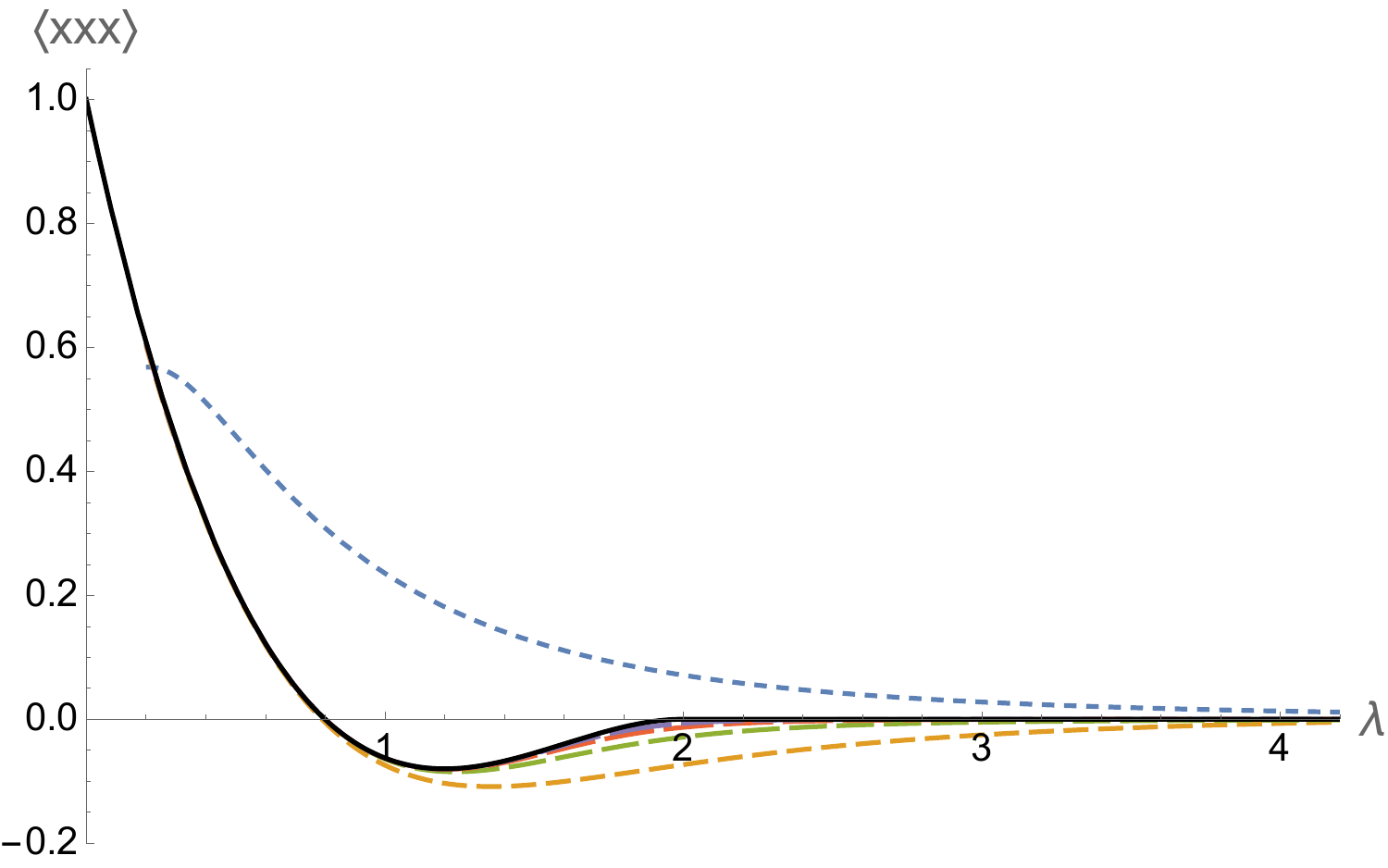}
\includegraphics[scale=0.35]{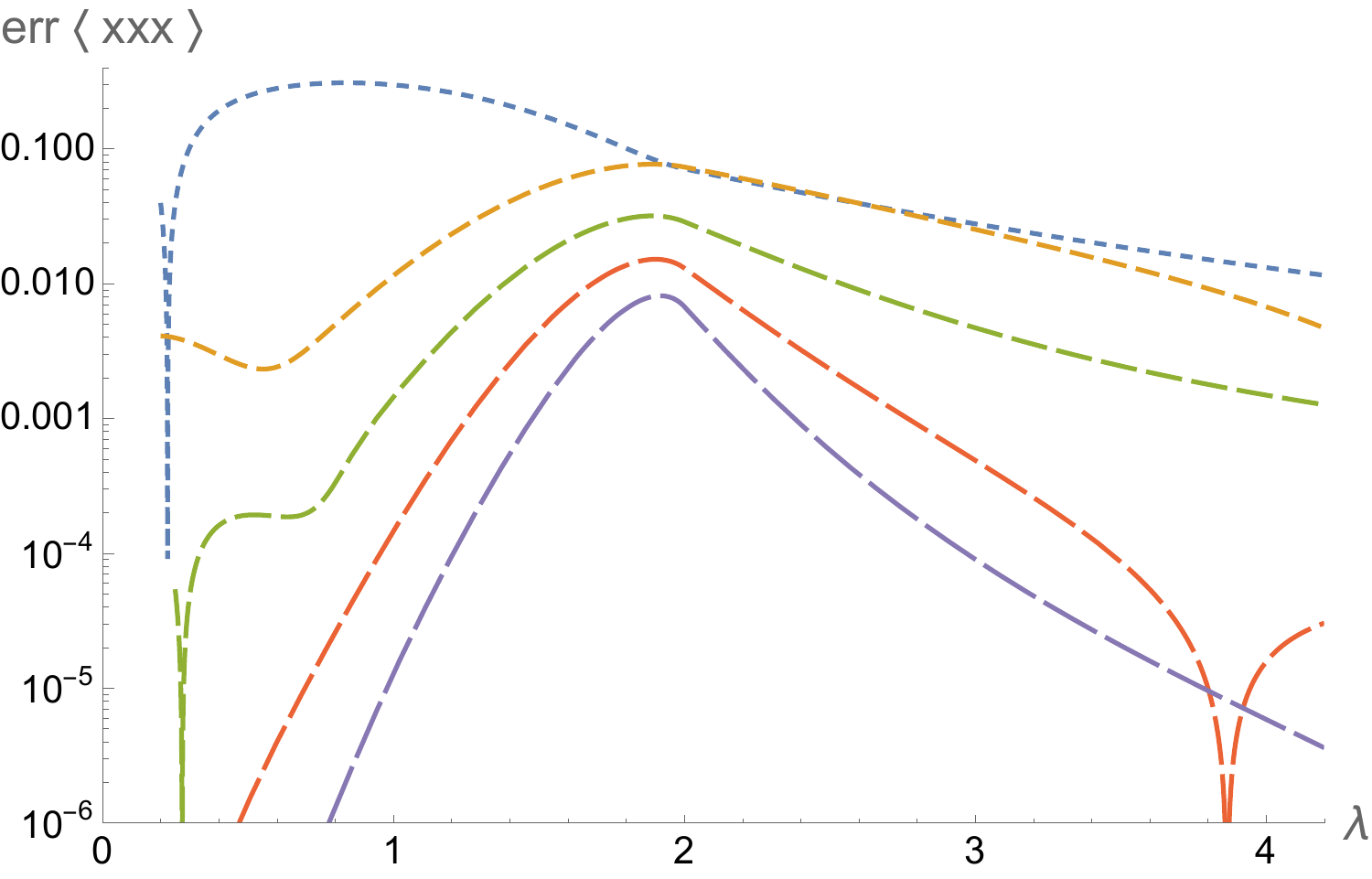}%
\hspace*{-1cm}
\vspace*{-8pt}
\caption
    {%
    Left: Winding three loop expectation $w_3 \equiv \langle xxx \rangle$ in
    the Euclidean one-plaquette model.
    Shown is the exact result (solid black line) and
    results from variational calculations using 1 through 5 generators.
    Results with four generators are barely visually distinguishable
    from the exact curve.
    Right: Semi-log plot of the absolute error magnitude between exact and variational results
    with the number of generators ranging from 1 to 5.
    (The same line styles as in Fig.~\ref{fig:ym1e-f} are used.)
    \label{fig:ym1e-w3}
    }
\end{figure}

Figure \ref{fig:ym1e-f} shows a plot of the exact large-$N$ free energy $f$
together with the results from variational calculations using coherence
group generators $\tr (E \, U^k)$ with maximal windings ranging from 1 up to 5.
Figure \ref{fig:ym1e-w1} shows a similar plot for the unit winding
loop expectation $w_1$,
while
Fig.~\ref{fig:ym1e-w3} shows results for the winding three expectation $w_3$.%
\footnote
    {%
    In all cases, loops with sufficiently high winding numbers are retained
    so that observable truncation error is negligible.
    Except at rather weak coupling, retaining two to three times the
    number of observables as generators is sufficient.
    But for couplings well below 1,
    the curvature matrix becomes increasingly poorly conditioned
    due to very small eigenvalues.
    This causes observable truncation effects to become more significant,
    with substantially larger truncations required to obtain good results.
    The development of a nearly singular curvature matrix
    is a consequence of the eigenvalue distribution (\ref{eq:ym1e-rho0})
    becoming increasingly peaked around the identity.
    This causes the actions of winding-$k$ generators on
    the eigenvalue distribution, given by
    $
	\delta\rho_0(\theta) \propto
	\cos k \theta \, (\partial \rho_0(\theta) / {\partial\theta})
    $,
    to become ever more nearly linearly dependent.
    }
The left hand side of these figures illustrate the approach of the variational
approximations to the exact results as the number of generators increases.
The logarithmic error plots on the right hand side of these figures
display the magnitudes of absolute errors between the exact results and the progressively
improving variational approximations.

The phase transition at $\lambda = 2$ is not visually apparent in the
left-hand plots of the free energy or winding-one loop expectation $w_1$,
and results with three generators for these observables
are barely distinguishable from the exact curves.
Nevertheless, the error plots on the right do show a clear change in character
across the phase transition.
For the triply winding loop,
whose expectation $w_3$ exactly vanishes above $\lambda = 2$,
the variational results inevitably smooth over the non-analytic behavior
at the phase transition.
But results with four or more generators deviate only very slightly,
just in the immediate neighborhood of the transition,
from the exact curve.
It is apparent that the variational results converge rather well to the
correct large-$N$ limits as the number of generators,
and corresponding variational parameters, increases.

\subsection{Hamiltonian one-plaquette model}
\label{sec:ym1h}

The Hamiltonian one-plaquette model is defined as
\begin{equation}
    H = N\left[ \lambda \, \tr E^2 
    + \lambda^{-1} \, \tr (2 - U - U^\dagger) \right] ,
\end{equation}
where $E$ and $U$ satisfy the specialization of the
commutation relations (\ref{eq:YMcomms}) to a single link.
The large-$N$ limit of this theory may be solved by
writing the ground state wavefunction as a function of
eigenvalues of the matrix $U$ and then factoring
out a Vandermonde determinant
of the eigenvalues, which has the effect of converting
the Schrodinger equation for eigenvalues into a theory
of free fermions
\cite{Wadia:1980cp,Jevicki:1980zq,Neuberger:1980qh}.

One finds a ground state density of eigenvalues given by
\begin{equation}
    \rho_0(\theta)
    =
    2 \sqrt{
	e + 2 \lambda^{-2} \cos\theta
    } \;
    \Theta( e + 2 \lambda^{-2} \cos\theta) \,,
\label{eq:ym1e-rho0}
\end{equation}
where $e$ is a Lagrange multiplier enforcing the normalization
constraint
$
    1 = \int \frac {d\theta}{2\pi} \, \rho_0(\theta)
$
and $\Theta(z)$ is again the unit step function.
When $\lambda > \lambda_{\rm c} \equiv 8/\pi$
(corresponding to $e > 2/\lambda^2$) the eigenvalue density
$\rho_0$ is strictly positive for all $\theta$, while for
$\lambda < \lambda_{\rm c}$ the density $\rho_0$
vanishes for some range of angles $|\theta | \ge \theta_{\rm max}(\lambda)$.

The large-$N$ expectation value of a
winding-$k$ Wilson loop is given by
\begin{equation}
    w_k \equiv \lim_{N\to\infty}  \left\langle \tfrac 1N \tr U^k \right\rangle
    = \int \frac{d\theta}{2\pi} \> \rho_0(\theta) \, \cos(k \theta) \,,
\end{equation}
where the final one-dimensional integral must be evaluated numerically.
The large-$N$ kinetic energy expectation value may be shown to be given by
\begin{equation}
    \lim_{N\to\infty}
    \left\langle \tfrac 1N \tr E^2 \right\rangle
    = \int \frac{d\theta}{2\pi} \> \rho_0(\theta)^3 \,,
\end{equation}
and the ground state energy may be expressed as
\begin{equation}
    \epsilon_{\rm g.s.}
    \equiv
    \lim_{N \to \infty}
    \left\langle H \right\rangle / N^2
    =
    2\lambda^{-1}
    + \lambda \Bigl[
	e(\lambda) - \tfrac 1{12}
	- \tfrac 16
	\int \frac{d\theta}{2\pi} \> \rho_0(\theta)^3
    \Bigr] .
\end{equation}
The energy gap, or excitation energy to the lowest excited state,
has a finite large-$N$ limit given by 
\cite{Neuberger:1980qh}
\begin{equation}
    \mu \equiv \lim_{N \to\infty} E_1 - E_0
    =
    \lambda
    \Bigl[
	\left(1 + \Theta(\lambda_{\rm c} {-} \lambda)\right) 
	\int_{-\theta_{\rm max}(\lambda)}^{\theta_{\rm max}(\lambda)}
	\frac {d\theta}{2\pi} \> \rho_0(\theta)^{-1}
    \Bigr]^{-1} .
\end{equation}
Excitation energies to higher excited states are just integer multiples of $\mu$
\cite{Neuberger:1980qh}.
For the $k$'th even-parity excited state,
\begin{subequations}
\begin{equation}
    \delta E_k^+ \equiv \lim_{N\to\infty} E_k^+ - E_0
    =
    \begin{cases}
	k \, \mu \,, & \lambda > \lambda_{\rm c} ;
	\\
	2 k \, \mu \,, & \lambda < \lambda_{\rm c} ,
    \end{cases}
\end{equation}
while for the $k$'th parity odd excited state,
\begin{equation}
    \delta E_k^- \equiv \lim_{N\to\infty} E_k^- - E_0
    =
    \begin{cases}
	k \, \mu \,, & \lambda > \lambda_{\rm c} ;
	\\
	(2 k - 1) \, \mu \,, & \lambda < \lambda_{\rm c} .
    \end{cases}
\end{equation}
\end{subequations}
The ground state energy $\epsilon_{\rm g.s.}$ is only
twice-differentiable at $\lambda = \lambda_{\rm c}$,
signaling a third order phase transition.
The energy gap $\mu$ vanishes at this phase transition
in the highly singular fashion,
\begin{equation}
    \mu \sim 2\pi \, (1 + \Theta(\lambda{-}\lambda_{\rm c})) \bigm/
	    \ln [\lambda_{\rm c} / (\lambda {-} \lambda_{\rm c})] \,.
\end{equation}

Applying the coherent state variational algorithm to this Hamiltonian model
is again straightforward.
Within the framework of the program \textit{Gordion},
this theory is viewed as living on a one-dimensional spatial
lattice periodically compactified to a single lattice spacing, so
the link variable $U$ is the spatial Wilson loop around this compact direction.

\begin{figure}[tp]
\hspace*{-1cm}%
\includegraphics[scale=0.35]{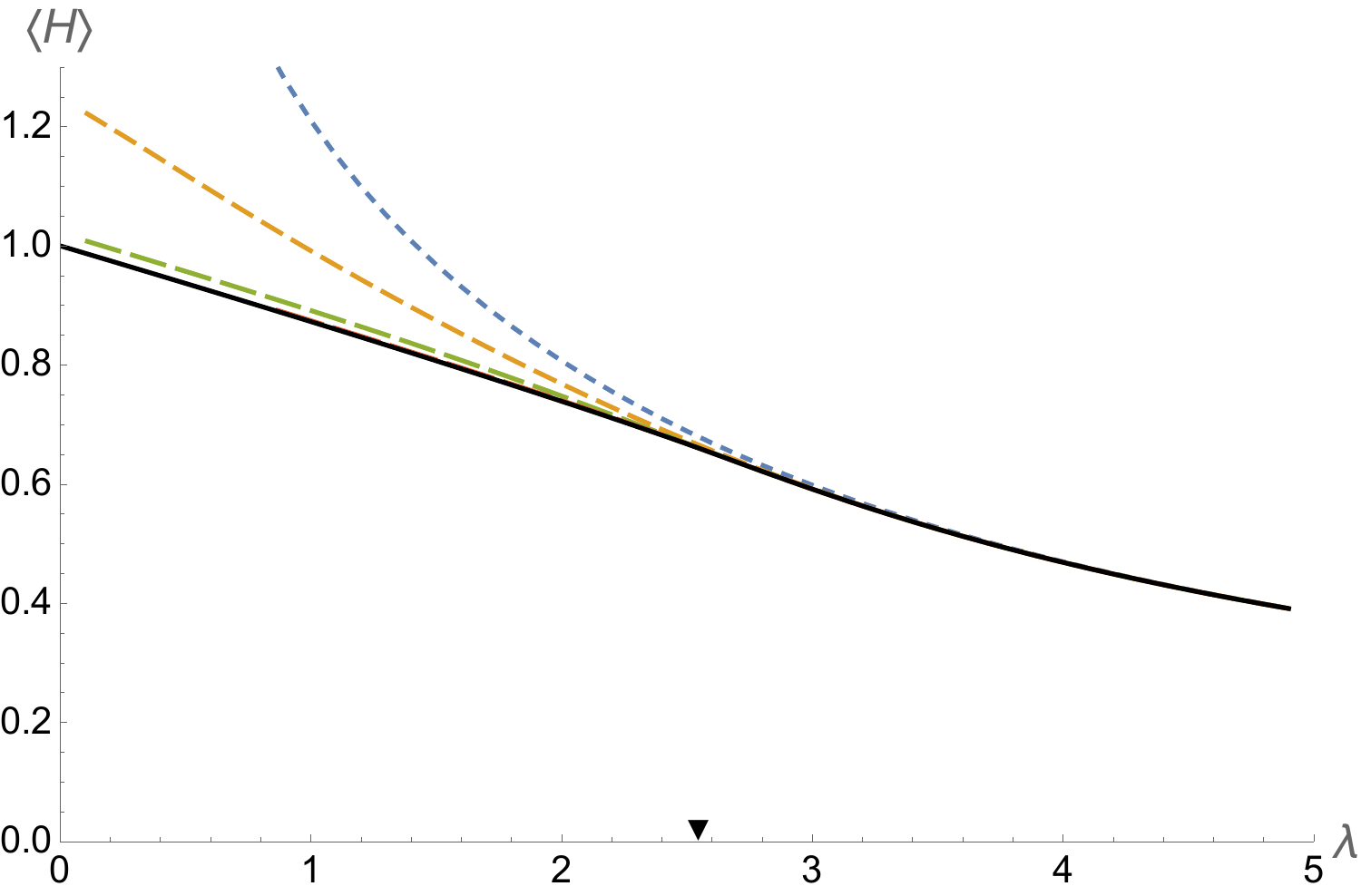}
\includegraphics[scale=0.35]{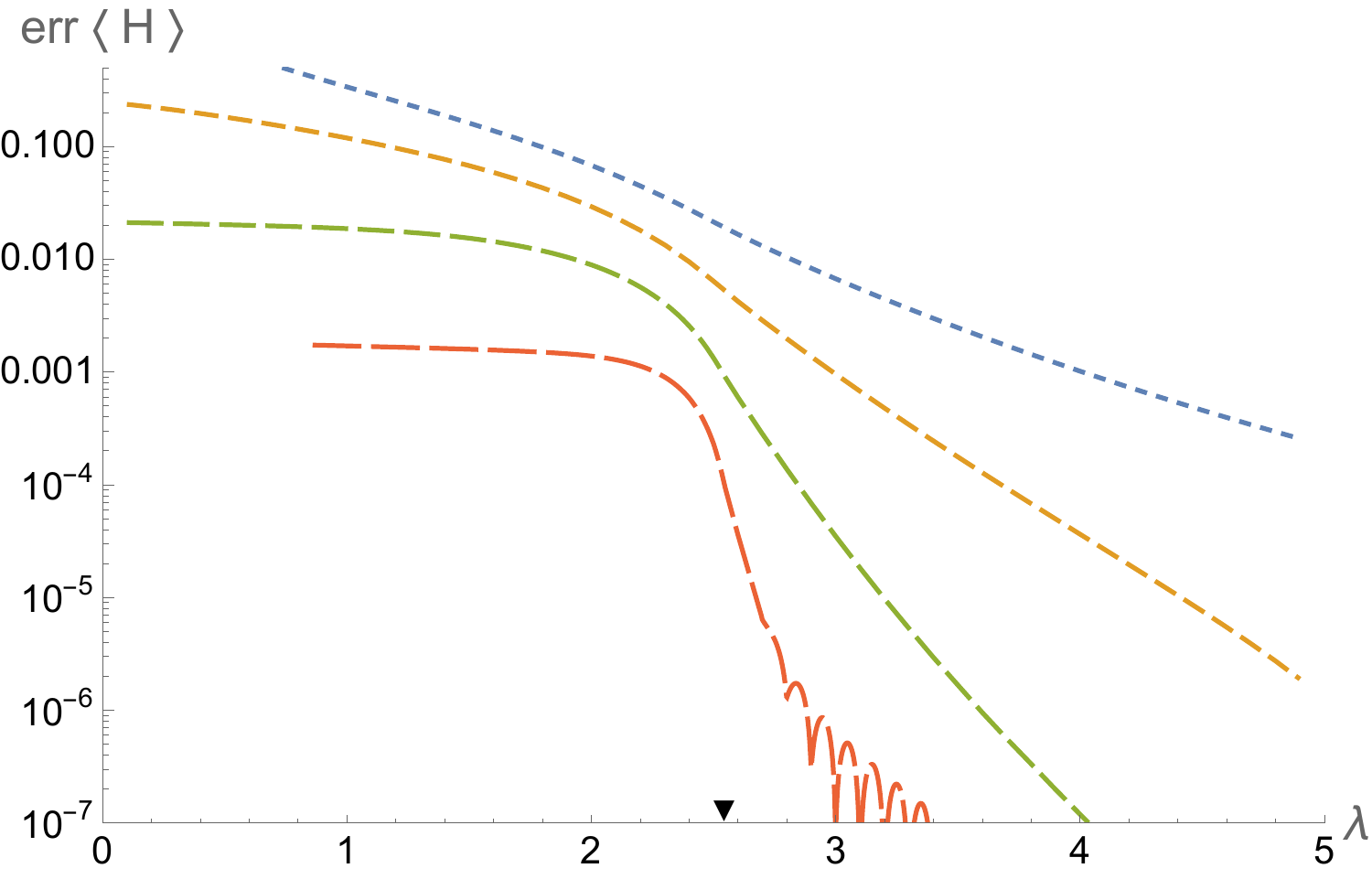}%
\hspace*{-1cm}
\vspace*{-8pt}
\caption
    {%
    Left: Ground state energy of the Hamiltonian one-plaquette model.
    Shown is the exact result (solid black line) and
    results from variational calculations using 1, 2, 4 and 8 generators.
    (The eight generator curve is indistinguishable from the exact curve.)
    Right: Semi-log plot of the absolute error magnitude between exact and variational results
    with 1 (blue), 2 (orange), 4 (green) and 8 (red) generators;
    curves with progressively longer dashes have an increasing number of generators.
    The black triangle on the $x$-axis marks the position of the phase transition.
    \label{fig:ym1h-h}
    }
\end{figure}

\begin{figure}[tp]
\hspace*{-1cm}%
\includegraphics[scale=0.35]{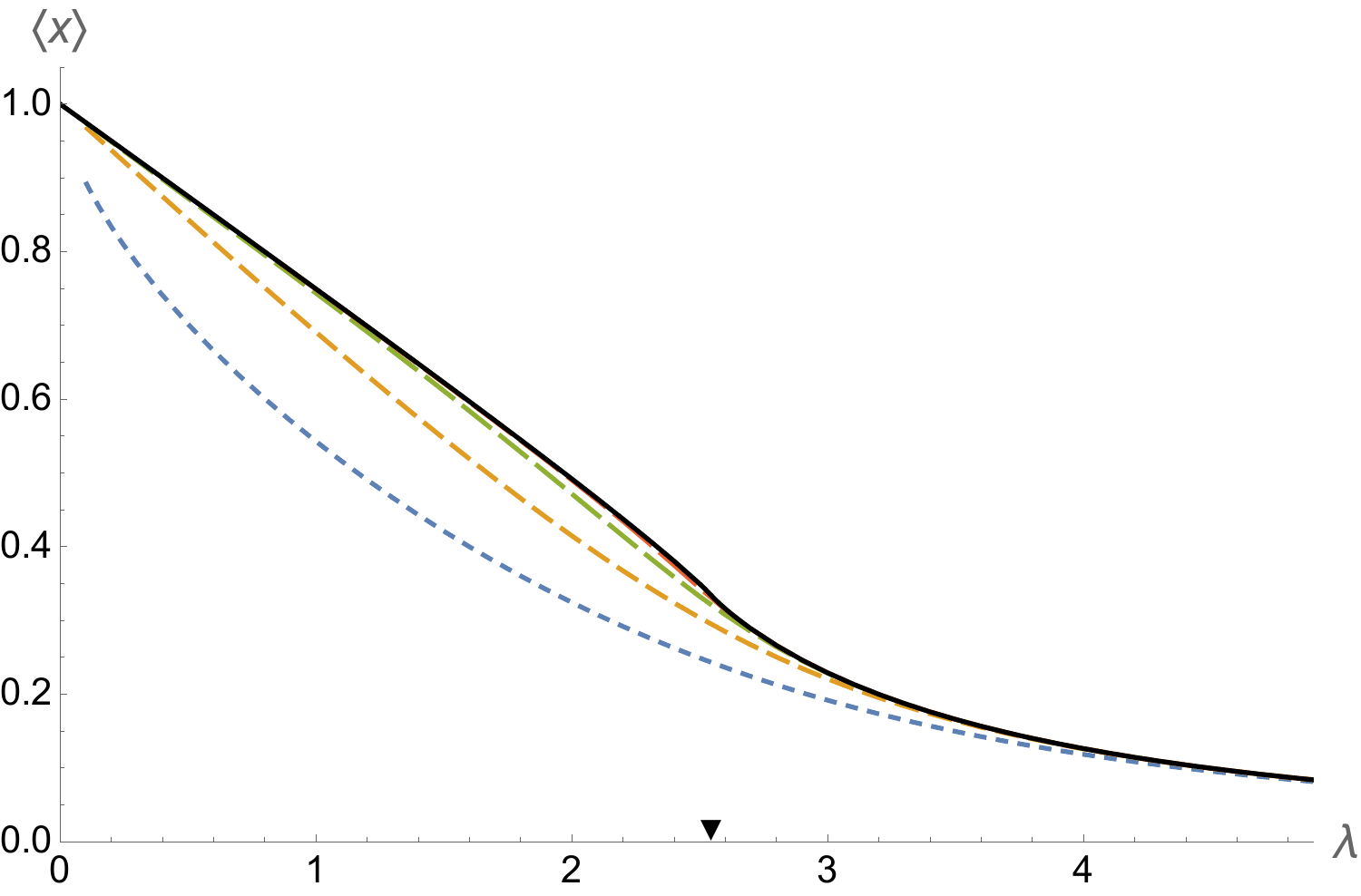}
\includegraphics[scale=0.35]{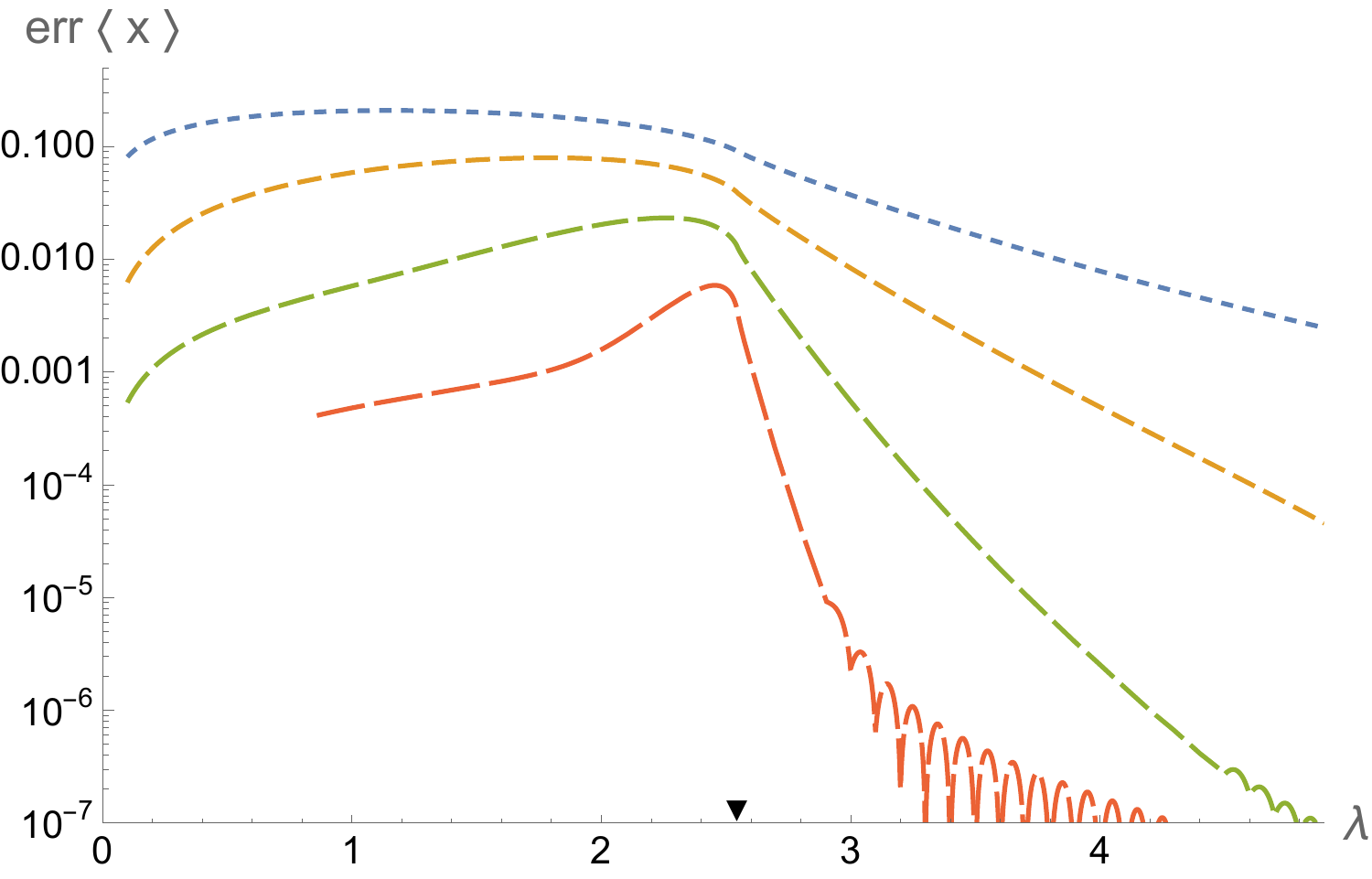}%
\hspace*{-1cm}
\caption
    {%
    Left: Single winding loop expectation value $w_1 \equiv \langle x \rangle$ in
    the Hamiltonian one-plaquette model.
    Shown is the exact result (solid black line) and
    results from variational calculations using 1, 2, 4 and 8 generators. 
    (The eight generator curve is indistinguishable from the exact curve.)
    Right: Semi-log plot of the absolute error magnitude between exact and variational results
    with 1, 2, 4 and 8 generators.
    (The same line styles as in Fig.~\ref{fig:ym1h-h} are used.)
    \label{fig:ym1h-w1}
    }
\end{figure}

\begin{figure}[tp]
\hspace*{-1cm}%
\includegraphics[scale=0.35]{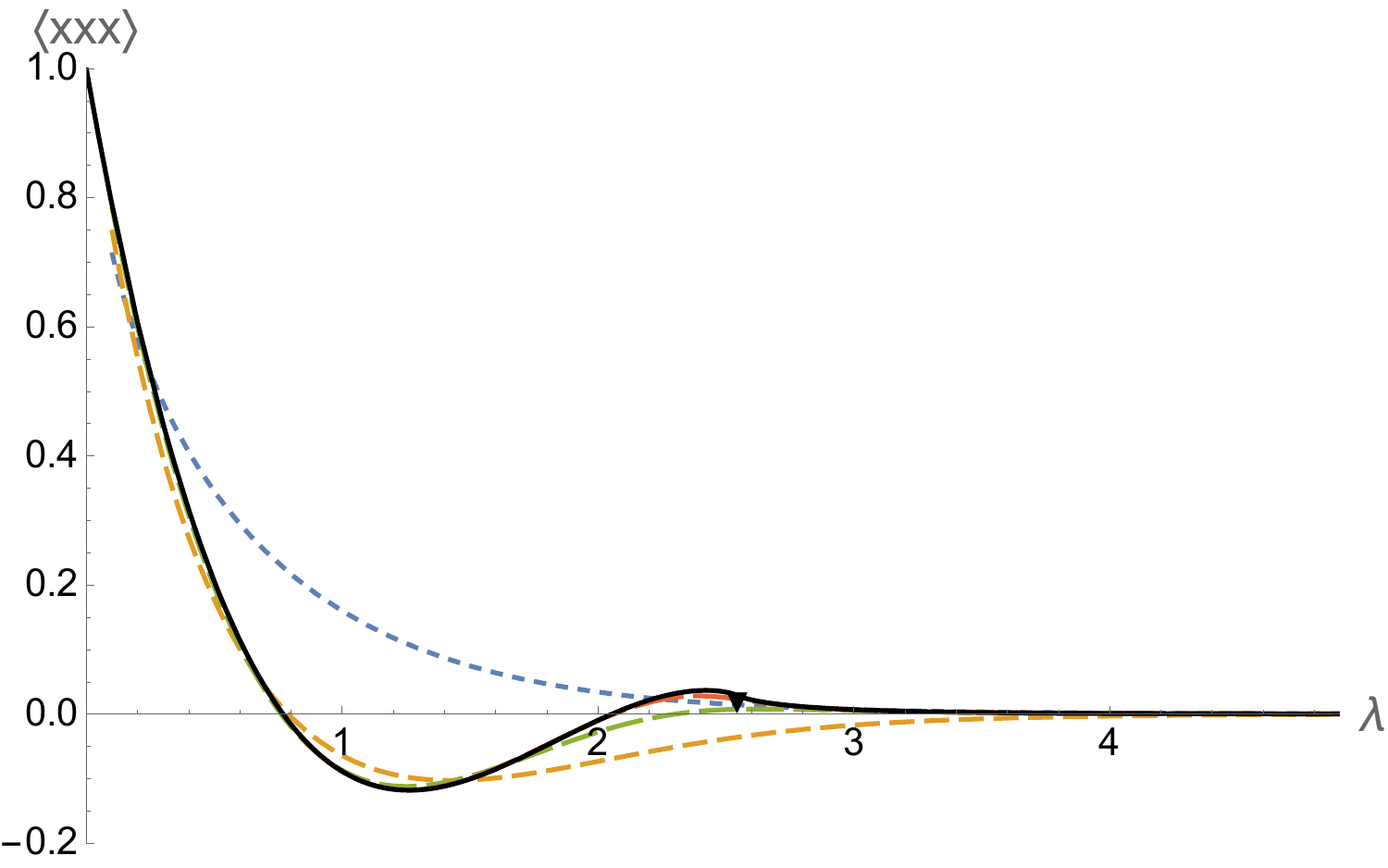}
\includegraphics[scale=0.35]{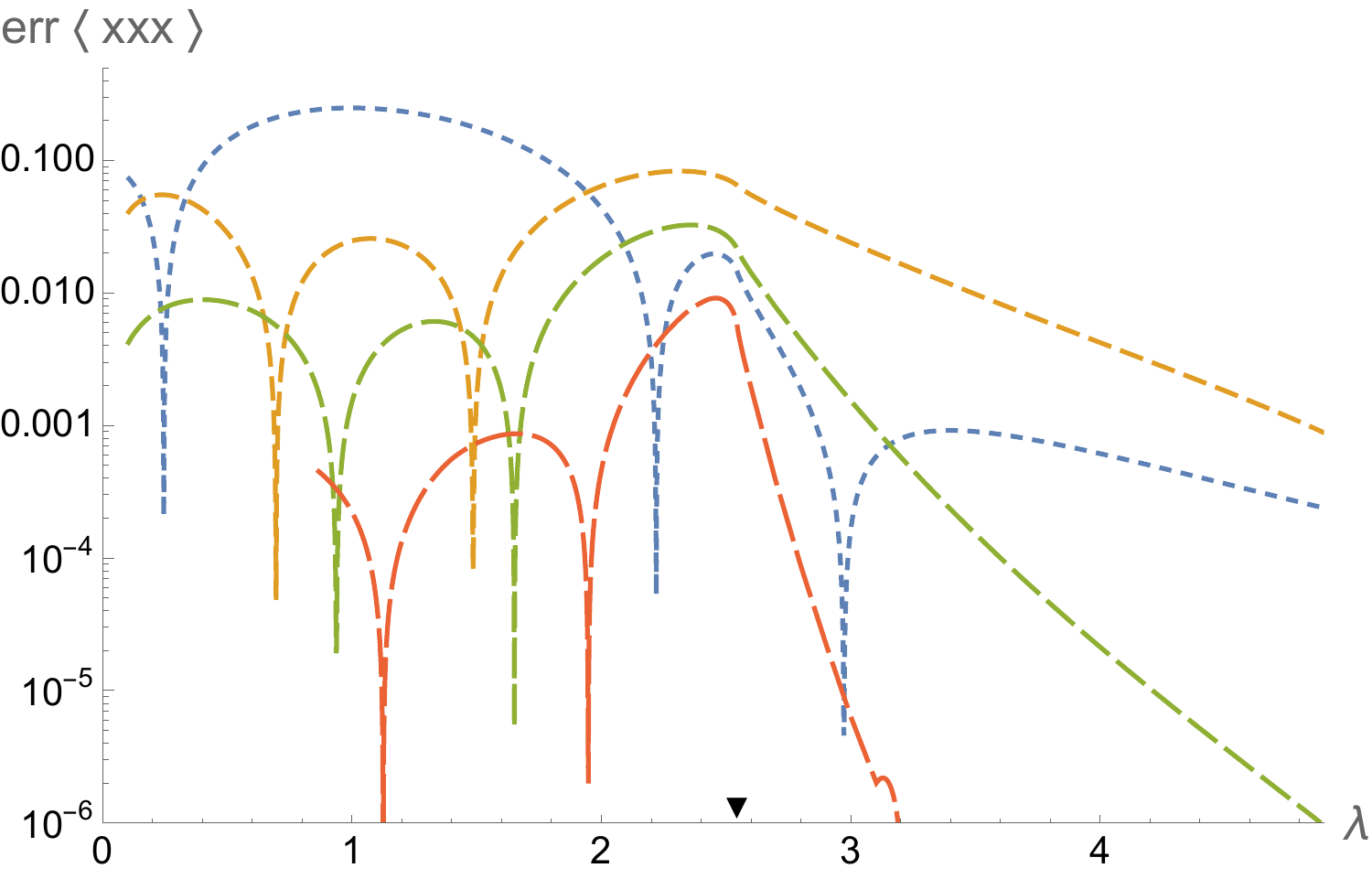}%
\hspace*{-1cm}
\vspace*{-8pt}
\caption
    {%
    Left: Triply winding loop expectation value $w_3 \equiv \langle xxx \rangle$ in
    the Hamiltonian one-plaquette model.
    Shown is the exact result (solid black line) and
    results from variational calculations using 1, 2, 4 and 8 generators.
    (The eight generator curve is only barely indistinguishable from the exact curve.)
    Right: Semi-log plot of the absolute error magnitude between exact and variational results
    with 1, 2, 4 and 8 generators.
    (The same line styles as in Fig.~\ref{fig:ym1h-h} are used.)
    \label{fig:ym1h-w3}
    }
\end{figure}

Figure \ref{fig:ym1h-h} shows a plot of the exact large-$N$ ground state energy
$\epsilon_{\rm g.s.}$
together with the results from variational calculations using coherence
group generators with maximal windings of 1, 2, 4 and 8.
Figure \ref{fig:ym1h-w1} shows a similar plot for the unit winding
loop expectation $w_1$,
while Fig.~\ref{fig:ym1h-w3} shows results for the winding three expectation $w_3$.%
\footnote
    {%
    In all cases, sufficiently many loops are retained
    so that observable truncation error is negligible.
    }
The black triangle on the $x$-axis of these plots marks the position of the phase transition.

Just as in the Euclidean one-plaquette model,
these results show good convergence as the number of generators increases.
Unsurprisingly, the non-monotonic behavior of the triply-winding expectation $w_3$
requires more variational parameters to accurately reproduce.

\begin{figure}[tp]
\vspace*{10pt}
\hspace*{-1cm}%
\includegraphics[scale=0.35]{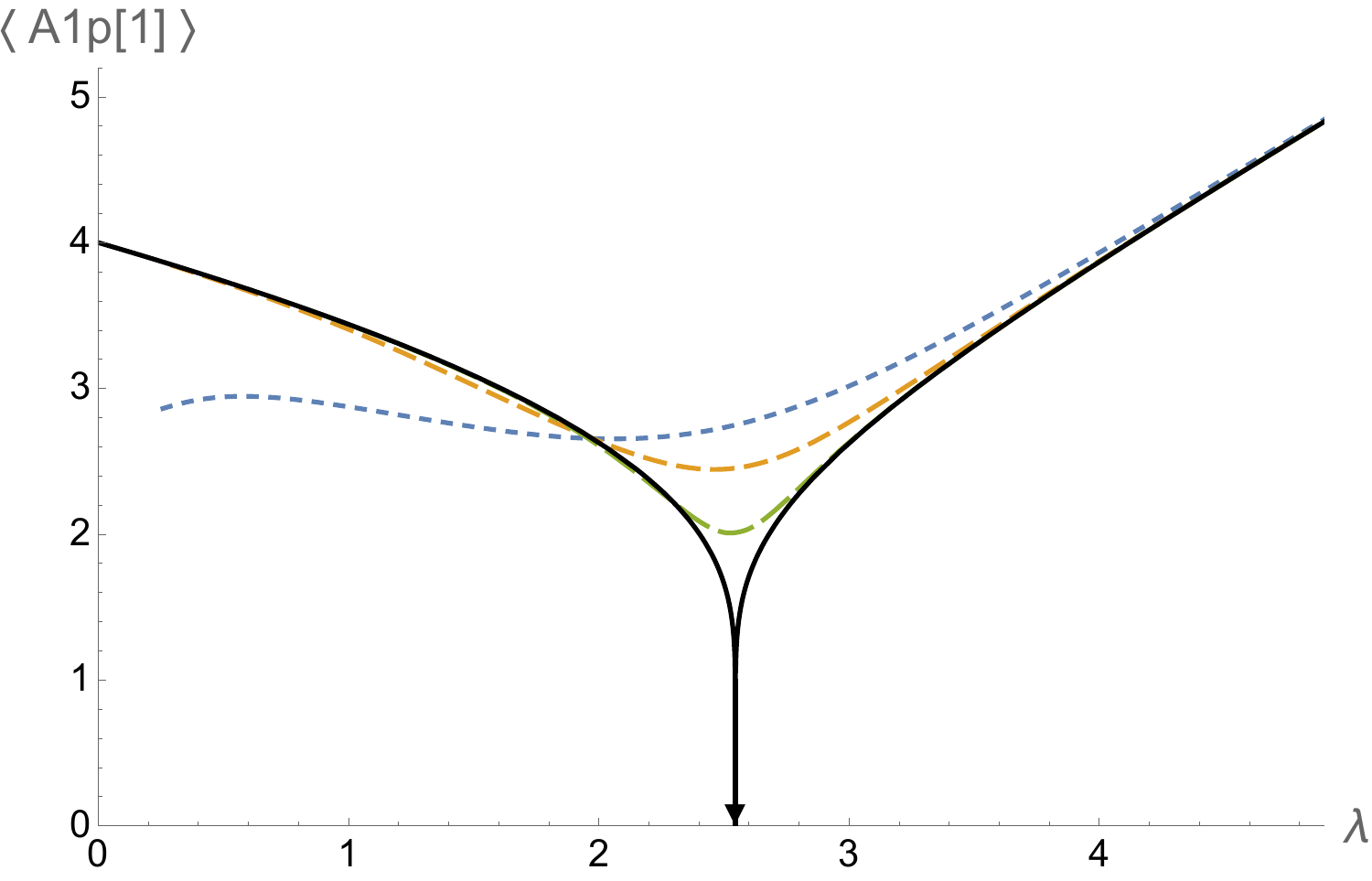}
\includegraphics[scale=0.35]{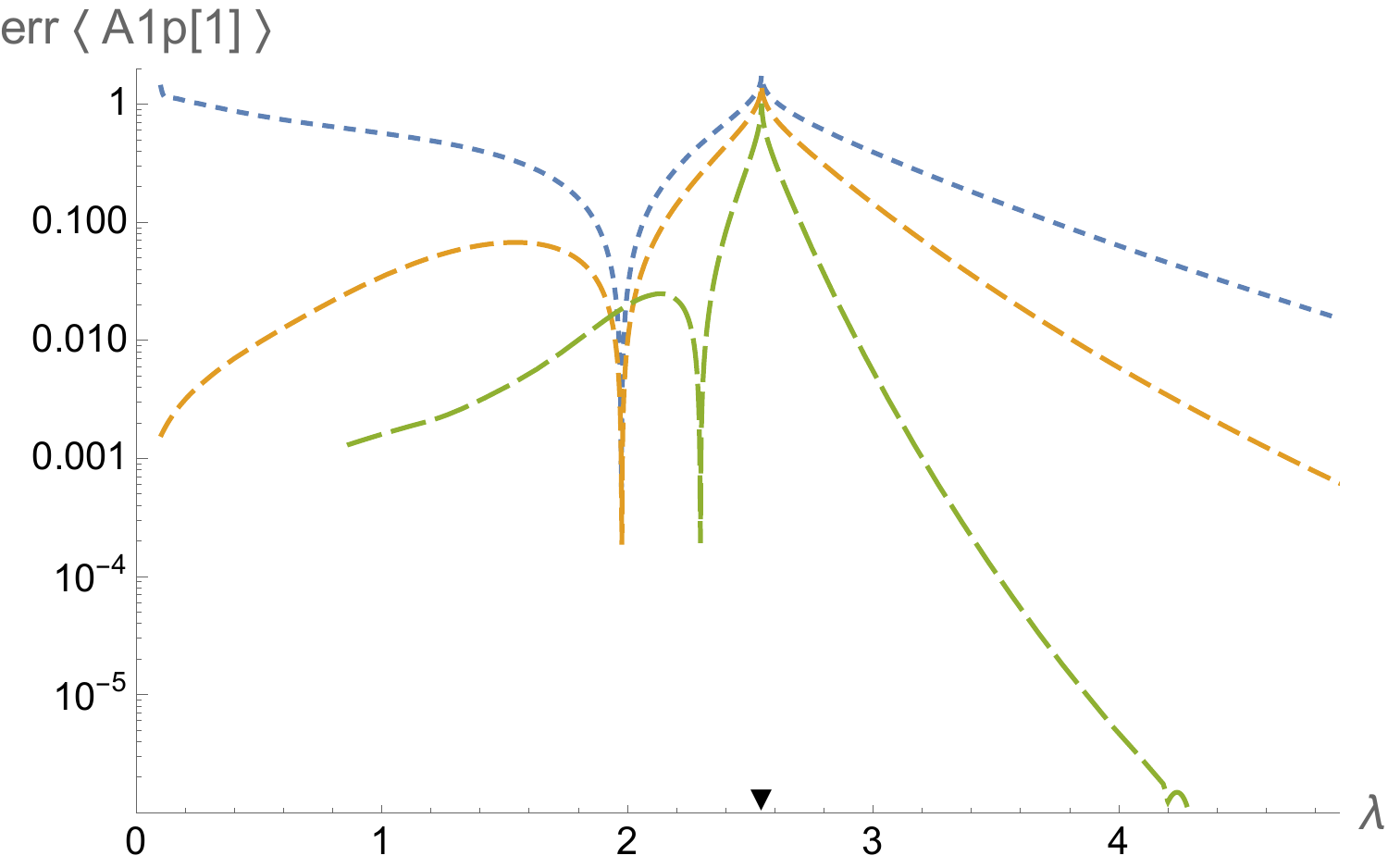}%
\hspace*{-1cm}
\vspace*{10pt}
\caption
    {%
    Left: Excitation energy to the first parity-even excited state, $\delta E^+_1$, in
    the Hamiltonian one-plaquette model.
    Shown is the exact result (solid black line) and
    results from variational calculations using 2, 4 and 8 generators.
    Right: Semi-log plot of the absolute error magnitude between exact and variational results
    with 2 (blue), 4 (orange) and 8 (green) generators;
    curves with progressively longer dashes have an increasing number of generators.
    The black triangle on the $x$-axis marks the position of the phase transition.
    \label{fig:ym1h-a1p1}
    }
\end{figure}

\begin{figure}[tp]
\vspace*{10pt}
\hspace*{-1cm}%
\includegraphics[scale=0.35]{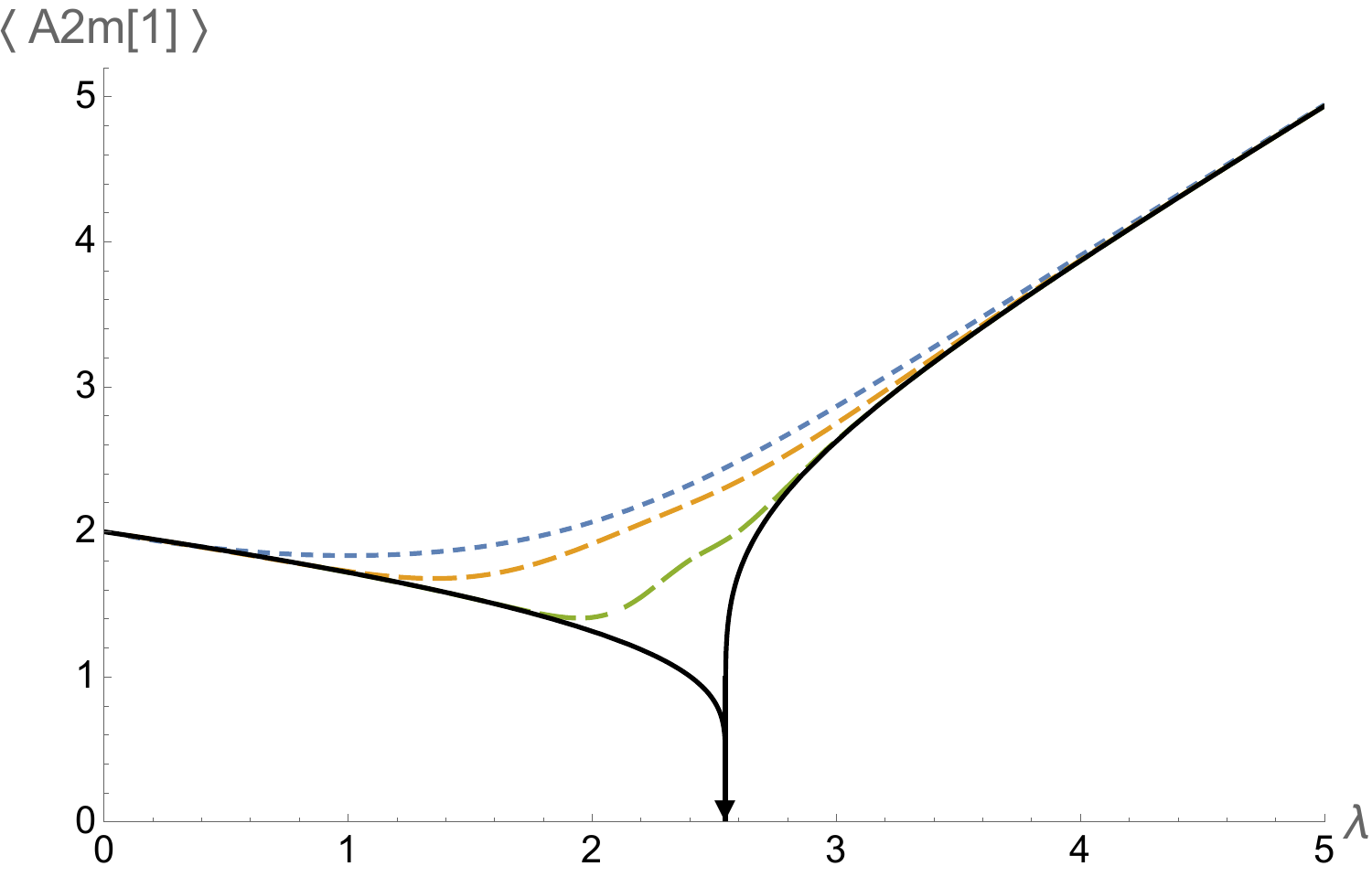}
\includegraphics[scale=0.35]{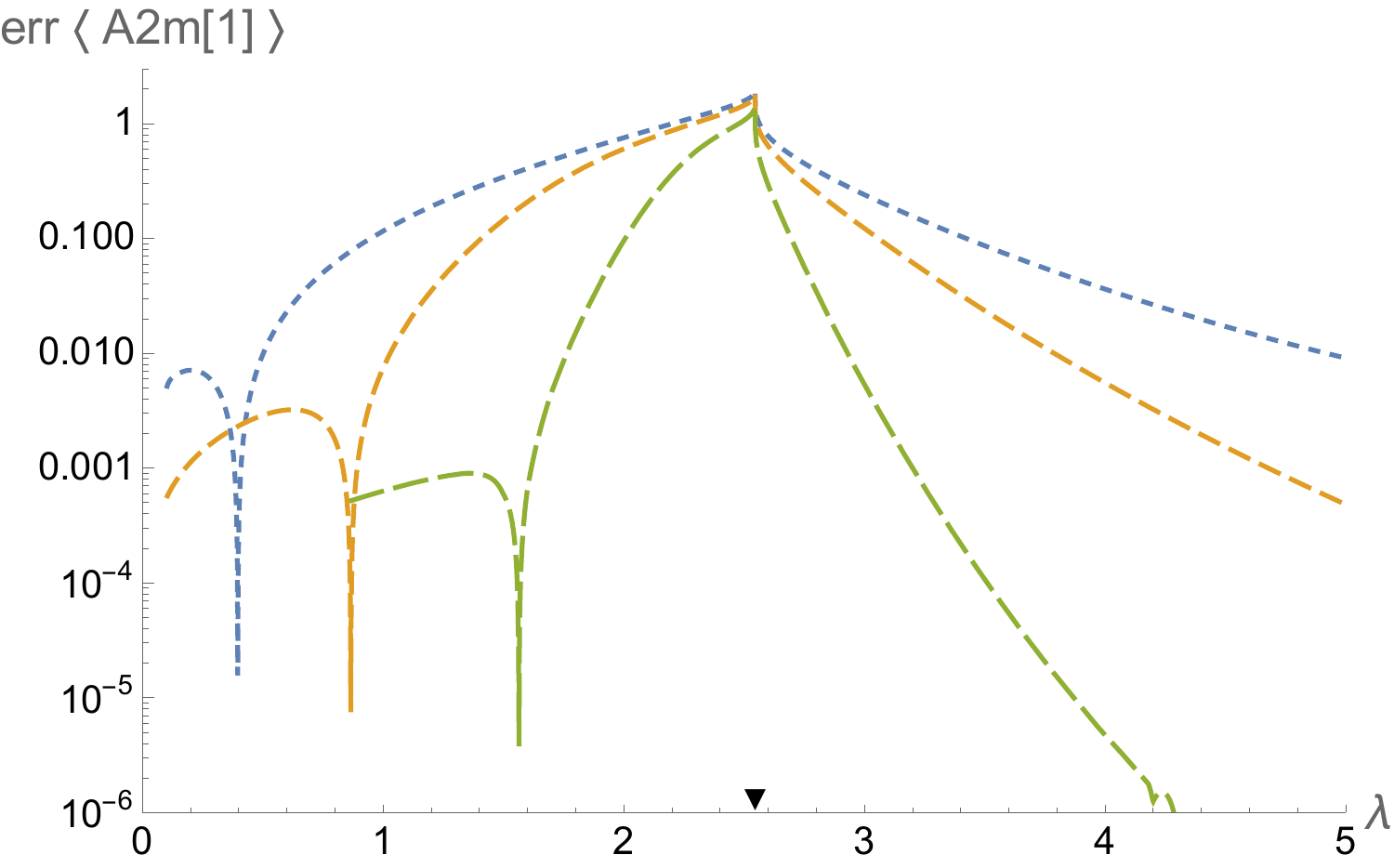}%
\hspace*{-1cm}
\vspace*{10pt}
\caption
    {%
    Left: Excitation energy to the first parity-odd excited state, $\delta E^-_1$, in
    the Hamiltonian one-plaquette model.
    Shown is the exact result (solid black line) and
    results from variational calculations using 2, 4 and 8 generators.
    Right: Semi-log plot of the absolute error magnitude between exact and variational results
    with 2, 4 and 8 generators.
    (The same line styles as in Fig.~\ref{fig:ym1h-a1p1} are used.)
    \label{fig:ym1h-a2m1}
    }
\end{figure}

Figure \ref{fig:ym1h-a1p1} shows a plot of the excitation energy to the first parity-even
excited state,
$\delta E^+_1$, together with variational results using 2, 4 or 8 generators,
while Fig.~\ref{fig:ym1h-a2m1} shows a plot of the excitation energy to the first parity-odd 
excited state,
$\delta E^-_1$, together with variational results using 2, 4 or 8 generators.

\newpage

The extremely abrupt vanishing of these excitation energies at the phase transition is
unavoidably smoothed over by the variational results.
Increasing numbers of variational parameters are needed to obtain results which
do a better job of revealing the dramatic dip in the excitation energy around the transition.

\clearpage

\subsection{2D Euclidean Yang-Mills}
\label{sec:ym2e}

As shown in Ref.~\cite{Gross:1980he}, in two-dimensional Euclidean Yang-Mills theory
on an infinite cubic lattice one may perform a change of variables from 
link variables to plaquette variables and, because there are no Bianchi identity
constraints in two dimensions, each resulting plaquette variable is a completely
independent group element.
Consequently, the partition function reduces to an uncorrelated product of identical
single plaquette integrals,
\begin{equation}
    Z = \int \prod_\ell \> dU_\ell \;
	e^{-\frac N\lambda \sum_p \tr (2 - U_{\partial p} - U_{\partial p}^\dagger)}
    =
    \prod_{i,j} \int dP_{i,j} \; e^{-\frac N\lambda \, \tr (2 - P_{i,j} - P_{i,j}^\dagger)} \,.
\end{equation}
Here $P_{i,j} \equiv U^x_{ij} \, U^y_{i+1,j} \, U^{x \, \dagger}_{i,j+1} \, U^{y \, \dagger}_{i,j}$
is the plaquette variable starting at lattice site $(i,j)$.
The inverse relation is easiest to write in an axial gauge where all $y$-directed links
are set to the identity, $U^y_{i,j} = 1$. Then (up to a $y$-independent gauge transformation):
\begin{equation}
    U^x_{i,j}
    =
    \begin{cases}
	1 \,, & j = 0 \,;
	\\
	P^\dagger_{i,j-1} \,
	P^\dagger_{i,j-2}
	\cdots
	P^\dagger_{i,0} \,, & j > 0 \,;
	\\
	P_{i,j} \,
	P_{i,j+1}
	\cdots
	P_{i,-1} \,, & j < 0 \,.
    \end{cases}
\end{equation}

One can, of course, apply the coherent state variational algorithm to each of these
independent one-plaquette models, in exactly the manner discussed above.
But coherence group generators which act on a single plaquette while leaving all other
plaquettes unchanged are, when expressed in terms of the original link variables and
single link derivatives (or electric field operators), extremely non-local.%
\footnote
    {%
    A ``plaquette'' generator $e_{i,j}^m$ satisfying
    $
	[e_{i,j}^m ,\, P_{k,l}] = \delta_{i,k} \, \delta_{j,l} \, P_{i,j}^{1-m}
    $,
    when written in terms of the original link variables
    is an infinite sum.
    One may show that
    $
	e_{i,j}^m =
	\sum_{p=0}^\infty
	N \, \tr [E^x_{i,j-p} (U^y)^p (U^y U^x U^{y\,\dagger} U^{x\,\dagger})^m (U^y)^{-p}]
    $
    is a valid form. (The site indices of the link variables are suppressed, but are
    uniquely determined by the requirement that the operator be a sensible closed loop.)
    }

A test of the coherent state variational method
which is far more challenging that the previous applications to
single plaquette models
is to attempt to solve two-dimensional Euclidean Yang-Mills theory
using coherence group generators of bounded extent
\textit{without} transforming to plaquette variables.
This mimics what can be done in higher dimensional theories.
An initial effort to do just this was made in Ref.~\cite{CSVAII}, but with
observable truncations which were more limited than what is possible to handle today.

Variational calculations were performed with generator truncations at order 2, 4 and 6,
corresponding to one, two and three plaquette operators,
and sets of observables truncated at strong-coupling orders up to 32.
These are calculations with, respectively, 1, 4 and 13 variational parameters,
and up to 115 million observables.
Table \ref{tab:ym2e} shows various statistics of the performed calculations,
including the number of terms in the resulting sets of geodesic equations.

\begin{table}
\begin{tabular}{c|c|c|c|c|c}
    ~generator~ & ~observable~ & ~\# variational~ & ~total \#~ & ~\# geodesic~ & ~~info file~~
\\
    order limit & order limit & parameters & ~observables~ & terms & size
\\
\hline
    2 & 16 & 1 & 219 & --- & ---%
\footnote
    {
    These calculations used order 4 system information
    files with forth order generators turned off during minimization.
    }
\\
    2 & 20 & 1 & 4,208 & --- & ---\footnotemark[1]
\\
    2 & 24 & 1 & 113,662 & --- & ---\footnotemark[1]
\\
    2 & 28 & 1 & 3,493,718 & --- & ---\footnotemark[1]
\\
\hline
    4 & 16 & 4 & 219 & $3.4 \times 10^3$ & 116\, KB
\\
    4 & 20 & 4 & 4,208 & $9.8 \times 10^4$ & 1.98\, MB
\\
    4 & 24 & 4 & 113,662 & $3.3 \times 10^6$ & 62.6\, MB
\\
    4 & 28 & 4 & 3,493,718 & $1.2 \times 10^8$ & 2.22\, GB
\\
    4 & 32 & 4 & 115,452,663 & $4.5 \times 10^9$ & 80.6\, GB
\\
\hline
%    6 & 16 & 13 & 219 & $8.8 \times 10^3$ & 736\, KB
%\\
    6 & 20 & 13 & 4,208 & $2.5 \times 10^5$ & 6.91\, MB
\\
    6 & 24 & 13 & 113,662 & $8.1 \times 10^6$ & 158\, MB
\\
    6 & 28 & 13 & 3,493,718 & $2.9 \times 10^8$ & 4.7\, GB
\\
    6 & 32 & 13 & 115,452,663 & $1.1 \times 10^{10}$ & 175\, GB
\end{tabular}
\caption
    {%
    Statistics of variational calculations performed
    in two-dimensional Euclidean Yang-Mills theory.
    The penultimate column gives the total number of terms
    in the complete set of geodesic equations for the
    given calculation, while the final column gives the
    size of the system information file which records the
    selected sets of observables and generators,
    and the resulting expressions for the free energy
    gradient, curvature, and geodesic equations.
    \label{tab:ym2e}
    }
\end{table}

\begin{figure}[tp]
\hspace*{-1cm}%
\includegraphics[scale=0.35]{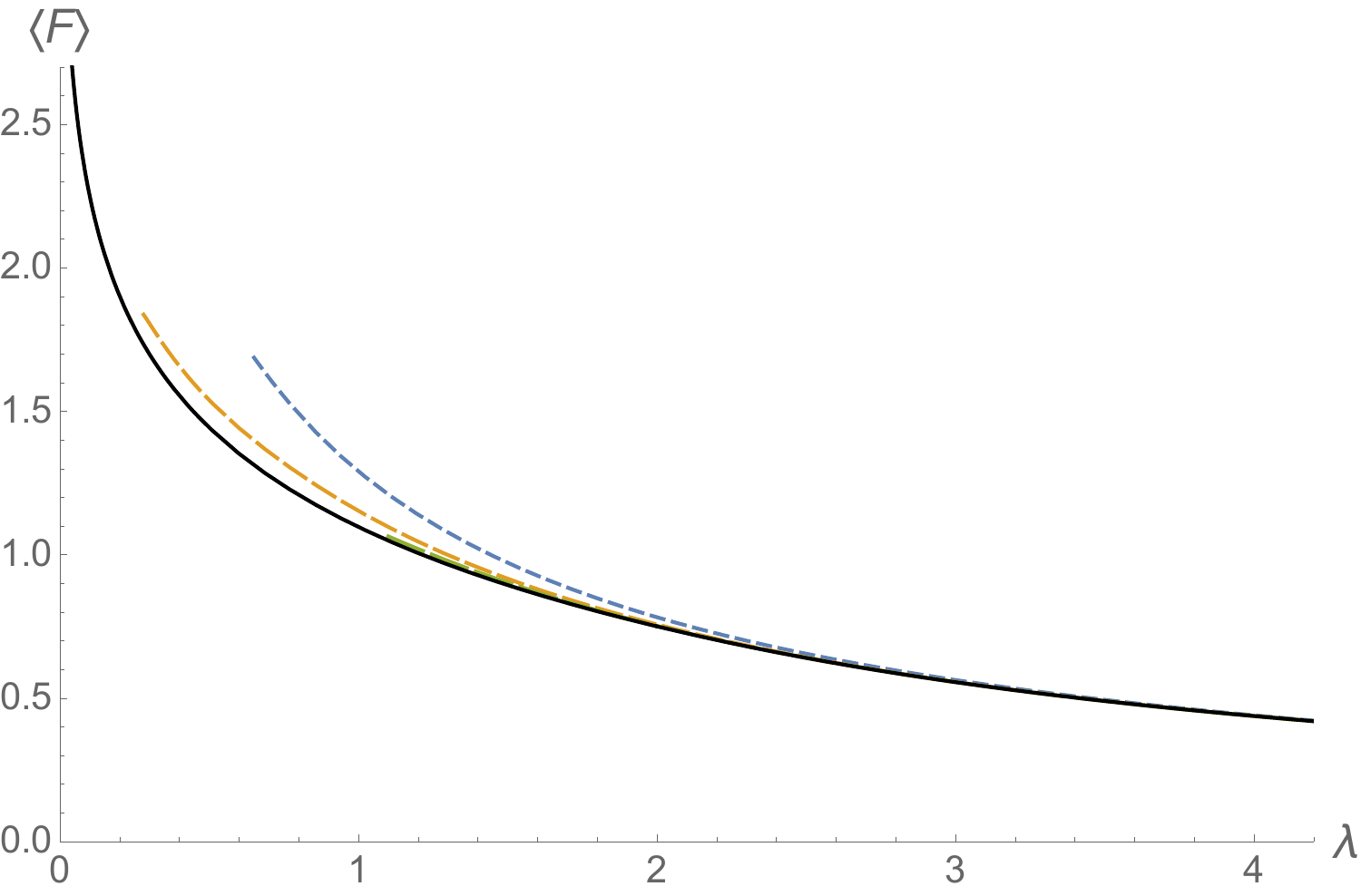}%
\includegraphics[scale=0.35]{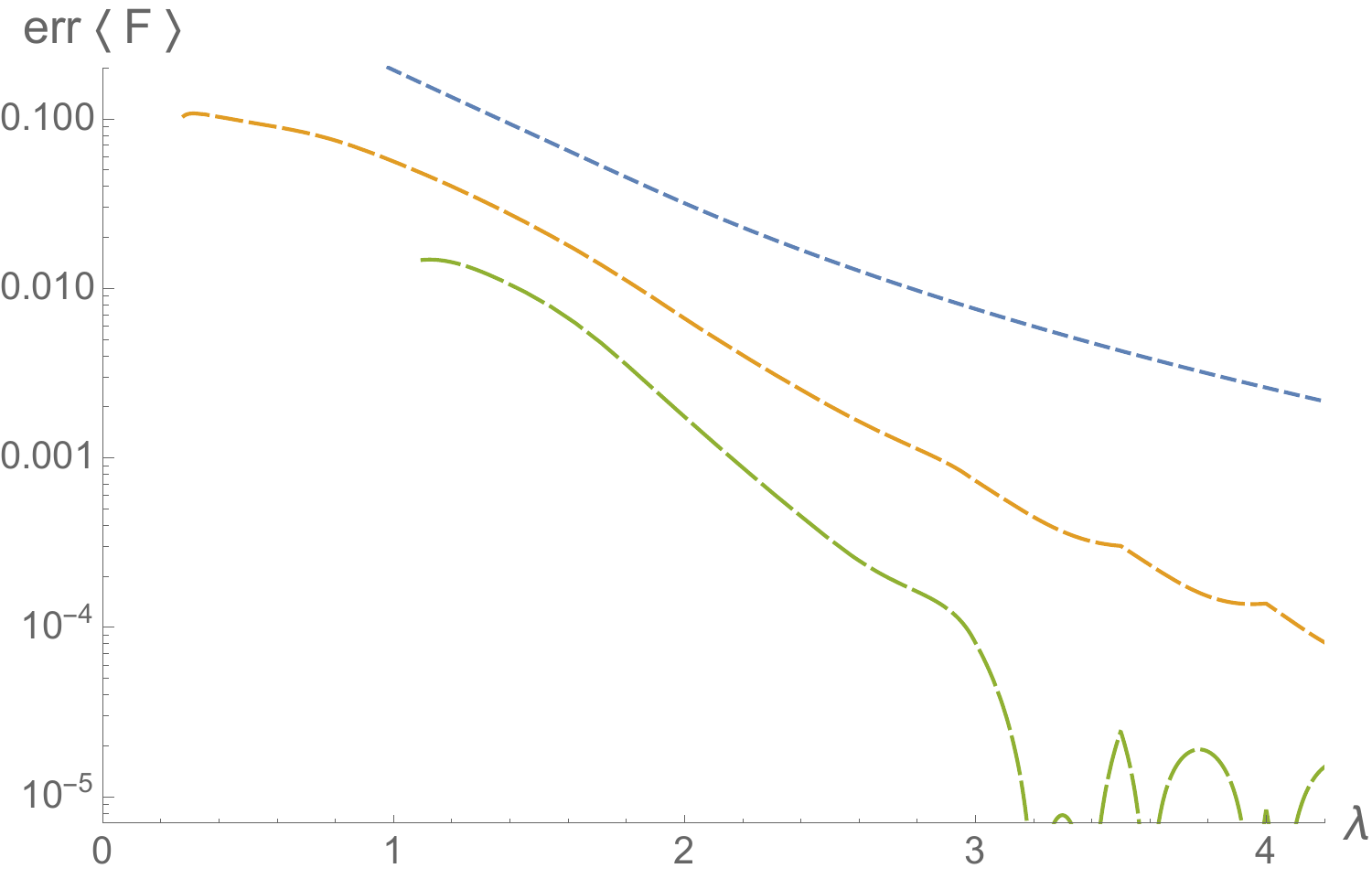}%
\hspace*{-1cm}
\caption
    {%
    Left: Free energy of two-dimensional Euclidean Yang-Mill theory.
    Shown is the exact result (solid black line) and
    results from variational calculations using order 2, 4 and 6 generators
    and observable truncation at order 28.
    Right: Semi-log plot of the absolute error magnitude between exact and variational results
    with order 2, 4 and 6 generators.
    (Line styles are the same as in earlier figures.)
    \label{fig:ym2e-F}
    }
\end{figure}

Figure \ref{fig:ym2e-F} shows the results for the free energy obtained from
variational calculations with order 2, 4 and 6 generators
and observable truncation at order 28.
In the left-hand plot of the free energy, the green order 6 curve (with longest
dashes) is barely distinguishable from the exact result.

\begin{figure}[tp]
\hspace*{-1cm}%
\includegraphics[scale=0.35]{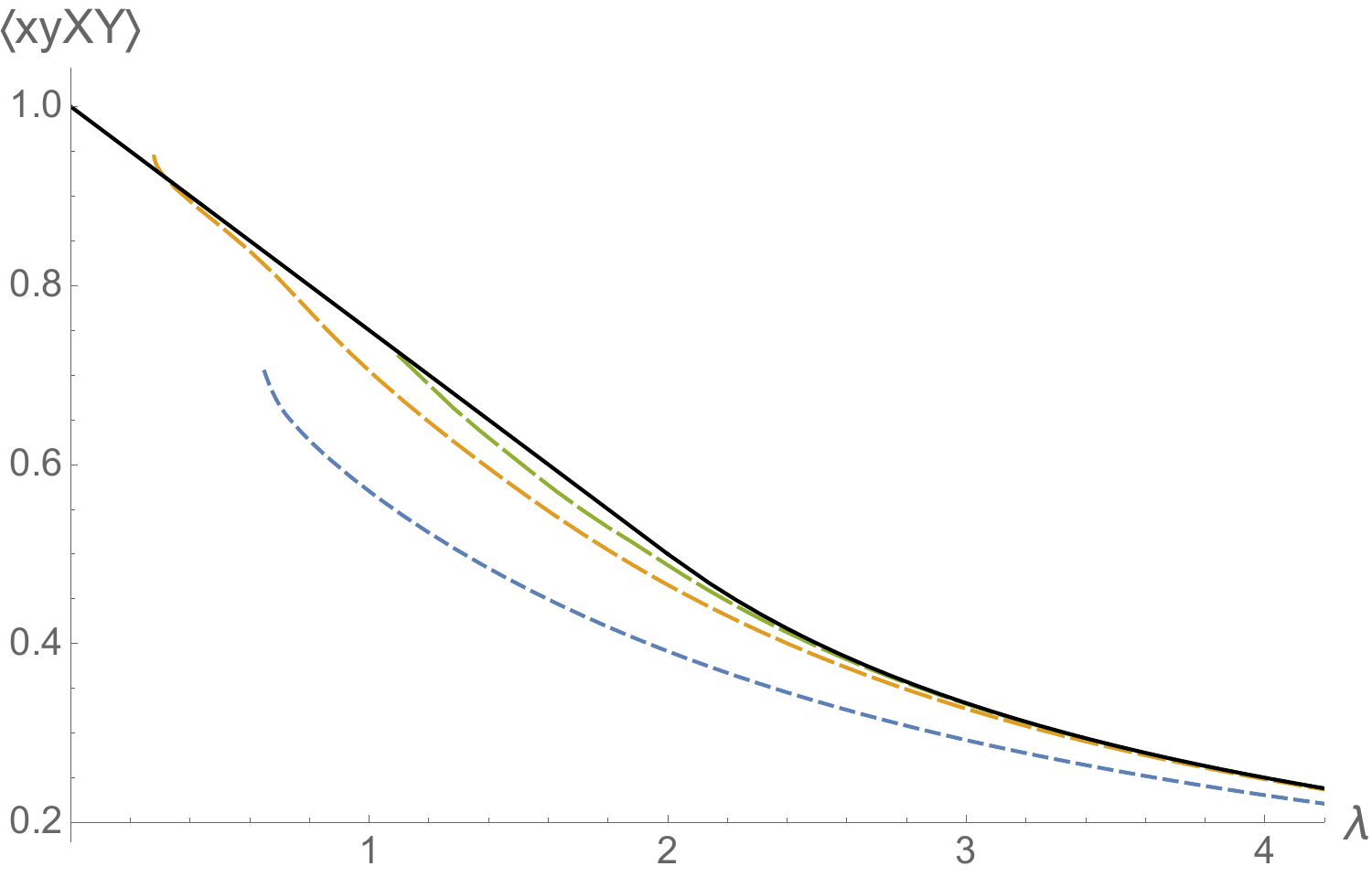}%
\includegraphics[scale=0.35]{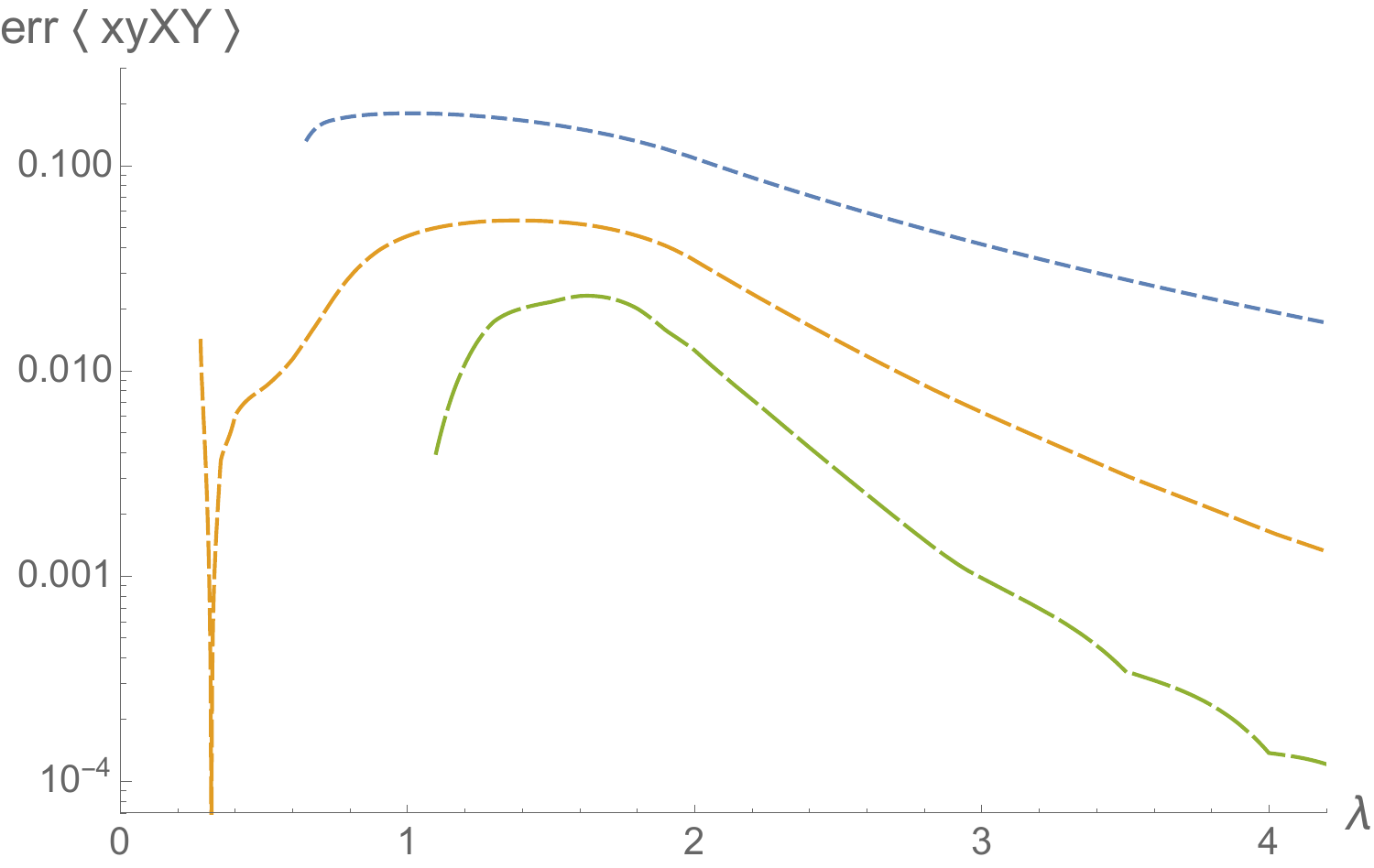}%
\hspace*{-1cm}
\caption
    {%
    Left: Expectation value of the single plaquette
    $\langle \textit{xyXY} \rangle$
    in two-dimensional Euclidean Yang-Mill theory.
    Shown is the exact result (solid black line) and
    results from variational calculations using order 2, 4 and 6 generators
    and observable truncation at order 28.
    Right: Semi-log plot of the absolute error magnitude between exact and variational results
    with order 2, 4 and 6 generators.
    (Line styles as in earlier figures.)
    \label{fig:ym2e-xyXY}
    }
\end{figure}

\begin{figure}[tp]
\hspace*{-1cm}%
\includegraphics[scale=0.35]{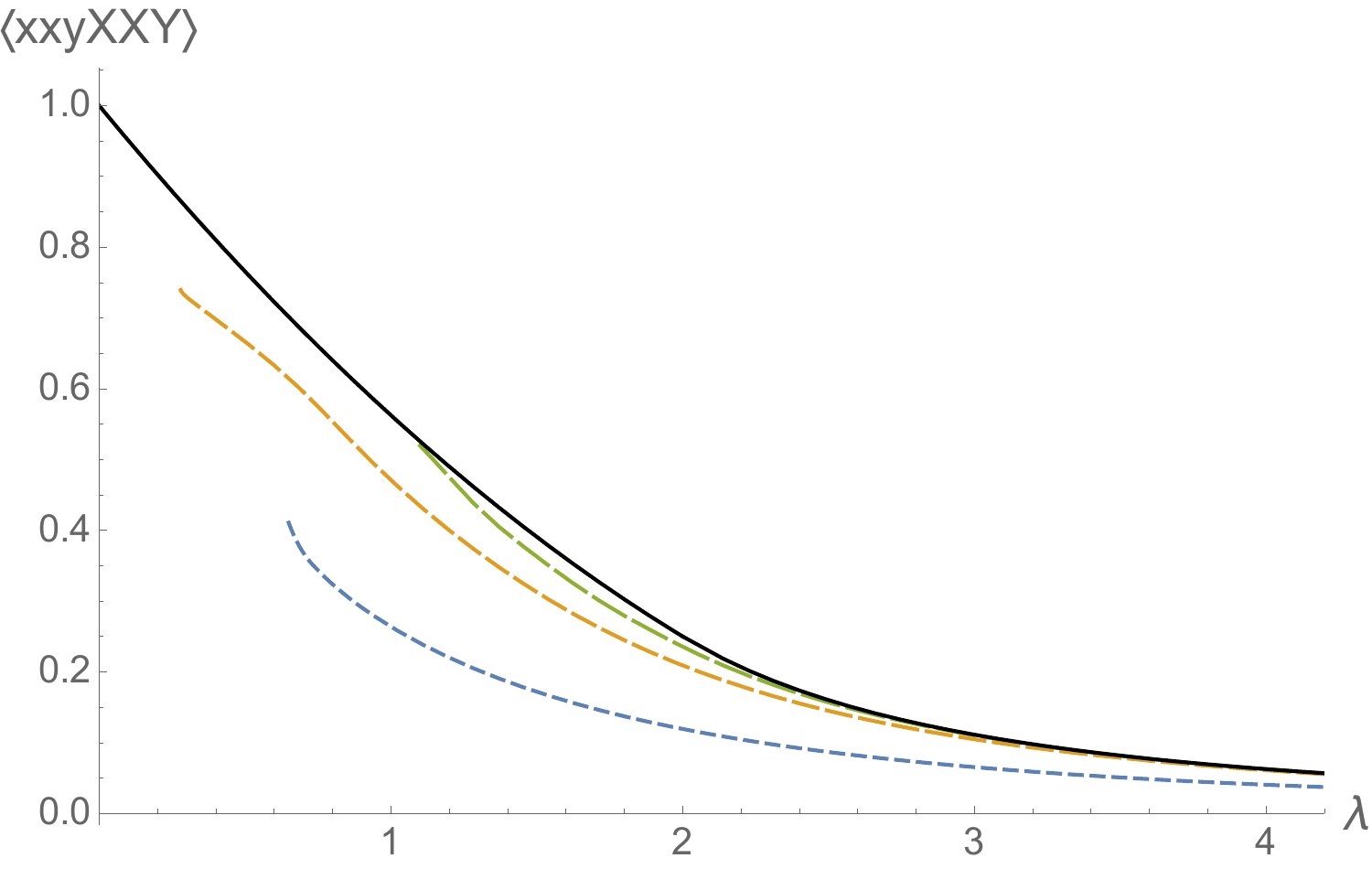}%
\includegraphics[scale=0.35]{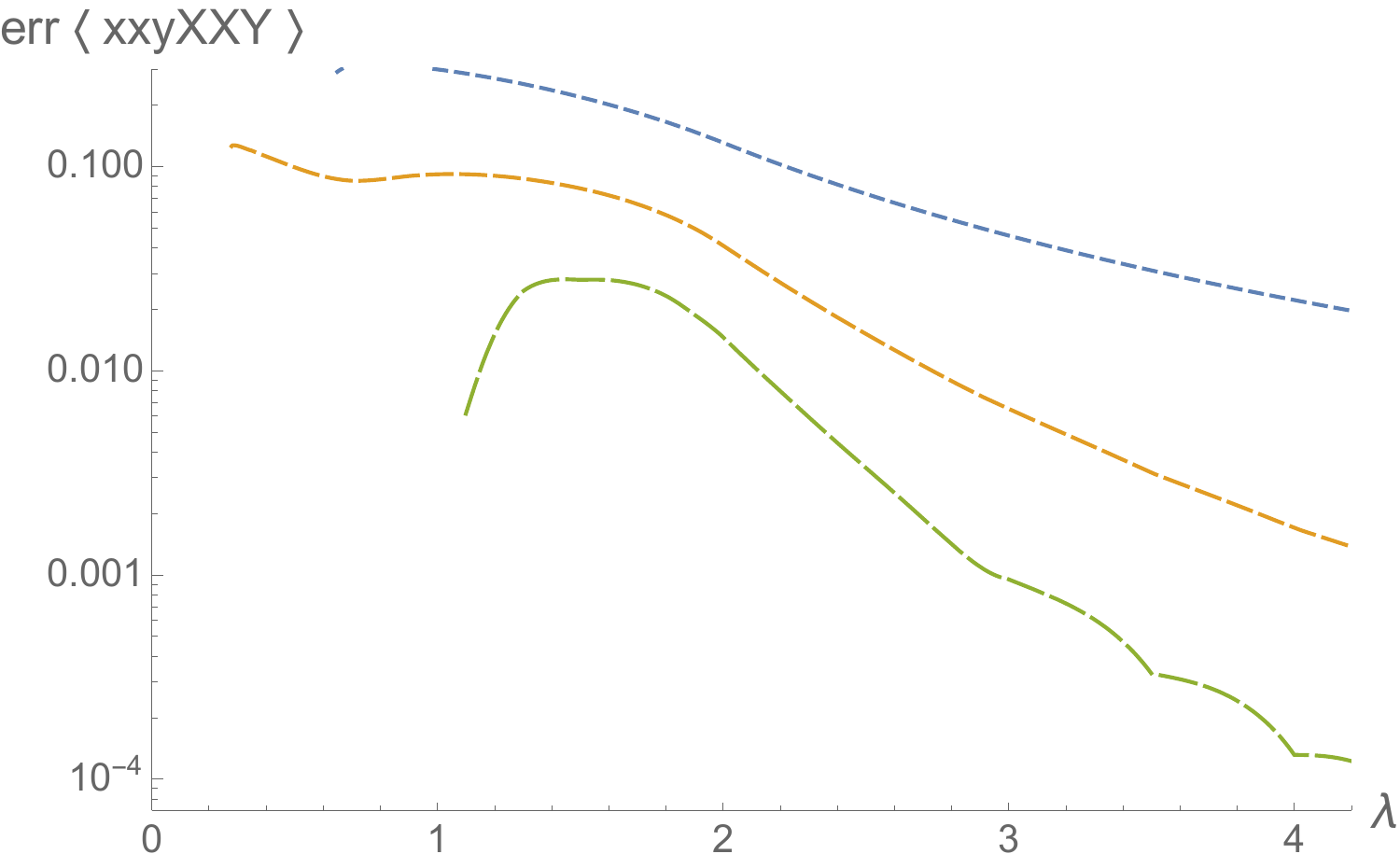}%
\hspace*{-1cm}
\caption
    {%
    Left: Expectation value of the $2 \times 1$
    rectangular loop $\langle \textit{xxyXXY} \rangle$ in
    two-dimensional Euclidean Yang-Mill theory.
    Shown is the exact result (solid black line) and
    results from variational calculations using order 2, 4 and 6 generators
    and observable truncation at order 28.
    Right: Semi-log plot of the absolute error magnitude between exact and variational results
    with order 2, 4 and 6 generators.
    (Line styles as in earlier figures.)
    \label{fig:ym2e-xxyXXY}
    }
\end{figure}

\begin{figure}[tp]
\vspace*{-7pt}
\hspace*{-1cm}%
\includegraphics[scale=0.35]{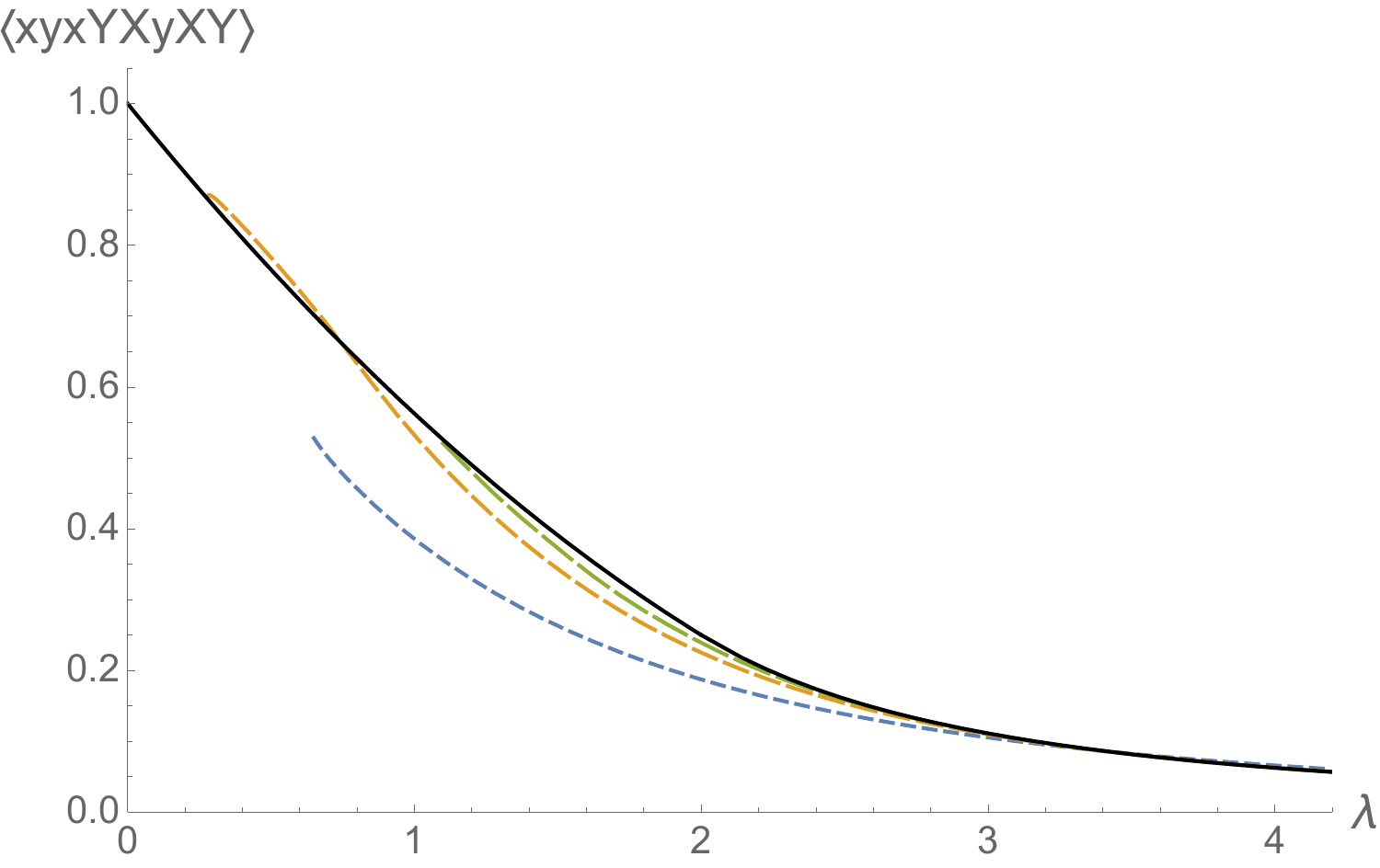}%
\includegraphics[scale=0.35]{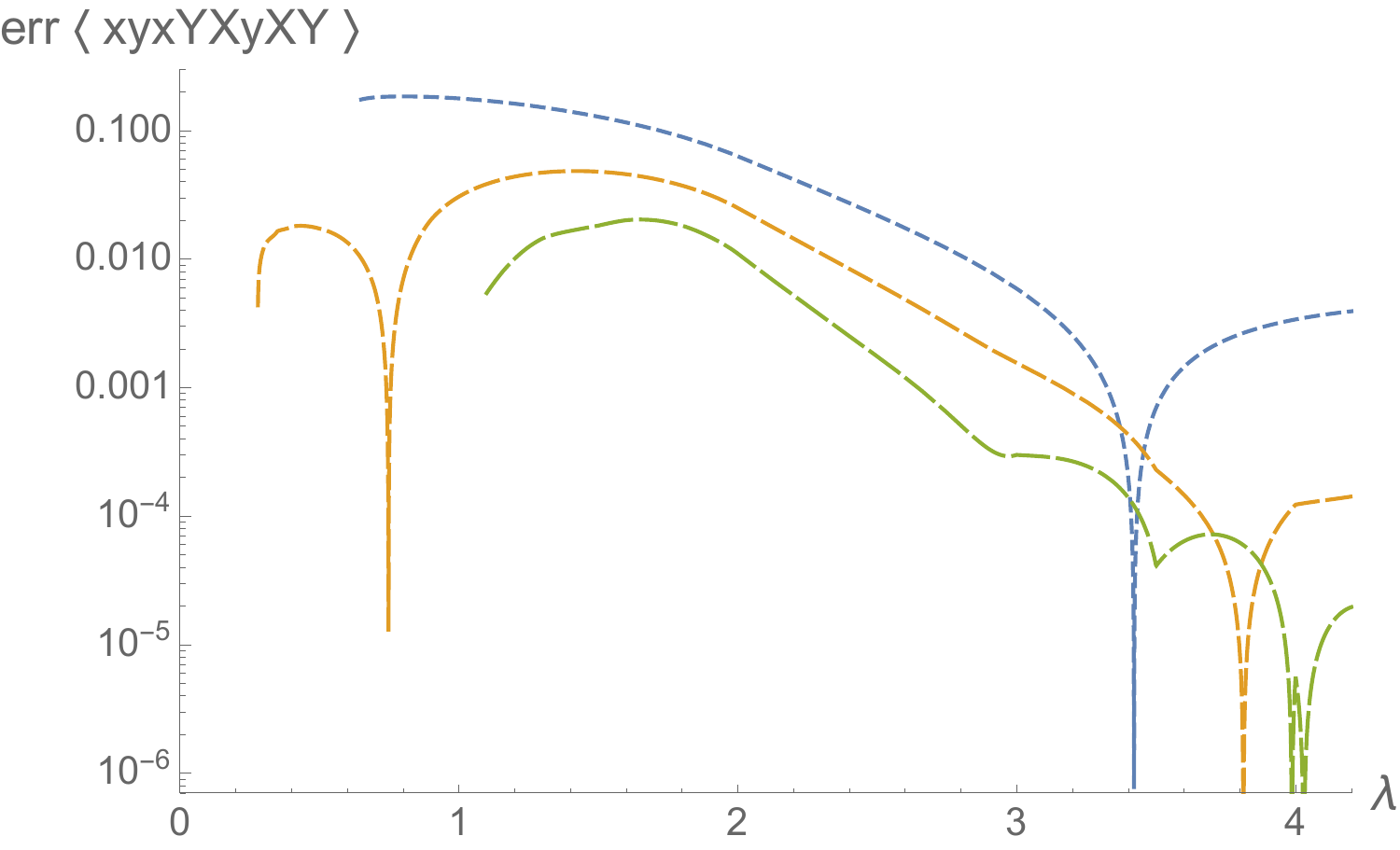}%
\hspace*{-1cm}
\caption
    {%
    Left: Expectation value of the two plaquette figure eight loop
    $\langle \textit{xyXYXyXY} \rangle$
    in two-dimensional Euclidean Yang-Mills theory.
    Shown is the exact result (solid black line) and
    results from variational calculations using order 2, 4 and 6 generators
    and observable truncation at order 28.
    Right: Semi-log plot of the absolute error magnitude between exact and variational results
    with order 2, 4 and 6 generators.
    (Line styles as in earlier figures.)
    \label{fig:ym2e-fig8}
    }
\vspace*{-7pt}
\end{figure}

\begin{figure}[tp]
\hspace*{-1cm}%
\includegraphics[scale=0.35]{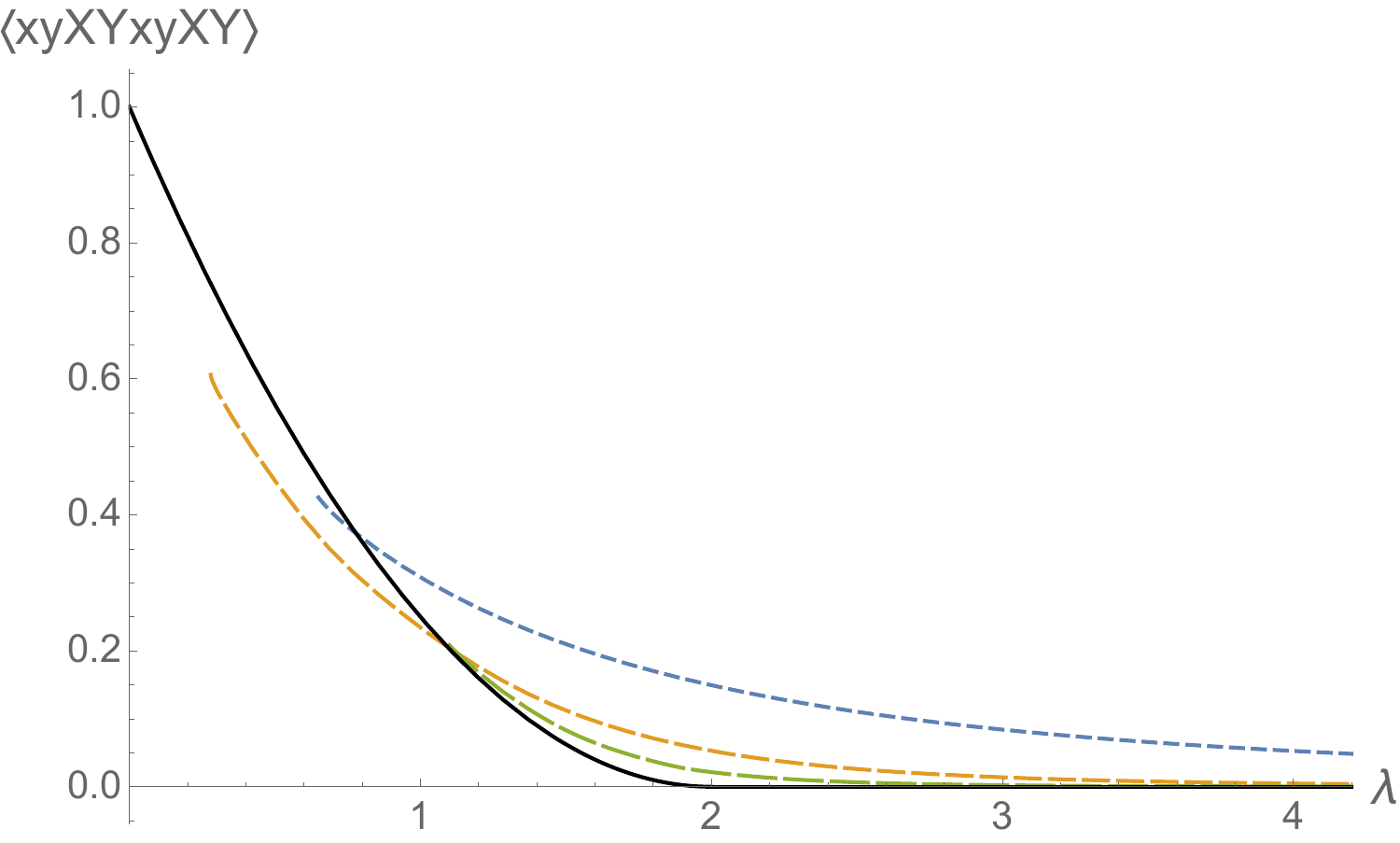}%
\includegraphics[scale=0.35]{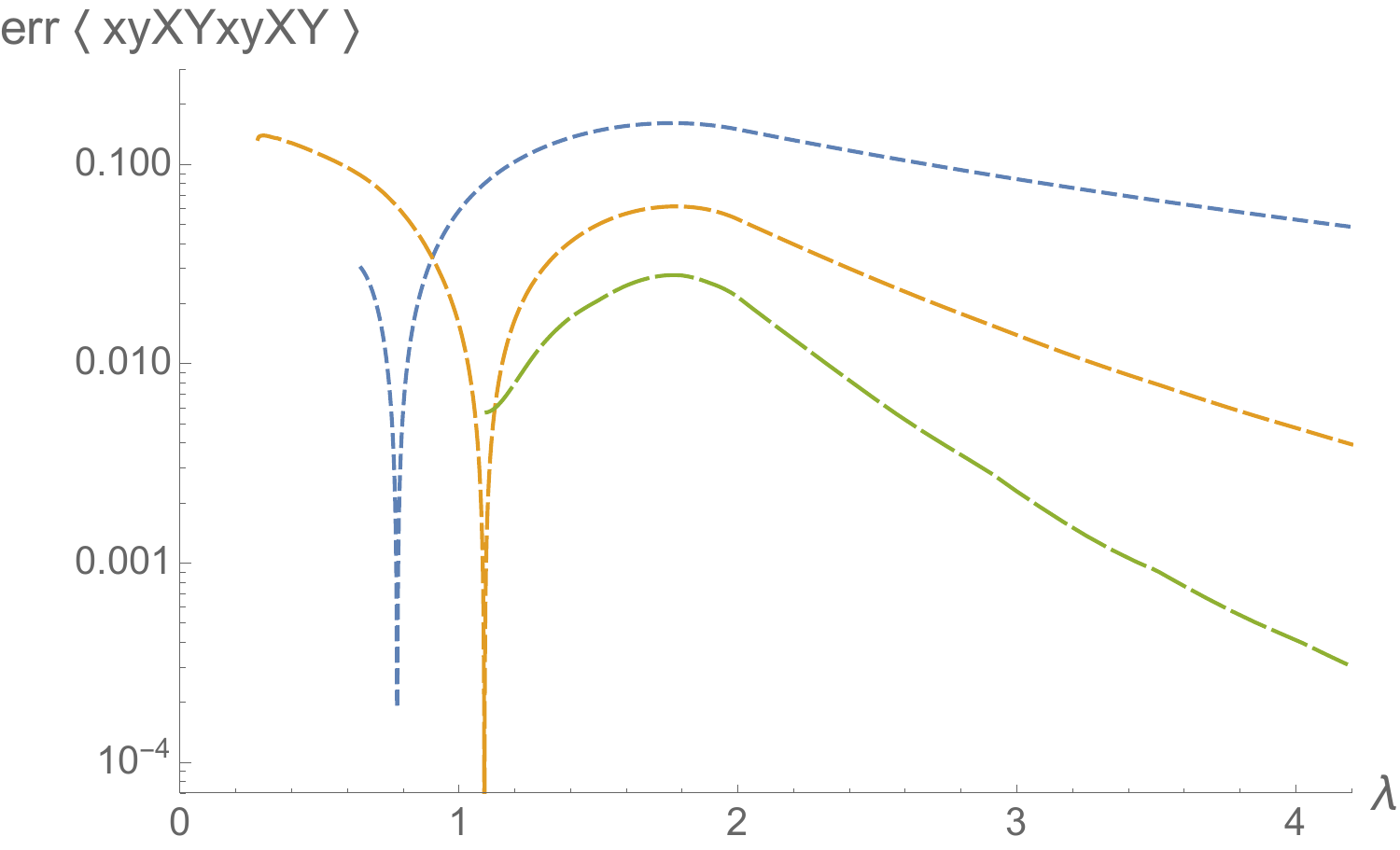}%
\hspace*{-1cm}
\caption
    {%
    Left: Expectation value of the winding two plaquette
    $\langle \textit{xyXYxyXY} \rangle$ in
    two-dimensional Euclidean Yang-Mill theory.
    Shown is the exact result (solid black line) and
    results from variational calculations using order 2, 4 and 6 generators
    and observable truncation at order 28.
    Right: Semi-log plot of the absolute error magnitude between exact and variational results
    with order 2, 4 and 6 generators.
    (Line styles as in earlier figures.)
    \label{fig:ym2e-xyXYxyXY}
    }
\vspace*{-7pt}
\end{figure}

The subsequent figures \ref{fig:ym2e-xyXY}--\ref{fig:ym2e-xyXYxyXY} show,
respectively, the corresponding results for the expectation values of a single plaquette,
a $2 \times 1$ rectangular loop, a $2 \times 1$ figure eight loop,
and a winding two plaquette.
The exact answer for both rectangular and figure eight $2 \times 1$ loops
is just the square of the single plaquette expectation value.

From these plots, it is apparent that the coherent state variational algorithm,
with a modest number of variational parameters, is working quite well down
to values of the gauge coupling well into the weak coupling regime.

\begin{figure}[tp]
\hspace*{-1cm}%
\includegraphics[scale=0.34]{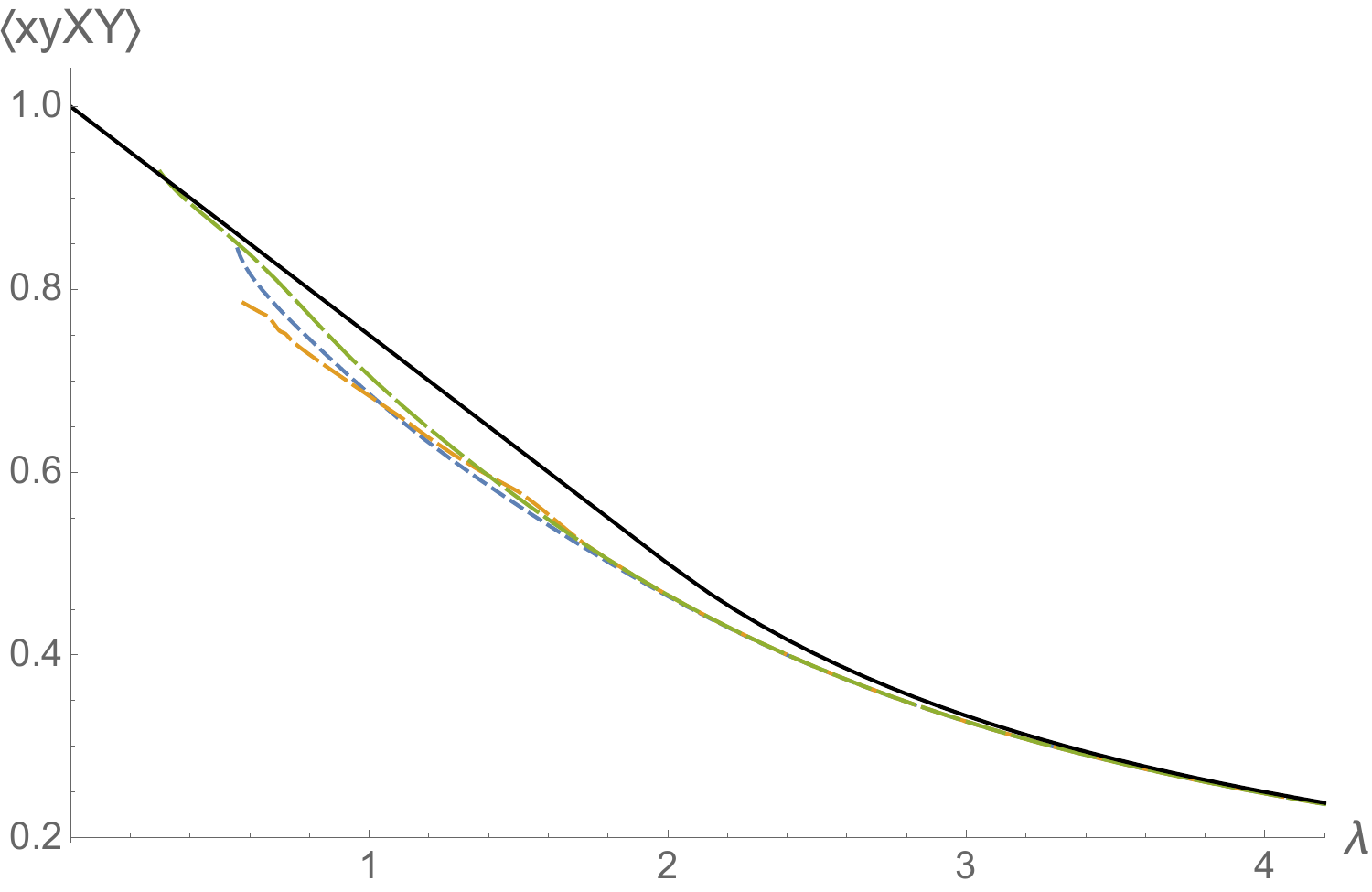}%
\includegraphics[scale=0.34]{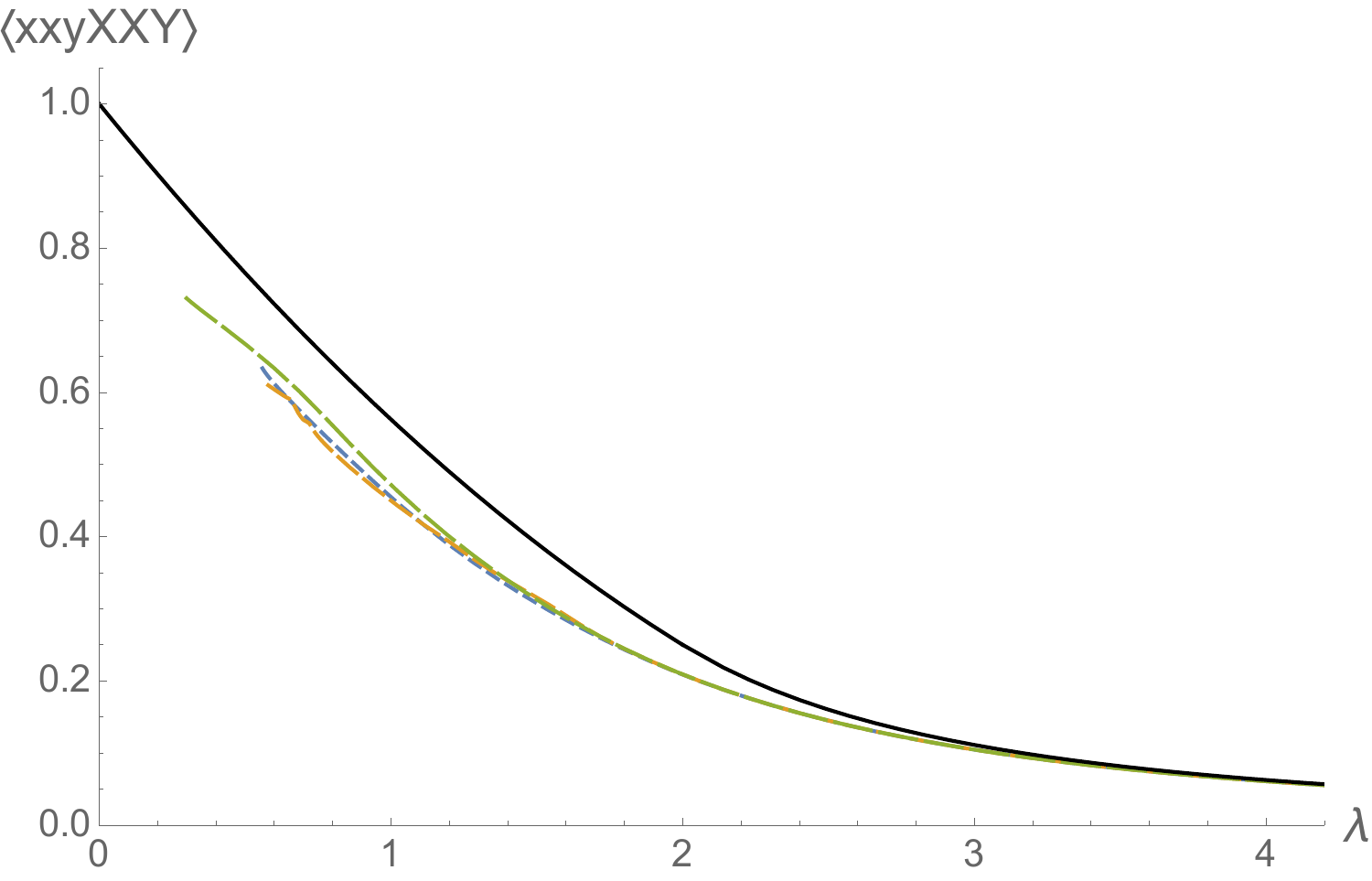}%
\hspace*{-1cm}
\\%
\hspace*{-1cm}%
\includegraphics[scale=0.34]{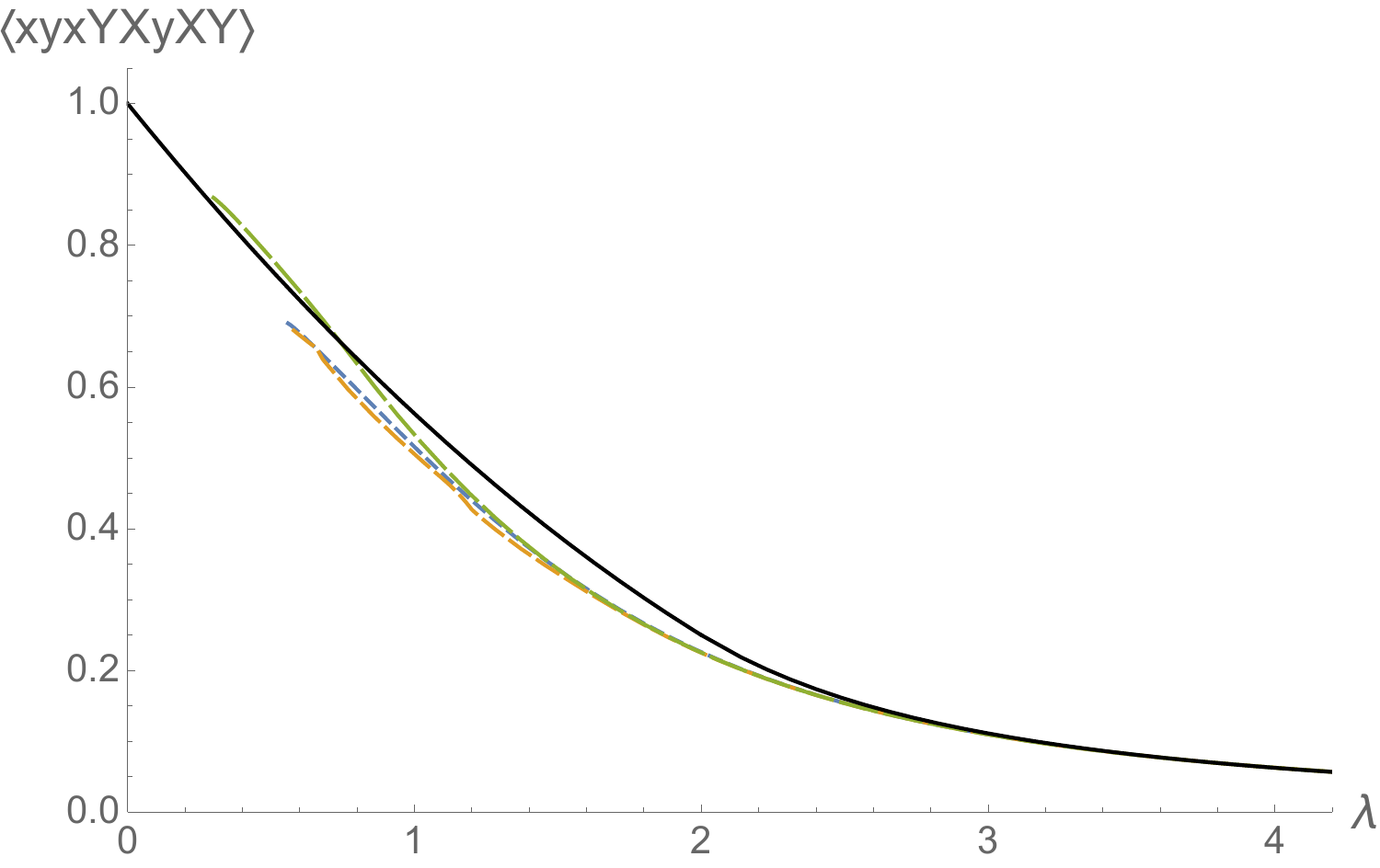}%
\includegraphics[scale=0.34]{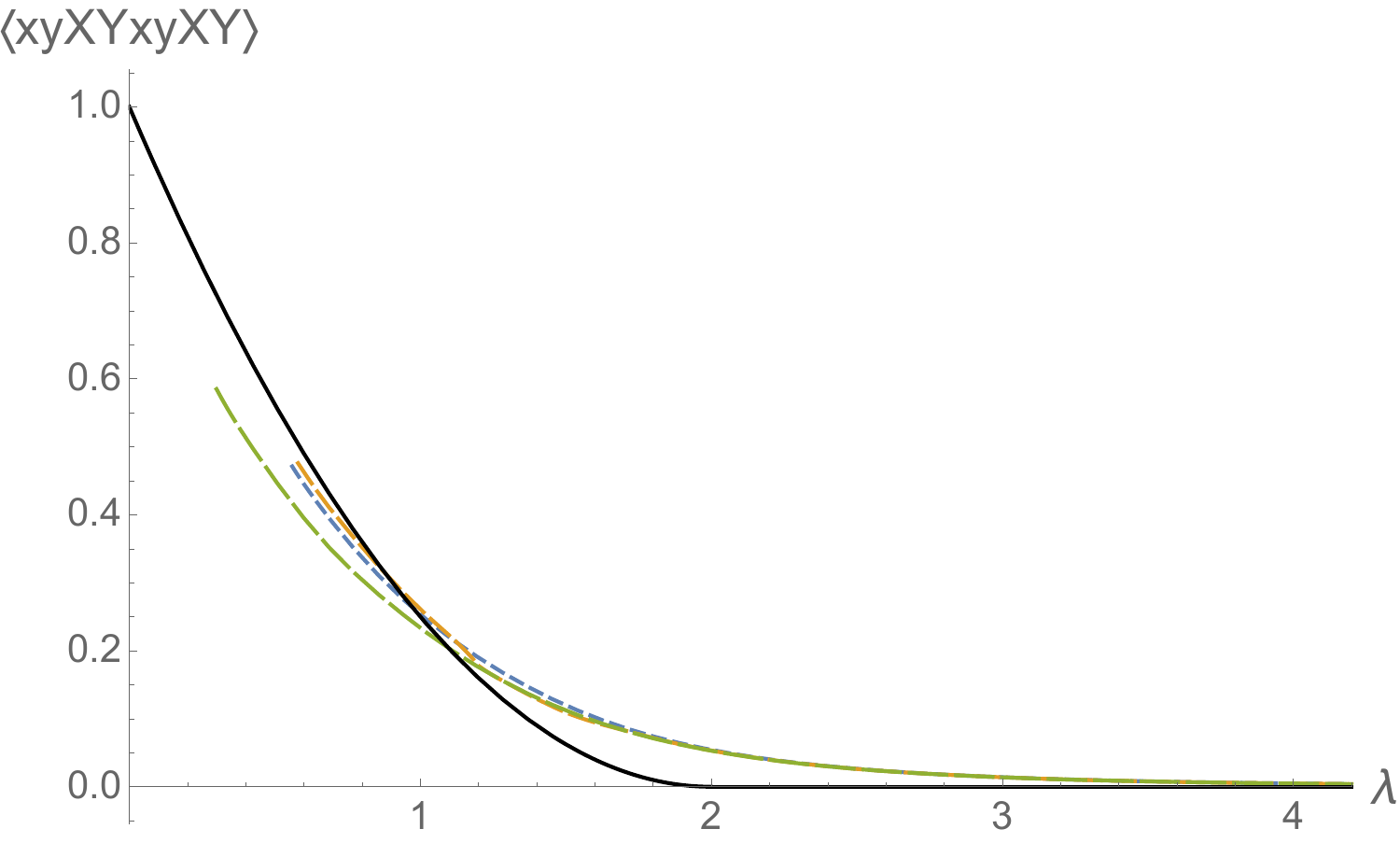}%
\hspace*{-1cm}
\caption
    {%
    Illustration of observable truncation effects:
    comparisons of results for expectation values of the indicated loops
    using order 4 generators and observable truncations at order 20, 24 and 28.
    (Line styles as in earlier figures.)
    \label{fig:obstrunc}
    }
\end{figure}

Figures \ref{fig:obstrunc} illustrate the dependence on the observable truncation order.
Shown are results for expectation values of the elementary plaquette, $2 \times 1$
rectangular and figure eight loops and the winding two plaquette,
from variational calculations with order 4 generators and observable truncations
at order 20, 24 and 28.
It is evident that, for these calculations, the observable truncation effects are
quite small when $\lambda \gtrsim 1$ but start becoming larger around
$\lambda \approx 0.7$.
With the same set of observable truncations
and order 6 generators,
calculations show somewhat larger observable truncation effects,
as one would expect,
with good agreement between the different truncations down to $\lambda \approx 1.5$
and more noticeable divergence below $\lambda \approx 1.2$.

\subsection{2+1D Hamiltonian Yang-Mills}
\label{sec:ym2h}

The 2+1 dimensional Yang-Mills Hamiltonian is given by
Eq.~(\ref{eq:YM-ham}), with an infinite two-dimensional cubic lattice.
Symmetry representations which will classify zero-momentum excited states,
and onto which coherence algebra generators will be projected,
are given by irreducible representations of the $C_{4v}$ crystallographic group,
conventionally named $A_1$, $A_2$, $B_1$, $B_2$ and $E$,
augmented by a $\pm$ superscript (or a $p$ or $m$ suffix in plots)
indicating the sign change
of the representation under charge conjugation.

Variational calculations were performed with generator truncations at order 2, 4 and 6,
corresponding to one, two and three plaquette operators,
and sets of observables truncated at strong-coupling orders up to 24.
These are calculations with, respectively, 1, 4 and 13 variational parameters,
and up to 44 million observables.
Table \ref{tab:ym2h} shows various statistics of the performed calculations,
including the number of terms in the resulting sets of geodesic equations.

\begin{table}
\begin{tabular}{c|c|c|c|c|c}
    ~generator~ & ~observable~ & ~\# variational~ & ~total \#~ & ~\# geodesic~ & ~~info file~~
\\
    order limit & order limit & parameters & ~observables~ & terms & size
\\
\hline
    2 & 16 & 1 & 22,987 & --- & ---%
\footnote
    {
    These calculations used order 4 system information
    files with forth order generators turned off during minimization.
    }
\\
    2 & 20 & 1 & 991,145 & --- & ---\footnotemark[1]
\\
    2 & 24 & 1 & 44,328,221 & --- & ---\footnotemark[1]
\\
\hline
    4 & 16 & 4 & 22,987 & $2.0 \times 10^6$  & 48.9\, MB
\\
    4 & 20 & 4 & 991,145 & $1.2 \times 10^8$ & 2.83\, GB
\\
    4 & 24 & 4 & 44,328,221 & $6.3 \times 10^9$ & 154\, GB
\\
\hline
    6 & 16 & 13 & 22,987 & $9.5 \times 10^6$ & 241\, MB
\\
    6 & 20 & 13 & 991,145 & $6.4 \times 10^8$ & 15.6\, GB
\\
    6 & 24 & 13 & 44,328,221 & $3.9 \times 10^{1\mathrlap{0}}$ & 937\, GB
\end{tabular}
\caption
    {%
    Statistics of variational calculations performed
    in 2+1 dimensional Hamiltonian Yang-Mills theory.
    The penultimate column gives the total number of terms
    in the complete set of geodesic equations for the
    given calculation, while the final column gives the
    size of the system information file which records the
    selected sets of observables and generators,
    and the resulting expressions for the Hamiltonian
    gradient, curvature, and geodesic equations.
    \label{tab:ym2h}
    }
\end{table}

\begin{figure}[tp]
\hspace*{-1.1cm}%
\includegraphics[scale=0.34]{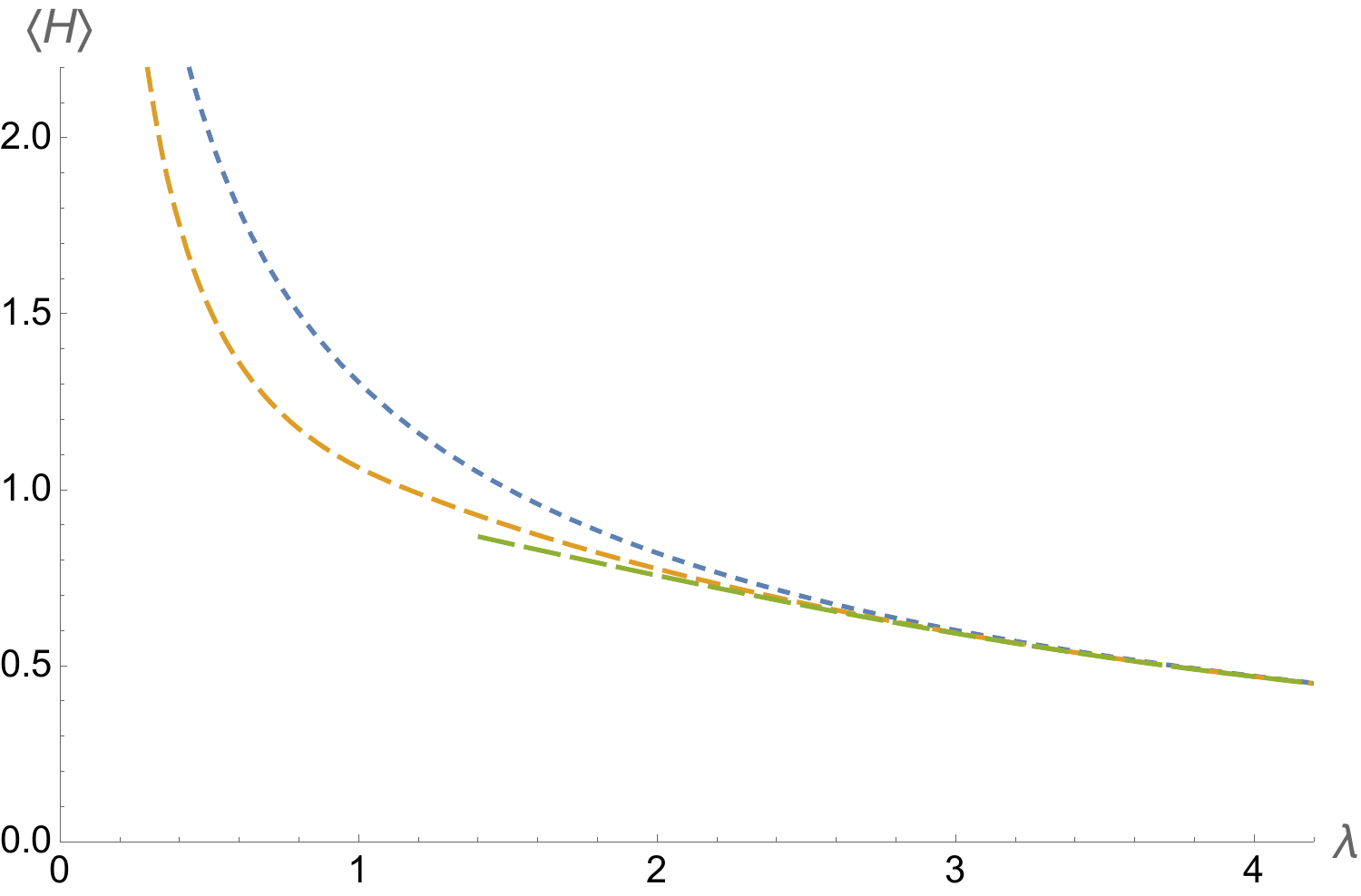}~~%
\includegraphics[scale=0.34]{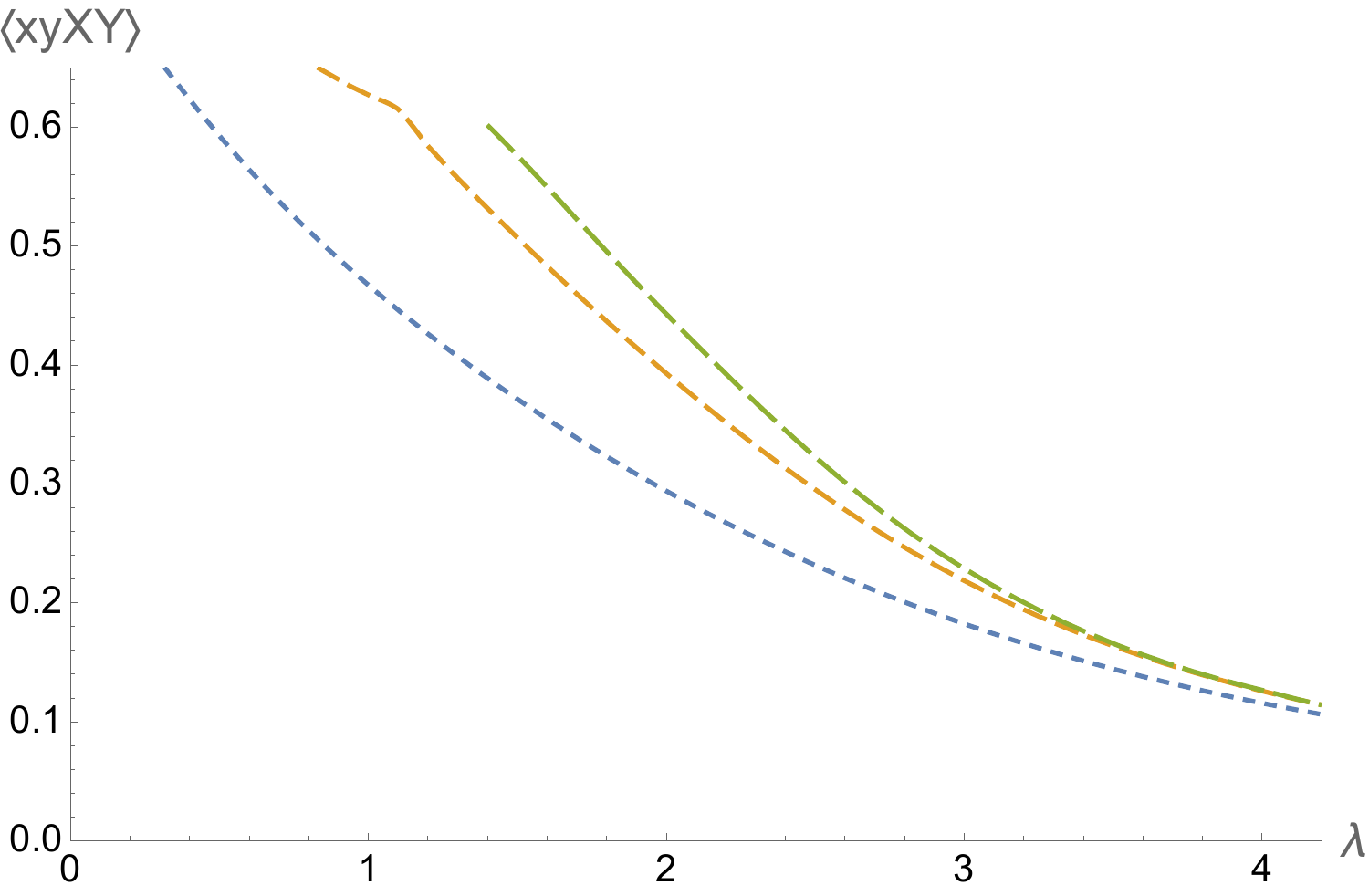}%
\hspace*{-1.5cm}
\\[5pt]
\hspace*{-1.3cm}%
\includegraphics[scale=0.35]{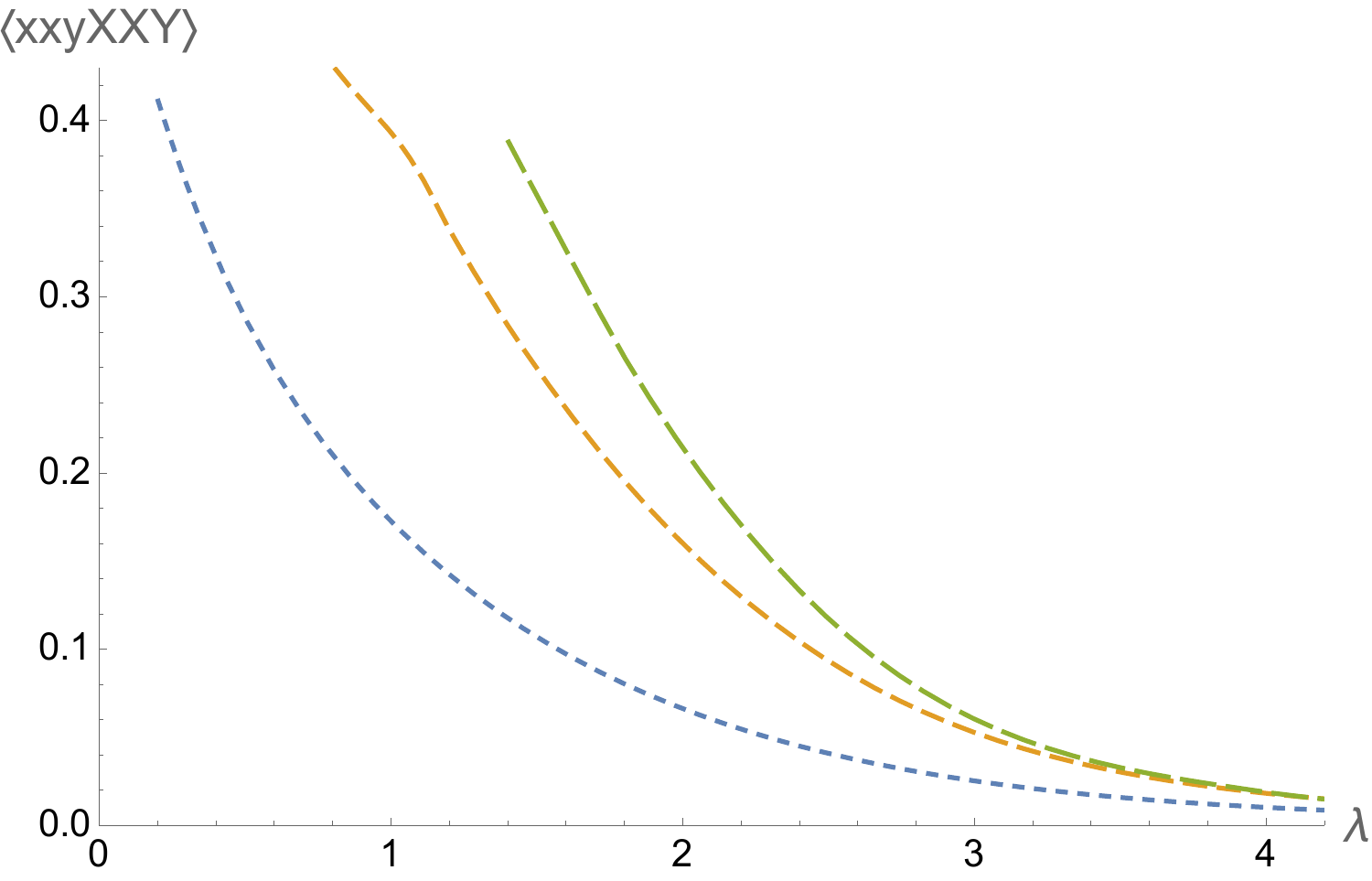}%
\includegraphics[scale=0.35]{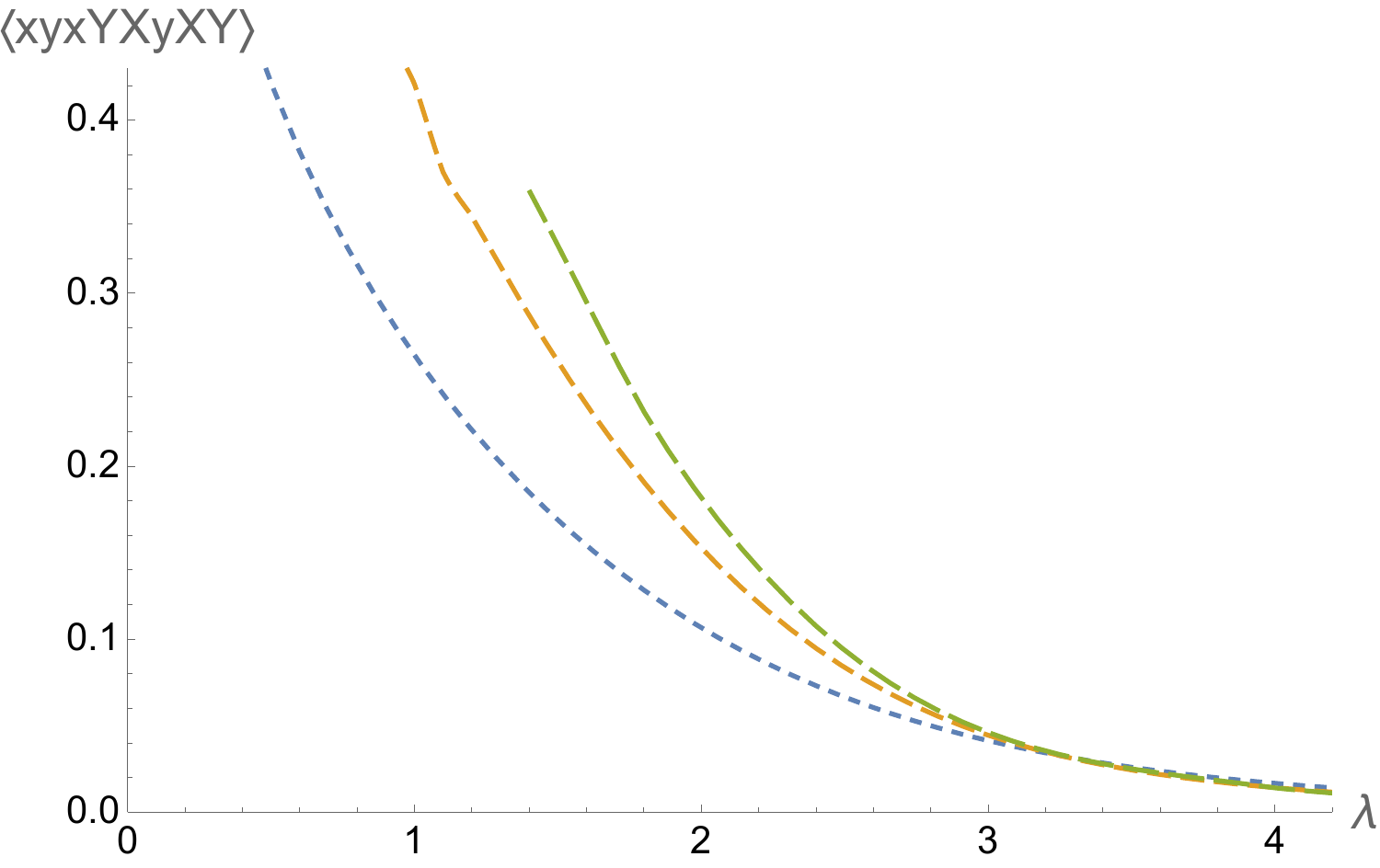}%
\hspace*{-1.5cm}
\vspace*{-0.2cm}
\caption
    {%
    Ground state energy (upper left)
    and expectation values of a
    single plaquette $\langle \textit{xyXY}\rangle$ (upper right),
    $2 \times 1$ loop $\langle\textit{xxyXXY}\rangle$ (lower left),
    and winding-two plaquette $\langle\textit{xyXYxyXY}\rangle$
    (lower right)
    in 2+1 dimensional Hamiltonian Yang-Mills theory.
    Shown are results from
    from variational calculations using order 2, 4 and 6 generators
    and observable truncation at order 24.
    (Line styles are the same as in earlier figures, with longer dashed curves
    corresponding to higher order truncations.)
    \label{fig:ym2h-A}
    }
\end{figure}

Figure \ref{fig:ym2h-A} shows the results for the ground state energy
and the expectation values of a single plaquette,
a $2 \times 1$ loop, and a winding-two plaquette
obtained from
variational calculations with order 2, 4 and 6 generators
and observable truncation at order 24.

The following figure \ref{fig:ym2h-B} shows results for the
lowest glueball mass in the $A_2^-$ and $A_1^+$ symmetry channels,
using order 2, 4 and 6 generators and observables truncated
at strong-coupling order 24.
The $A_2^-$ glueball is the lightest excitation, followed by
the $A_1^+$ glueball.
These glueball masses asymptote to $4\lambda$ at strong coupling.

The subsequent figure \ref{fig:ym2h-C} shows results for the
lowest glueball mass in the $B_1^+$, $B_2^-$ and $E^-$
symmetry channels obtained from
variational calculations with 4 and 6 generators
and observable truncation at order 24.
Order 2, or single plaquette, generators do not project
onto these representations.
The $B_1^+$ and $B_2^-$
glueball masses asymptote to $6\lambda$ at strong coupling,
while the $E^-$ glueball asymtotes to $8\lambda$.

\begin{figure}[tp]
\vspace*{-0.4cm}
\hspace*{-1.5cm}
\includegraphics[scale=0.34]{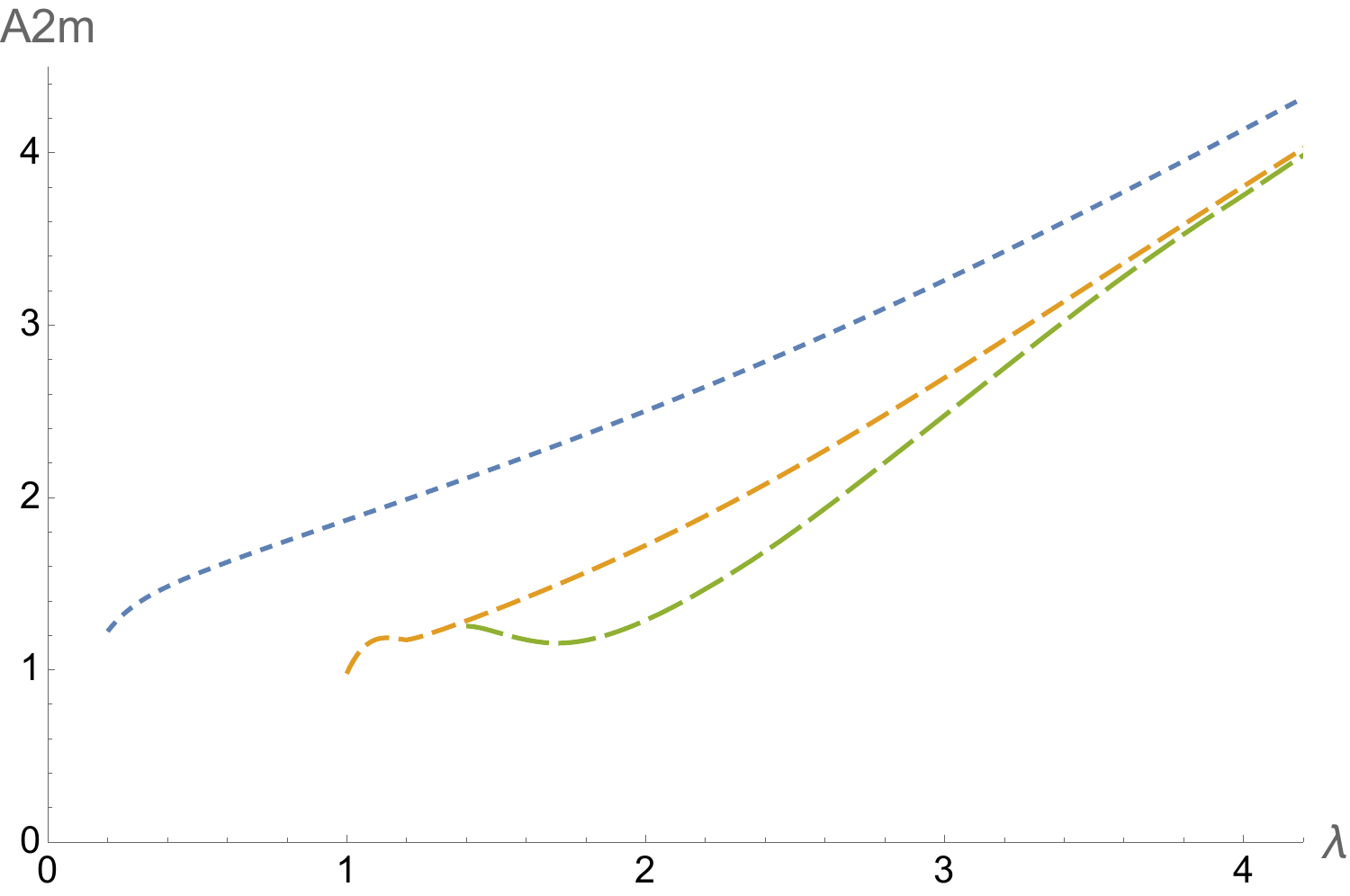}~~%
\includegraphics[scale=0.34]{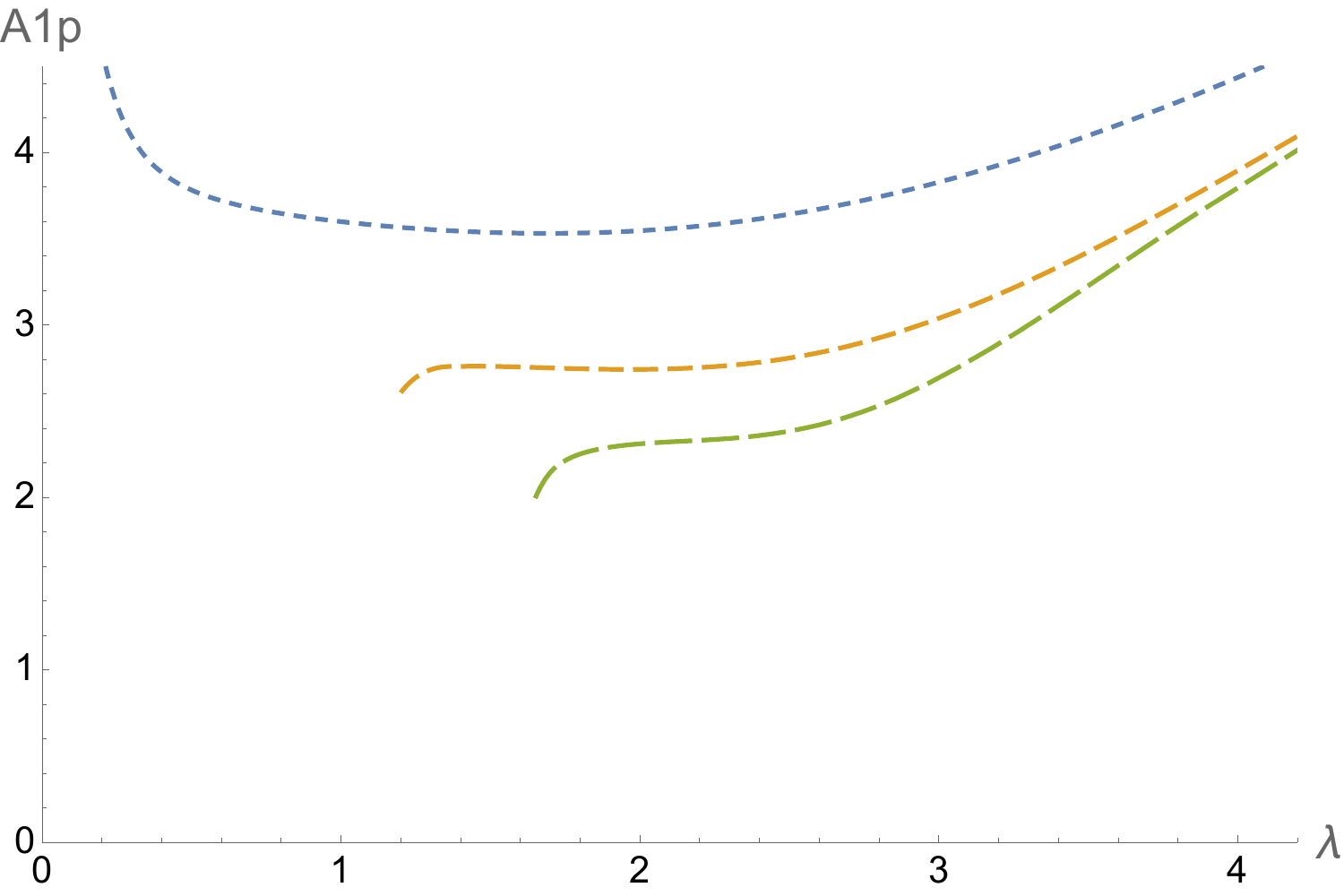}%
\hspace*{-1.5cm}
\vspace*{-0.2cm}
\caption
    {%
    Results
    for the lowest glueball mass in the $A_2^-$ and $A_1^+$ symmetry channels
    from variational calculations using order 2, 4 and 6 generators
    and observable truncation at order 24.
    (Line styles as in earlier figures.)
    \label{fig:ym2h-B}
    }
\end{figure}

\begin{figure}[tp]
\vspace*{-0.6cm}
\hspace*{-0.5cm}%
\includegraphics[scale=0.40]{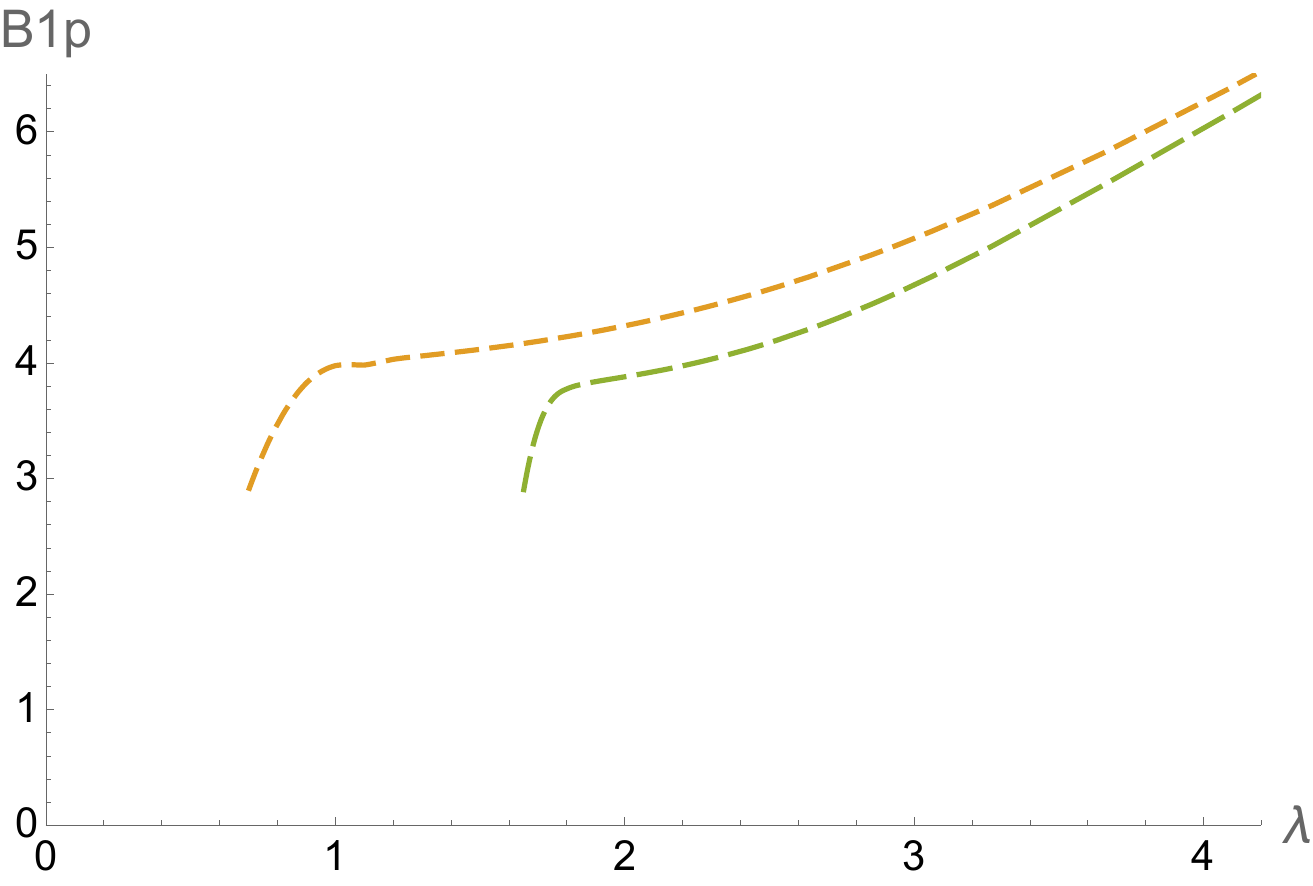}~~%
\includegraphics[scale=0.40]{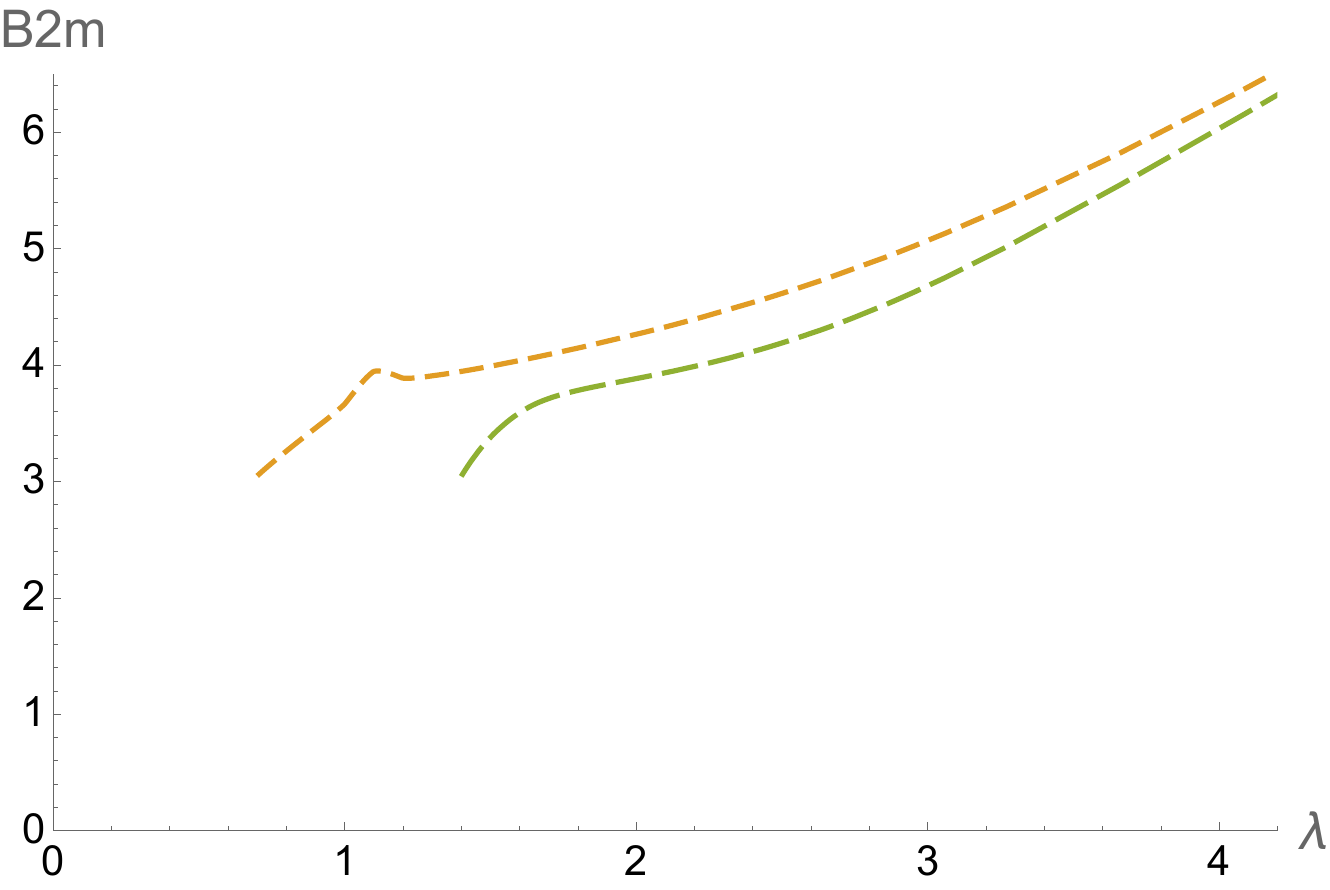}%
\hspace*{-0.5cm}
\\[5pt]
\includegraphics[scale=0.40]{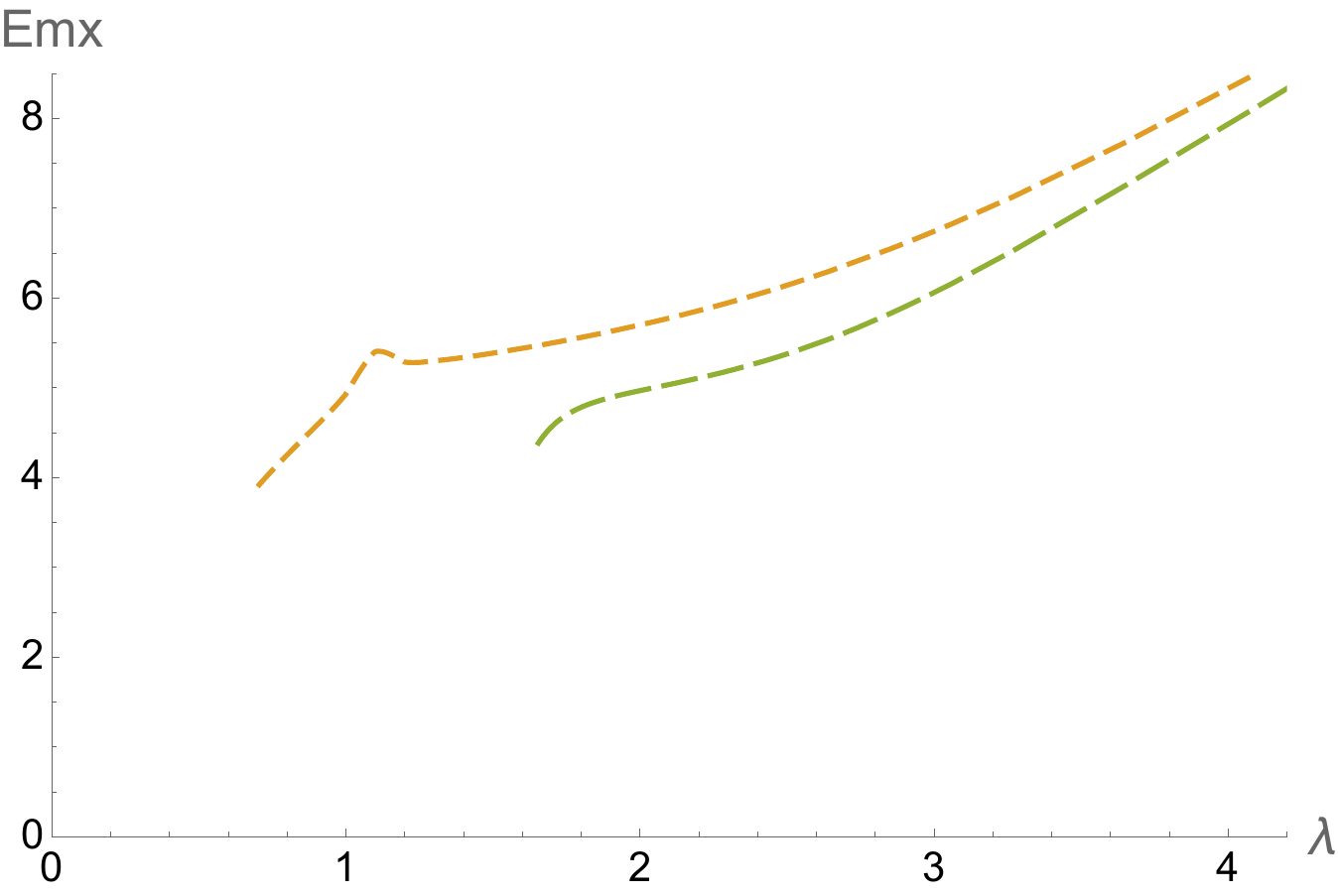}
\hspace*{-1.2cm}
\vspace*{-0.2cm}
\caption
    {%
    Results
    for the lowest glueball mass in the $B_1^+$, $B_2^-$ and $E^-$ symmetry channels
    from variational calculations using order 4 and 6 generators
    and observable truncation at order 24.
    (Line styles as in earlier figures.)
    Note that order 2 
    generators cannot project onto these representations.
    \label{fig:ym2h-C}
    }
\end{figure}

\begin{figure}[tp]
\vspace*{-0.2cm}
\includegraphics[scale=0.29]{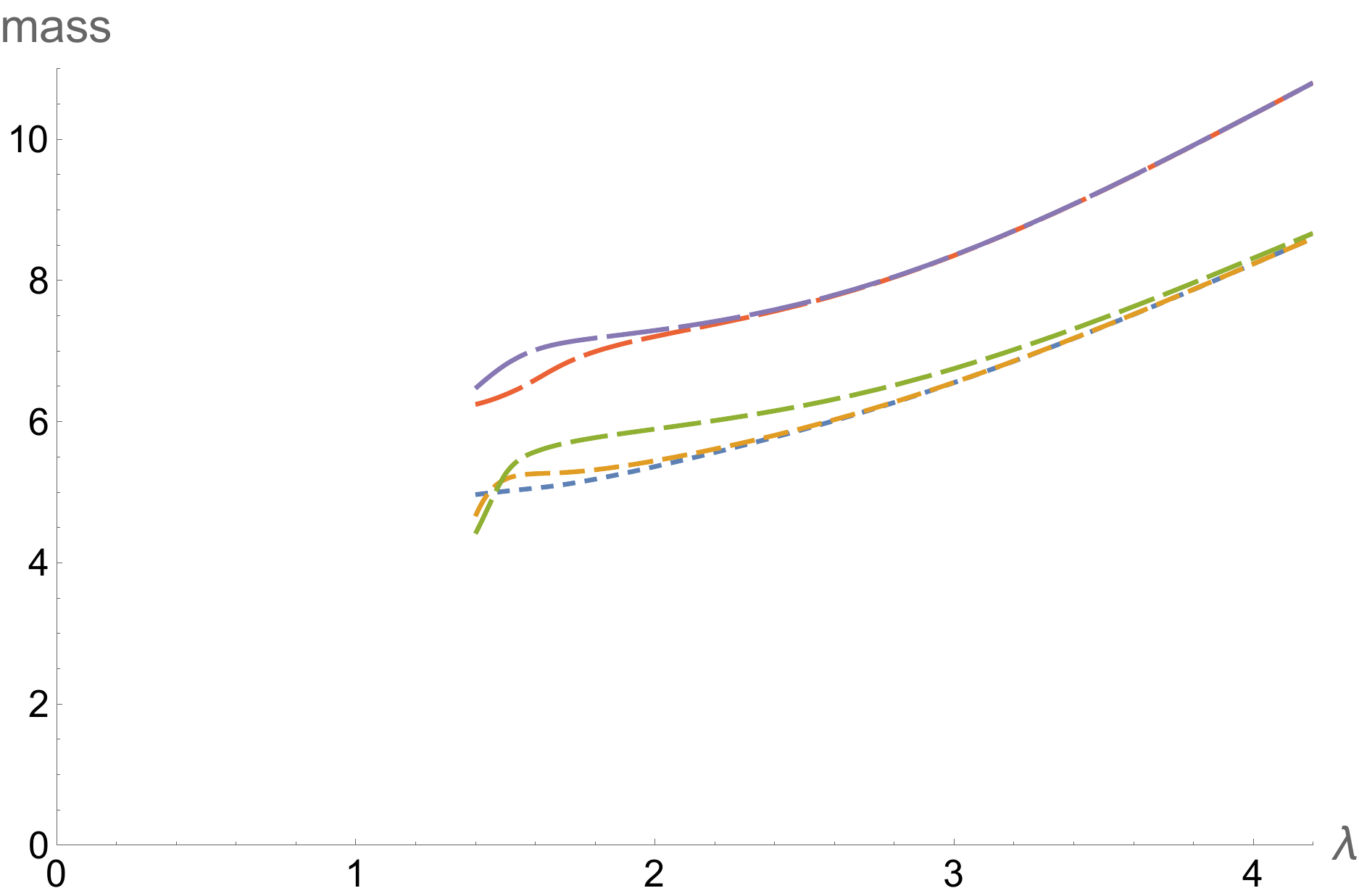}
\hspace*{-1.2cm}
\vspace*{-0.2cm}
\caption
    {%
    Results
    for the lowest glueball mass in
    the $B_2^+$ (shortest dash, blue),
    $B_1^-$ (orange),
    $E^+$ (green),
    $A_1^-$ (red)
    and $A_2^+$ (longest dash, purple) representations
    from variational calculations using order 6 generators
    and observable truncation at order 24.
    Note that order 6 is the minimal order for which generators can
    project onto these representations,.
    \label{fig:ym2h-D}
    }
\end{figure}

The final figure \ref{fig:ym2h-D} in this subsection
shows results for the lowest glueball mass in each of the
remaining symmetry channels:
$B_2^+$, $B_1^-$, $E^+$, $A_1^-$ and $A_2^+$.
The results shown come from variational calculations with
order 6 generators and observable truncation at order 24.
The lowest order for which a generator can project onto
any of these representations is order 6.
In the figure, the lowest, shortest dash line (for $\lambda \gtrsim 1.5$)
is the $B_2^+$ curve, while the progressively increasing curves
show the results for representations $B_1^-$, $E^+$, $A_1^-$ and $A_2^+$,
in that order.

Discussion of these results is postponed to section \ref{sec:discussion}.

\section {Observable approximation}\label{sec:approx}

In the results presented above, as well as
prior work exploring applications of the coherent state variational algorithm
\cite{CSVAI,CSVAII,
Lindqwister:1988xc,
Dickens:1987ih,
Somsky:1989},
any term in a geodesic equation or curvature matrix element involving an
observable \emph{not} in the retained set was simply dropped.
In other words, the expectation values of all non-retained observables
were approximated by zero.
With observable selection based on the strong-coupling order classification,
this provides a good approximation at least for some range of sufficiently strong coupling.
Inevitably, however, for any given truncation of the set of observables,
errors in the state representation caused by this truncation 
will grow as the lattice gauge coupling is decreased and the
ground state correlation length grows.

A natural question to consider is whether it is possible to do a
better approximation for the expectation values of observables outside
the truncation set.
Can the expectation values of non-retained observables be usefully
approximated by some simple functions of the expectation values of
retained observables?
This idea is motivated by the observation that nearly all Wilson loops
are self-intersecting (with the small fraction of non-self intersecting loops
dropping rapidly with increasing strong-coupling order).
From studies of confinement, it is known that expectation values of
simple, planar non-self intersecting Wilson loops are well-described by
a combination of area-law and perimeter-law behavior at both
strong and weak coupling,
\begin{equation}
    W_\Gamma \approx e^{ -\gamma \, |\Gamma| - \sigma \, \mathrm{area}(\Gamma) } ,
\end{equation}
where $|\Gamma|$ is the perimeter of a loop $\Gamma$,
area($\Gamma$) is the minimal area of a surface spanning $\Gamma$,
$\sigma$ is the string tension and $\gamma$ is a (UV sensitive)
perimeter-law coefficient.
To the extent that such a combination of area-law and perimeter-law terms provides
a good approximation for most Wilson loops
--- which is a major untested assumption ---
then the expectation value of a self-intersecting loop which is the composition of
two (individually closed) sub-loops, $\Gamma = \Gamma_1 \, \Gamma_2$,
should approximately satisfy the factorization relation
\begin{equation}
    W_\Gamma \approx W_{\Gamma_1} \times W_{\Gamma_2} \,,
\label{eq:approx}
\end{equation}
since the sum of the sub-loop perimeters equals the whole loop perimeter,
and the composition of the sub-loop spanning surfaces is a spanning surface of the whole loop.%
\footnote
    {%
    In 2D Euclidean Yang-Mills, this relation is exact for self-intersecting loops whose
    minimal spanning surfaces are disjoint and unfolded.
    More generally, however, there are cases where the sum of the sub-loop areas
    does not equal the \emph{minimal} area of a spanning surface of the full loop.
    }

For any geodesic equation term involving a loop $\Gamma$
which is not in the retained set of observables, if that loop can be factored
into sub-loops which are found in the retained set then approximating the
expectation value $W_\Gamma$ as a product of sub-loop expectations may reduce the
truncation error in the state representation.
This, it must be emphasized, is merely a hypothesis,
but one which seems worth exploring.

The current \emph{Gordion} code optionally implements such an observable
approximation scheme.
The relevant routine examines all self-intersections of an observable
appearing in a commutator result which is found to be
outside the truncation set.
Each self-intersection is assigned a numeric score
which is used to select preferred possible factorizations when there
are multiple possibilities.
The particulars of this intersection scoring
are chosen to favor factorizations which split the observable
into comparably-sized pieces, as this should maximize the probability
that both sub-observables do lie within the truncation set.
Factorization is attempted not just for Wilson loops, but also for
Wilson loops containing electric field insertions
(as well as fermion bilinears).
The precise details of this scoring and factorization algorithm
are unquestionably rather ad-hoc and are described more fully
in Ref.~\cite{Gordion}.

One unavoidable consequence of applying this observable approximation
scheme is a major increase in the number of terms in the resulting
geodesic equations, and consequent storage requirements,
as detailed in Tables \ref{tab:ym2e-approx} and \ref{tab:ym2h-approx}.

\begin{table}
\begin{tabular}{c|c|cc|cc}
      generator
    & observable
    & \multicolumn{2}{|c|}{geodesic terms}
    & \multicolumn{2}{c}{info file size}
\\
      ~order limit~
    & ~order limit~
    & ~w/o approx~
    & ~with approx~
    & ~w/o approx~
    & ~with approx~
\\
\hline
    4 & 16 & $3.4 \times 10^3$ & $2.3 \times 10^4$ & 116 KB & 433 KB
\\
    4 & 20 & $9.8 \times 10^4$ & $9.5 \times 10^5$ & 1.98 MB & 15.6 MB
\\
    4 & 24 & $3.3 \times 10^6$ & $4.1 \times 10^7$ & 62.6 MB & 659 MB
\\
    4 & 28 & $1.2 \times 10^8$ & $1.7 \times 10^9$ & 2.22 GB & 28.1 GB
%\\
%    4 & 32 & $4.5 \times 10^9$ & --- & 80.6 GB & [[---]]
\\
\hline
    6 & 20 & $2.5 \times 10^5$ & $7.9 \times 10^6$ & 6.91 MB & 132 MB
\\
    6 & 24 & $8.1 \times 10^6$ & $3.7 \times 10^8$ & 158 MB & 5.98 GB
\\
    6 & 28 & $2.9 \times 10^8$ & $1.7 \times 10^{10}$ & 5.07 GB & 270 GB
%\\
%    6 & 32 & $1.1 \times 10^{10}$ & --- & 186 GB &  [[---]]
\end{tabular}
\caption
    {
    2D Euclidean Yang-Mills:
    Comparisons of the number of geodesic equation terms,
    and resulting system information file sizes,
     with and without approximation of non-retained observables,
    for varying orders of generator and observable truncation.
    \label {tab:ym2e-approx}
    }
\end{table}

\begin{table}
\begin{tabular}{c|c|cc|cc}
      generator
    & observable
    & \multicolumn{2}{|c|}{geodesic terms}
    & \multicolumn{2}{c}{info file size}
\\
      ~order limit~
    & ~order limit~
    & ~w/o approx~
    & ~with approx~
    & ~w/o approx~
    & ~with approx~
\\
\hline
    4 & 16 & $2.0 \times 10^6$ & $9.7 \times 10^6$ & 48.9 MB &  233 MB
\\
    4 & 20 & $1.2 \times 10^8$ & $6.3 \times 10^8$ & 2.83 GB & 15.3 GB
%\\
%    4 & 24 & $6.3 \times 10^9$ & --- & 154 GB & ---
%\\
%\hline
%    6 & 20 & $9.5 \times 10^6$ & --- & 15.6 GB & ---
%\\
%    6 & 24 & $6.4 \times 10^8$ & --- & 937 GB & ---
\end{tabular}
\vspace*{-4pt}
\caption
    {
    2+1D Hamiltonian Yang-Mills:
    Comparisons of the number of geodesic equation terms,
    and resulting system information file sizes,
    with and without observable approximation
    of non-retained observables,
    for order 4 generator and observable truncation at order 16 or 20.
    \label {tab:ym2h-approx}
    }
\end{table}

\begin{figure}[tp]
\hspace*{-0.5cm}%
\includegraphics[scale=0.33]{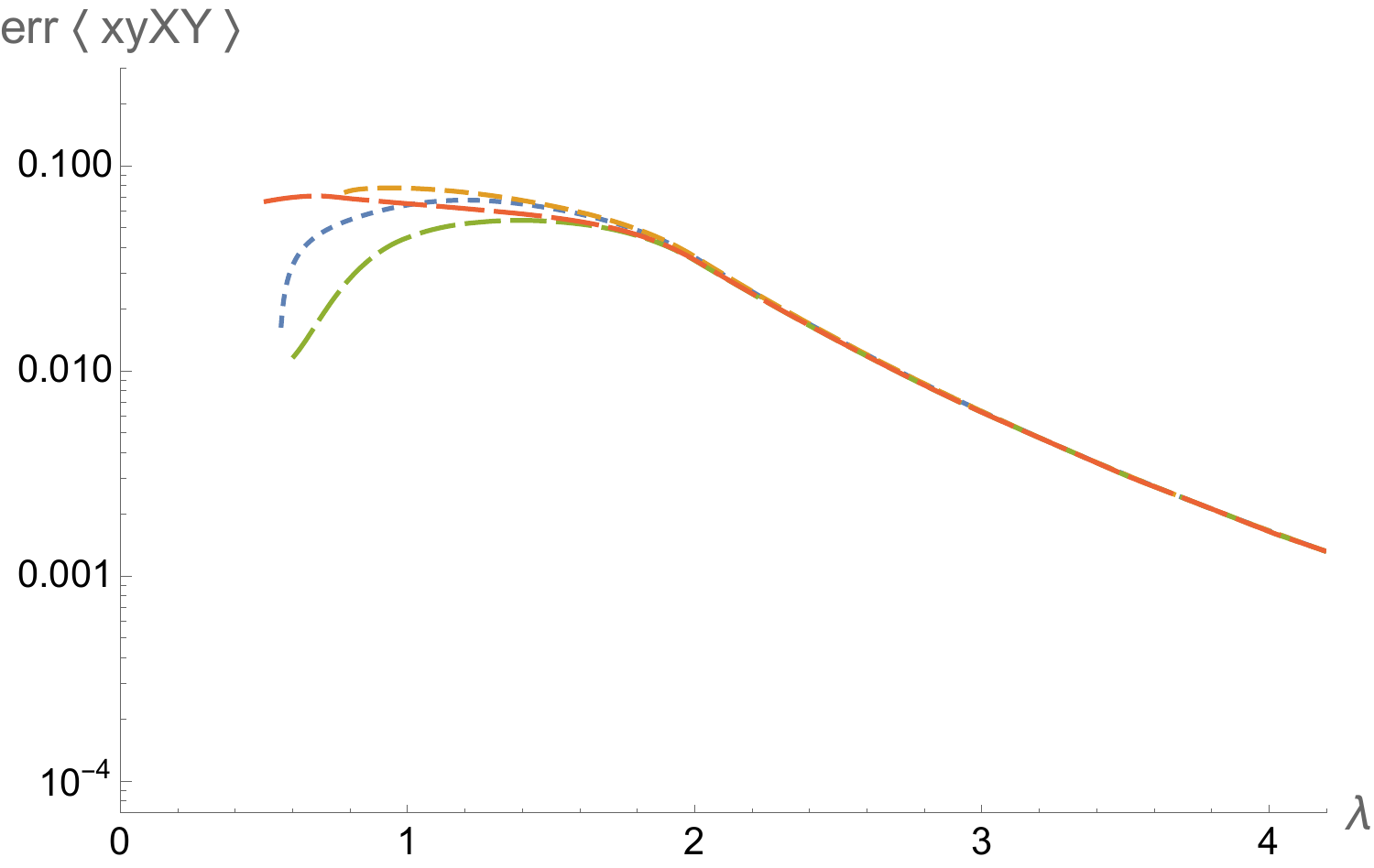}~~%
\includegraphics[scale=0.33]{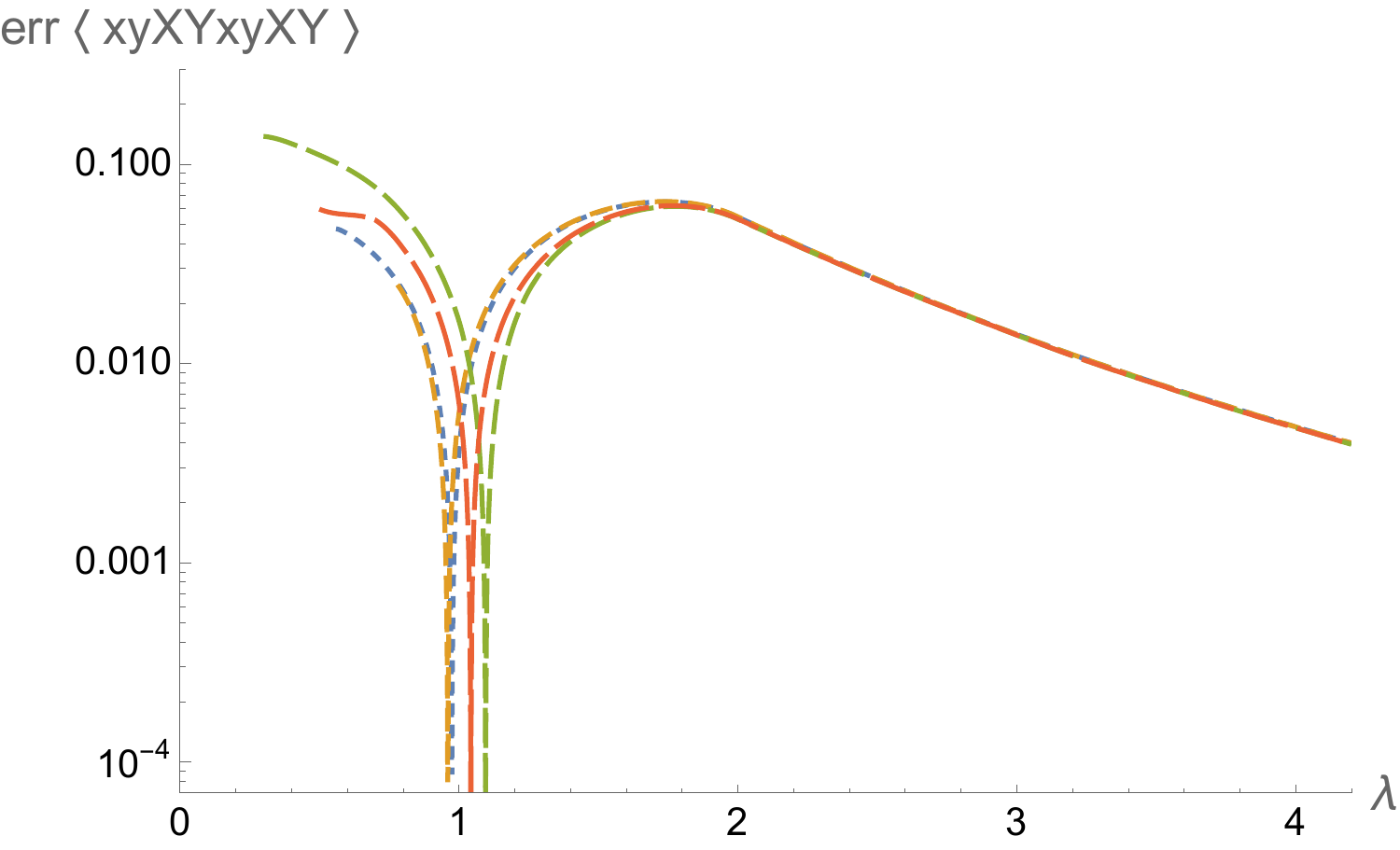}%
\hspace*{-0.5cm}
\vspace*{-5pt}
\caption
    {%
    Comparisons, in 2D Euclidean Yang-Mills,
    of the deviation from the exact results
    when observable approximation is, or is not, employed.
    Shown are the deviations from exact results for the
    single plaquette expectation $\langle \textit {xyXY}\rangle$
    (left panel)
    and the double winding plaquette expectation $\langle \textit {xyXYxyXY}\rangle$
    (right panel)
    from variational calculations with order 4 generators and
    observable truncation at orders 20 and 24.
    Blue (shortest dash) and orange curves (next shortest dash) are order 20
    observable truncation results, without and with observable approximation, respectively.
    Green and red (longest dash) curves are order 24 observable truncation results,
    without and with observable approximation, respectively.
    \label{fig:ym2e-approx}
    }
\end{figure}

\begin{figure}[tp]
\vspace*{-5pt}
\hspace*{-0.5cm}%
\includegraphics[scale=0.33]{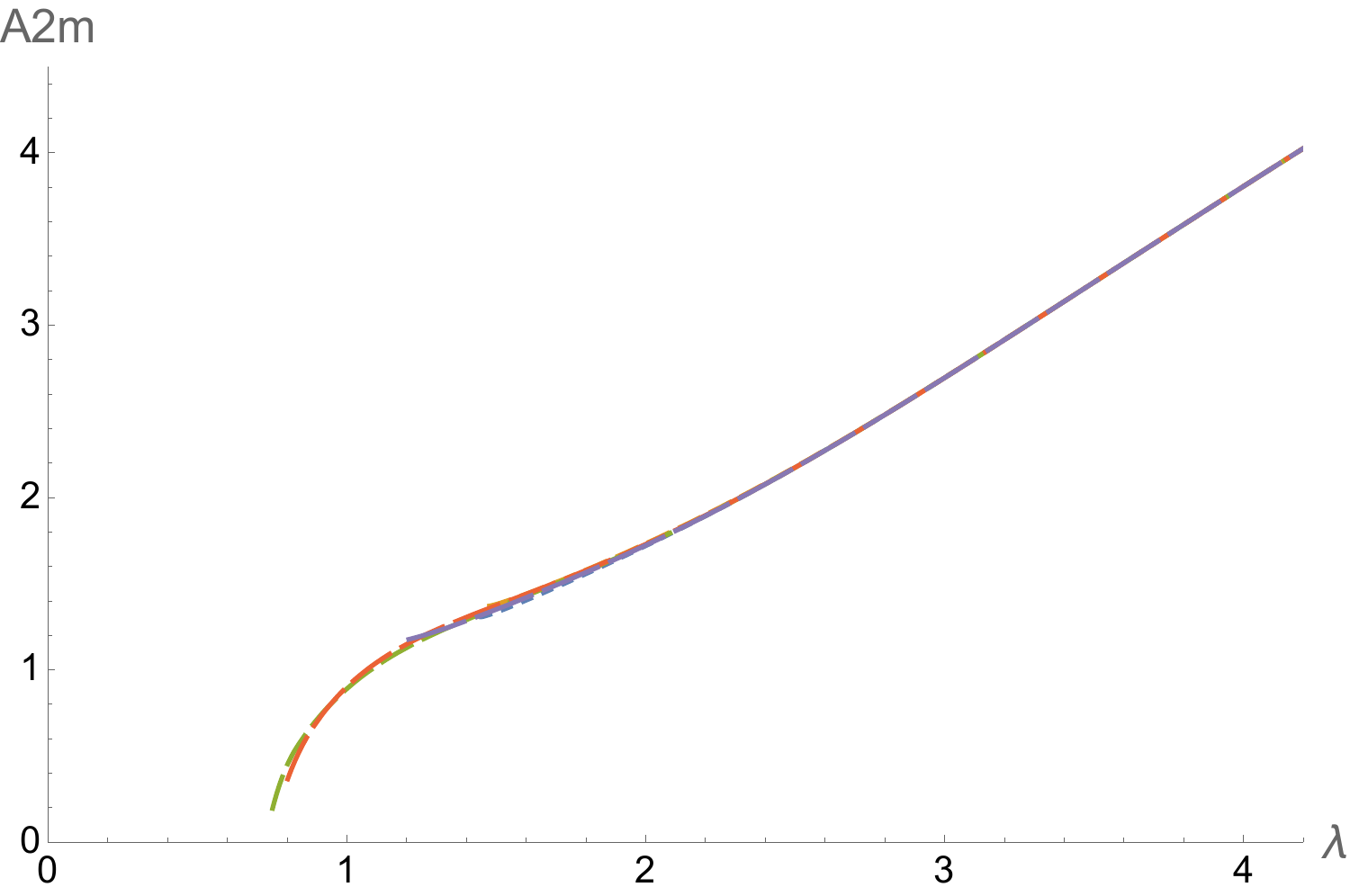}~~%
\includegraphics[scale=0.33]{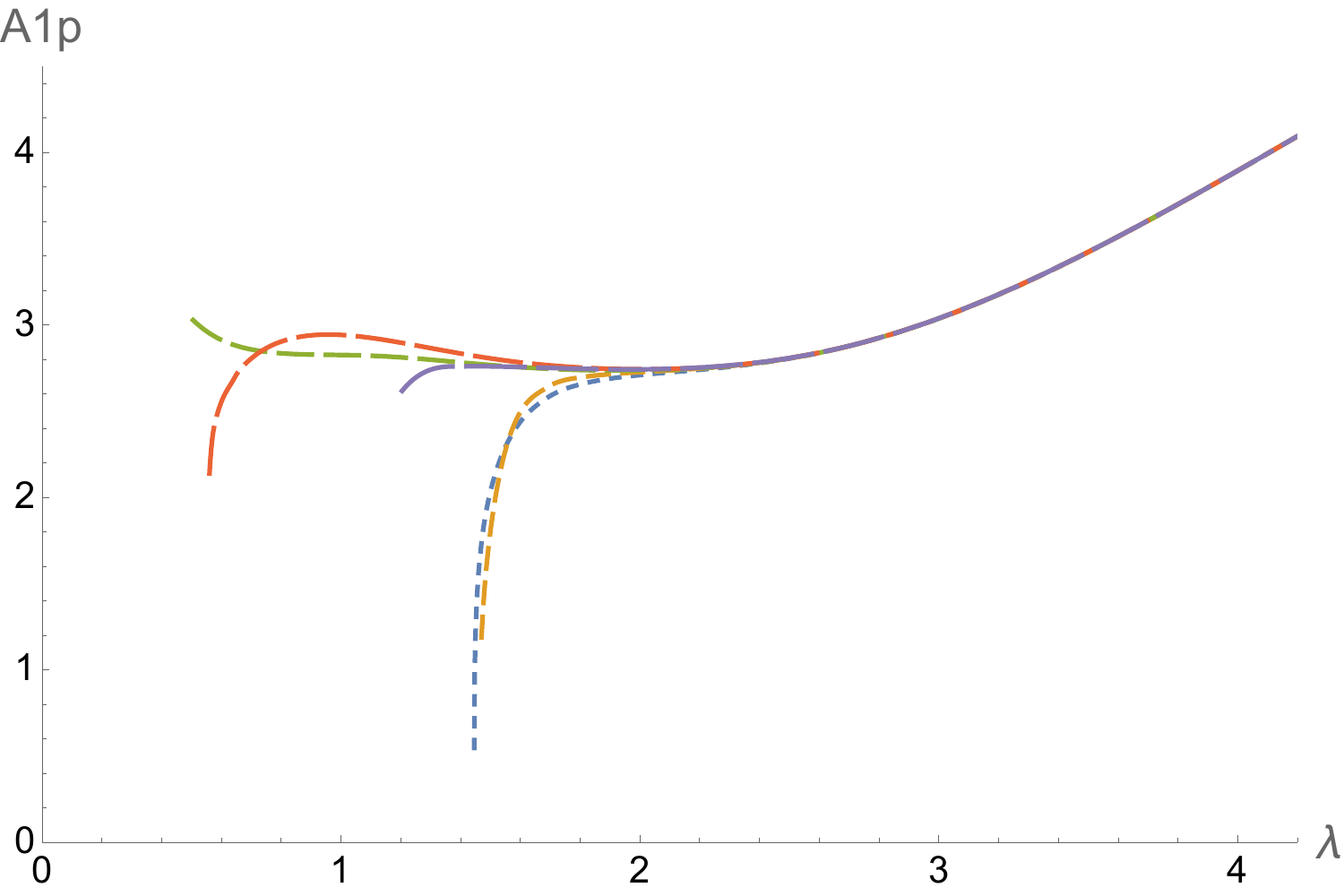}%
\hspace*{-0.5cm}
\vspace*{-5pt}
\caption
    {%
    Comparisons, in 2+1D Hamiltonian Yang-Mills,
    of results when observable approximation is, or is not, employed.
    Shown are results for the lowest
    $A_2^-$ and $A_1^+$ glueball masses
    from variational calculations with order 4 generators and
    observable truncation at orders 16, 20 and 24.
    Blue (shortest dash) and orange curves are for order 16 observable truncation,
    without and with observable approximation, respectively,
    while
    green and red (progressively longer dashes) curves are for order 20 observable truncation,
    without and with observable approximation, respectively,
    and purple (longest dash) shows observable 24 results without observable approximation.
    \label{fig:ym2h-approx}
    }
\end{figure}

Some results of this initial effort at observable approximation
are shown in Figs.~\ref{fig:ym2e-approx} and \ref{fig:ym2h-approx}.
Figure \ref{fig:ym2e-approx} compares deviations from the exact results
in 2D Euclidean Yang-Mills
for the winding-1 plaquette expectation value $\langle \textit {xyXY} \rangle$ and the
winding-2 plaquette expectation $\langle \textit{xyXYxyXY} \rangle$
from variational calculations with order 4 generators and
observable truncation, with and without observable approximation,
at orders 20 and 24.
Figure \ref{fig:ym2h-approx}
compares results in 2+1D Hamiltonian Yang-Mills
for the lowest
$A_2^-$ and $A_1^+$ glueball masses
from variational calculations with order 4 generators and
observable truncation, with and without observable approximation,
at orders 16 and 20, and without observable approximation at order 24.

These results are less promising than hoped for.
The most salient feature is the lack of any clear conclusion regarding 
the utility of this simplest observable approximation scheme.
In the 2D Euclidean results,
the order 20 curves
for $\langle \textit {xyXY} \rangle$
with observable approximation (orange curve of
the left panel of Fig.~\ref{fig:ym2e-approx})
do not have smaller deviations from the exact result,
or from the order 24 curves, than do
the unapproximated order 20 results (blue curve).
The order 24 results
for $\langle \textit {xyXY} \rangle$
with observable approximation (red curve)
deviates more from the exact result
than does the unapproximated order 24 curve (green).
However, this comparison flips
for the $\langle \textit {xyXYxyXY} \rangle$ results
shown on the right panel of Fig.~\ref{fig:ym2e-approx}:
the observable approximation result is somewhat more accurate than the unapproximated result.
Results for other observables, comparing order 24 observable truncations with 
and without observable approximation are similarly variable, with the observable
approximation results sometimes better and sometimes worse.

In the 2+1D Hamiltonian Yang-Mills results shown in Fig.~\ref{fig:ym2h-approx},
the most salient feature is again the lack of any clear conclusion regarding 
the effect of this simplest observable approximation scheme.
In results for the ground state energy, or for expectations of small loops
such as
$\langle \textit {xyXY} \rangle$,
$\langle \textit {xxyXXY} \rangle$, or
$\langle \textit {xyXYxyXY} \rangle$,
the dependence on the observable truncation order is quite small and
the effect of using observable approximation for non-retained observables
is nearly imperceptible in plots of these observables.
The same is true for the lowest $A_2^-$ glueball mass, shown on the left
of Fig.~\ref{fig:ym2h-approx}.
But results for the lowest $A_1^+$ glueball mass, shown on the right
of Fig.~\ref{fig:ym2h-approx},
display much larger dependence on the observable truncation order, with a
negligible difference between the order 16 results with and without
observable approximation,
and only rather small difference between the order 20 results with and without
such approximation above $\lambda \approx 0.7$.
In the $A_1^+$ mass results (as well as most other symmetry channels)
the results using observable approximation at one given
truncation order cannot be said to nearly mimic unapproximated results at a
higher observable truncation order.

Possibilities for improving this initial observable approximation scheme are
briefly discussed in the conclusions.

\section {Discussion}
\label{sec:discussion}

The various results presented in section \ref{sec:results} illustrate
both the potential and the challenges involved in applying the
coherent state variational algorithm to large $N$ gauge theories.
The results for single plaquette models,
both Euclidean and Hamiltonian,
in sections \ref{sec:ym1e} and \ref{sec:ym1h}
attest both to the correct functioning of the \textit{Gordion} code and to
the feasibility of obtaining good results from this variational approach
in theories having continuous phase transitions.
The results on two dimensional Euclidean Yang-Mills theory
in section \ref{sec:ym2e} show, completely unsurprisingly,
that without using the non-local reduction to independent plaquette variables
much larger observable truncation sets are needed to obtain good results.
Nevertheless, the results of section \ref{sec:ym2e} show that this
is quite feasible today.

The results for 2+1D Hamiltonian Yang-Mills theory
in section \ref{sec:ym2h} reveal how much more demanding 
the Hamiltonian theory on a two-dimensional spatial lattice is
in comparison to the Euclidean theory on the same lattice.
In part, this reflects the much larger size
of observable sets (of a given truncation order)
in Hamiltonian theories due to the need to
include Wilson loops with two electric field insertions.
But the plots of section \ref{sec:ym2h} also show notably slower
convergence of results with increasing order of the generator
selection (or number of variational parameters) as compared to the
analogous 2D Euclidean results.

In 2+1D lattice Yang-Mills theory, if one inserts a lattice
spacing $a$ to define the (inverse) cutoff scale, then
the dimensionless lattice coupling $\lambda = a \, g^2 N$,
with the 't Hooft coupling $g^2 N $ having
dimensions of mass.
In the lattice regulated theory
each glueball mass, in units of $a^{-1}$, is some
function of the lattice coupling, $m \, a = f(\lambda)$.
In the continuum limit, each such glueball mass must be
some pure number $c$ times $g^2 N$, as there is no other
relevant scale, and hence lattice glueball
masses, times $a$, must have the weak coupling form
$
    f(\lambda) = c \, \lambda + O(\lambda^2)
$
as $\lambda \to 0$.
Or in other words, lattice glueball masses must approach
a straight line through the origin,
Similarly, every Wilson loop expectation value, for any
fixed lattice loop $\Gamma$, must approach
a straight line with intercept one,
$
    \langle W_\Gamma \rangle = 1 + c_\Gamma \, \lambda + O(\lambda^2)
$.

The results reported in section \ref{sec:ym2h}
reach values of $\lambda$ at which the single plaquette
Wilson loop expectation value exceeds 0.5, and the
order 6 curves for the small loop expectation values
shown in Fig.~\ref{fig:ym2h-A}
are certainly consistent with a very smooth linear approach to
1 as $\lambda \to 0$ with small higher order corrections.
But the analogous statement cannot be made for the
various glueball mass curves plotted in figures
\ref{fig:ym2h-B}--\ref{fig:ym2h-D}.
The highest (sixth) order curves plotted in these figures
do not yet convincingly show linear approach to the origin.

Despite the truncation-induced limitations in the current results
for glueball masses in 2+1D Yang-Mills, one might nevertheless attempt to
extract an estimate (or crude best guess) for the continuum limit of
light glueball masses
by drawing straight lines from the origin which intersect with the
order 6 curves in figures \ref{fig:ym2h-B}--\ref{fig:ym2h-D}
at a point of tangency.
Regarding this as providing a serious estimate of continuum masses is,
of course, clearly premature.
First, the significant changes seen in Figs.~\ref{fig:ym2h-B}
and \ref{fig:ym2h-C} between the results of fourth and sixth order
generator truncations (or 4 to 13 variational parameters)
strongly suggests that inclusion of yet higher order generators will be
needed before decently converged results can be obtained.
Second, in the continuum limit
the $B_1^+$ and $B_2^+$ masses should become degenerate
as these are both components of a spin-2
continuum representation,
and likewise for $B_1^-$ and $B_2^-$ masses.
The plotted results do not yet show any such near degeneracy
between these channels.%
\footnote
    {%
    However, if one does boldly extrapolate by drawing a tangent
    through the origin to the order 6 curves, one finds
    $m/g^2 N \approx 0.9$ for the lightest $A_1^+$ glueball,
    reasonably close to the value of 0.81 obtained in
    Ref.~\cite{Athenodorou:2016ebg}
    from Euclidean lattice simulations.
    Results for other representations differ more substantially,
    with $m/g^2 N \approx 0.65$ for the lightest $A_2^-$ glueball,
    well below the value of 1.2 from Ref.~\cite{Athenodorou:2016ebg},
    while for other channels the crudely extrapolated slopes
    are well above the Euclidean simulation values.
    }

Results from any given variational calculation will cease to provide
a good approximation when the effects of the truncation in the selected
sets of either observables or generators become undesirably large,
and it is inevitable that such truncation effects will grow with 
decreasing coupling (or increasing correlation length).
The location, and manner, of the resulting breakdown varies with the
particular truncations employed.
This is reflected in the varying termination points of plotted results
in the various figures in section \ref{sec:results}.
Truncation induced problems can manifest in different ways as
the coupling is lowered:
\begin{enumerate}
\item
    The curvature matrix (\ref{eq:ddH}) 
    may cease to have purely real and positive eigenvalues,
    with either some eigenvalue going negative, or else
    complex conjugate pairs of eigenvalues appearing.
\item
    Certain Wilson loop expectations can move into the
    unphysical domain and, in particular, develop magnitudes
    exceeding unity.
\item
    The results may lack obvious internal inconsistencies but
    simply differ substantially from higher order calculations.
\end{enumerate}

Complex curvature eigenvalues can appear because the
coherence generator induced variations defining the
curvature matrix are not coordinate derivatives and
amount to using a non-coordinate set of basis vectors at
any point in the large $N$ phase space.
Consequently, the curvature matrix (\ref{eq:ddH})
is not exactly symmetric.
By the Jacobi identity, the antisymmetric part
of the curvature matrix equals the gradient
of the Hamiltonian in the direction of a
commutator of generators,
$
    (d^2H)_{[ij]} =
    \langle [ [H ,\, [e_i ,\, e_j]] \rangle
$.
Because only a finite set of generators can be retained,
this commutator of generators may lie outside the set of
selected generators.
The size of this antisymmetric part is negligible at
strong coupling, but grows as the coupling decreases.
At some point this can cause  pairs of
curvature eigenvalues which are initially real
to collide and move off into the complex plane.
This may indicate that the generator truncation is no longer sufficient
to obtain good results at a given value of coupling.

In Hamiltonian theories, the curvature matrix $d^2H$ (\ref{eq:ddH})
is block diagonal with one block involving double commutators with
pairs of time-reversal even generators and one block involving
double commutators with pairs of time-reversal odd commutators,%
\footnote
    {%
    For a detailed discussion of lattice symmetries and their implications
    for coherence group generators and observables, see
    appendix C of the implementation notes \cite{Gordion}.
    }
while the Lagrange bracket matrix (\ref{eq:L}) is block odd-diagonal,
with the non-vanishing components involving a commutator of a
$T$-even generator with a $T$-odd generator.

The iterative minimization (in Hamiltonian theories) is only sensitive
to the $T$-even block of the curvature matrix
in the completely symmetric lattice symmetry channel,
since the gradient in time-reversal odd directions automatically vanishes.
For Newton minimization to show the expected quadratic convergence,
it is essential that non-symmetric definition (\ref{eq:ddH}) of the curvature
be used when predicting the location of the minimum.

The small oscillation eigensystem (\ref{eq:spectra}),
in any given symmetry channel,
depends on both the $T$-even and $T$-odd blocks
of the curvature as well as the Lagrange bracket matrix $L$.
The curvature matrix may be symmetrized
for spectrum calculations,
guaranteeing purely real symmetrized curvature eigenvalues.
If both blocks of the curvature are positive definite 
then the small oscillation eigensystem (\ref{eq:spectra}) necessarily
yields real oscillation frequencies (in $\pm$ pairs).
But if either block of the curvature matrix develops a negative
eigenmode, then complex oscillation frequencies may appear in
the computed small oscillation spectrum.
Should this occur, this is a clear sign that one is beyond
the regime of utility of a given truncation for the particular symmetry channel.
For example, in the order 4 spectrum results for the
$A_1^+$, $A_2^-$ and $B_1^+$ channels,
shown as the orange curves in
Figs.~\ref{fig:ym2h-B} and \ref{fig:ym2h-C}, 
negative curvature eigenvalues develop around couplings of
1.2, 0.9 and 0.6, respectively, leading to the lowest oscillation
frequencies in these channels becoming unphysical.

For some truncations, a complex or negative curvature eigenvalue
which eventually appears may correspond to a variational direction in which
the gradient of the Hamiltonian (or free energy) is very small,
so that the presence of this non-positive eigenvalue below some value of
coupling may have very little impact on the iterative Newton extremization
(even though it indicates that the truncation is not well describing some
variation which is nearly transverse to the gradient).
But eigenvalues which are highly sensitive to the truncation and,
at some value of coupling, pass through zero can lead to completely
invalid predictions (\ref{eq:predict})
for the location of an extremum
and consequent complete failure of convergence
of the Newton iterative extremization.
This failure mode can sometimes (but not always) be tamed by
using a singular value pseudoinverse, i.e.,
performing a singular value decomposition of the curvature and
omitting contributions to the inverse curvature coming from
singular values whose magnitude falls below a chosen cutoff.
In, for example, the order 6 generator, order 28 observable results
for 2D Euclidean Yang-Mills,
appearing in Figs.~\ref{fig:ym2e-F}--\ref{fig:ym2e-xyXYxyXY},
a singular value cutoff of 0.6 was used below $\lambda = 2$.
No sign of the imposition of this singular value cutoff at $\lambda = 2$ is visually
apparent in the generator order 6 curves in these figures.
But quite often an eigenvalue going negative
(in the $T$-even block of the symmetric lattice symmetry channel)
does signal the end of utility of a given calculation.

For a given set of generators, if the observable truncation order
is sufficiently low then the resulting expressions for curvature matrix elements
may have expectation values either omitted or approximated for
observables which should have appeared but are above the truncation limit.
This can lead to negative or complex
curvature eigenvalues appearing even at rather large values of coupling.
For example, in 2+1D Yang-Mills, if one attempts to use order 6 generators
with observable truncation at order 16, a negative curvature eigenvalue
appears already at $\lambda \approx 2.15$.
Consequently, calculations with order $k$ observables should generally
retain at least up to order $4k$ observables.
Given the rapid growth in the size of observable sets, and storage
needed for the geodesic equations, this concern is the major issue
which prevented going to generator orders above 6 in the presented
2+1D Hamiltonian Yang-Mills calculations.

With a loop-list observable truncation scheme, a separate issue is that
the geodesic equations for observables of highest order within the retained set
are necessarily ``damaged'' by the omission (or imperfect approximation)
of observables of yet higher orders.
Consequently, integrating the finite, truncated set of geodesic equations
can, at some point, lead to unphysical expectation values of Wilson loops
which, for example, violate the basic unitarity bound $|W_\Gamma| \le 1$.
The good news, so to speak, is that such unitarity violations in high
order observables do not seem to rapidly feed down and immediately drive
unphysical behavior in lower order observables.
In, for example, the order 28 observable results shown in
Figs.~\ref{fig:ym2e-F}--\ref{fig:ym2e-xyXYxyXY} for 2D Euclidean Yang-Mills
theory,
the first appearance of unitarity bound violating Wilson loop expectations
occurs at $\lambda \approx 1.3$, 1.65, and 1.9 for generator orders
2, 4 and 6, respectively.
And yet nothing notable is visibly apparent at these coupling values in
the displayed plots of smaller Wilson loop expectations.
However,
in the 2+1D Hamiltonian Yang-Mills theory results,
the small oscillation spectrum is significantly more sensitive to 
higher order observables than low order observable expectation values
and, for example, the visible bumps in the fourth order
$B_2^-$ and $E^-$ curves of Fig.~\ref{fig:ym2h-C} are likely related to 
certain growing unphysical high order loop expectations which first exceed
unity around $\lambda = 1.3$.
Finally, in some cases
as one moves to lower and lower values of gauge coupling,
increasingly large (and unphysical) loop expectations can lead
to genuine run away behavior and
non-convergence of the ODE integration routine.
This completely stops the progression to smaller values of coupling,
and is the reason the generator order 6 results in 2+1D Yang-Mills
shown in Fig.~\ref{fig:ym2h-A} do not extend below $\lambda = 1.4$;
ODE convergence failed at $\lambda = 1.35$ after loop expectations
exceeding one first appeared at $\lambda = 1.6$.

The final potential truncation-induced limitation mentioned above ---
namely, decent looking, internally consistent but increasingly inaccurate results ---
is seen to occur in calculations with order 2 generators and a single
variational parameter.
It is unsurprising that such calculations, once beyond the
strong coupling regime, quickly become increasingly inaccurate
despite not experiencing negative curvature modes or breakdown in ODE integration.
With multiple variational parameters, one of the other issues above
eventually always seems to occur.

%The major take-away lesson, however, is that such truncation error
%induced breakdown does not prevent one from reaching values of coupling
%which are well into the weak coupling regime.

In any infinite dimensional variational scheme, it is inevitable that
understanding and minimizing truncation effects associated with the
state representation is a major issue.
As discussed in section \ref{sec:approx}, this motivated the effort
to explore a factorization-based approximation scheme for non-retained observables.
This initial effort to implement an observable approximation scheme turns out to be
rather disappointing.
For any given observable truncation order,
incorporating factorized approximations of non-retained observables
using the implemented algorithm
dramatically increases the number of terms in geodesic equations
(and consequent storage requirements) but,
as seen in Figs.~\ref{fig:ym2e-approx} and \ref{fig:ym2h-approx},
cannot be said to lead to results which effectively mimic results from
higher order but unapproximated observable truncations.

The basic idea motivating a factorization-based approximation scheme
was that the typical electric flux sheet spanning a self-intersecting Wilson
loop (in a confining theory) should truly look like the union of the
typical flux sheets which would span each of the sub-loops produced by
splitting the original loop at a self-intersection.
The simple factorization
$
    \langle W_\Gamma \rangle
    \approx
    \langle W_{\Gamma_1} \rangle \,
    \langle W_{\Gamma_2} \rangle 
$
for a self-intersecting loop $\Gamma = \Gamma_1\Gamma_2$
should be a decent approximation
if the flux sheets of the two sub-loops do not significantly interact,
and if the sign of $\langle W_\Gamma\rangle$ agrees with the
the product of signs of the sub-loop expectations.
But folded flux sheets do interact, as seen by the non-linear dependence
on winding number in the logarithm of Wilson loop expectation values
in one-plaquette models
(or in other 2D Yang-Mills loop expectations \cite{Kazakov:1980zj}).
Moreover, Wilson loop expectation values, while necessarily real,
are not always positive.
For these reasons, it was always clear that a simple factorization based
approximation scheme will never be exact.
The hope was that it could provide a good approximation for a sufficiently
large fraction of observables to be an overall improvement.
It should be the case that increasing lattice dimension
increases the fraction of loops for which
a simple factorization-based approximation works well,
as increasing dimensions make it less likely that a
generic loop will have a loop-spanning flux sheet with 
fold-induced flux interactions.
So while the goal of developing a useful observable approximation scheme
remains, more work is needed to explore possible schemes, especially
on three dimensional lattices.

\section {Conclusion}
\label{sec:conclusion}

It has always been clear that a numerical solution of large $N$ Yang-Mills
theory (or QCD) is a very tough computational task.
The results presented in this paper on 2+1 dimensional Hamiltonian Yang-Mills
theory illustrate what is currently practical using a desktop computer.
The current software \cite{Gordion} is capable of performing
analogous variational calculations in the lattice Hamiltonian formulation
of 3+1 dimensional Yang-Mills theory,
as well as calculations of the light meson spectrum
in Hamiltonian formulations of 2+1 and 3+1 dimensional QCD.
Initial results for these theories will be presented in a subsequent paper.

The 2+1 dimensional Yang-Mills results presented in section \ref{sec:ym2h}
may be viewed as a promising initial effort but are certainly
more limited than one would like regarding how far it was feasible to push
into the weak coupling regime.%
\footnote
    {%
    However, it should be noted that there are no alternative methods,
    currently available, which can even begin to study non-Abelian Hamiltonian
    lattice gauge theories on infinite two or three dimensional lattices.
    }
Answering a variety of open questions will help determine
the ultimate reach of this approach.
On the purely computational side, it will be very helpful to understand:
\begin{enumerate}
\item
    Can the current \textit{Gordion} program code be effectively implemented
    on a massively parallel high performance computing cluster?
    The challenge is whether inter-node data transfer rates
    in a large cluster with a hierarchical memory architecture
    will be sufficient to enable efficient integration
    of the massive set of coupled geodesic equations
    when these equations are partitioned across a great many nodes.

\item
    Can GPUs be effectively utilized in the integration of the
    geodesic equations?
    This has not yet been explored.
    One aspect of this involves floating point precision.
    All work to date has used 64-bit floating point arithmetic
    in the integration of geodesic equations.
    If only 32-bit floating point arithmetic is used,
    will the resulting precision loss ever become problematic?

\end{enumerate}

In terms of the achievable effectiveness of variational approximations,
more conceptual open questions include:
\begin{enumerate}
\setcounter{enumi}{2}
\item
    Are there feasible alternatives to the strong-coupling
    order classification of observables which would provide
    superior selection criterion in a loop-list truncation?
    Euclidean lattice bootstrap efforts
    \cite{Anderson:2016rcw,
    Kazakov:2022xuh}
    have used a simple cutoff on loop length
    to define selected subsets of Wilson loops.
    The strong-coupling order classification used in this
    work is more complicated to implement but, by design,
    produces superior results for a given truncation size
    in the strong coupling regime.
    Whether this remains true as one pushes into the
    weak coupling region, and whether some other selection
    criterion would be clearly superior, is unknown.

\item
    Similarly, is there a superior selection criterion for
    coherence group generators, other than simply using all
    generators up to a specified creation order (proportional to
    the number of plaquettes from which the generator is built)?
    The current code optionally implements a definition of
    generator normalization \cite{Gordion},
    which is a necessary first step
    for allowing meaningful comparisons of the relative
    importance of different generators as the minimization
    proceeds to smaller values of gauge coupling.
    But a detailed study of the relative importance of
    different generators is yet to be performed.

\item
    In Euclidean lattice simulations, much effort has been
    devoted to the development of improved lattice actions
    which speed up convergence to the continuum limit
    \cite{Symanzik:1983dc,
    Weisz:1982zw,
    Weisz:1983bn}.
    Analogous improved lattice Hamiltonians should be feasible
    to construct and implement in the coherent state approach.
    How much might this improve results?

\item
    Is it possible to formulate a more accurate, but practical,
    observable approximation scheme which performs better than the
    rather ad-hoc factorization algorithm described in section
    \ref{sec:approx}?
    As clearly seen in tables \ref{tab:ym2e} and \ref{tab:ym2h},
    the very rapid growth in the number of retained observables
    with increasing truncation order,
    and consequent growth in the amount of memory required to hold
    the geodesic equations, is a major limiting factor.
    Calculations to date 
    have found that using order $k$ generators with observables
    truncated at less than order $4k$ produce poor results.
    This is unsurprising as $4k$ is the minimal observable order for
    which all observable expectations appearing in curvature matrix
    elements (for the standard Kogut-Susskind Hamiltonian) are directly retained.
    Is it possible to formulate an observable approximation scheme
    sufficiently accurate to allow good results to be
    obtained when using order $k$ generators and observables
    truncated at less than order $4k$?
    Is it possible to identify, in a computationally efficient manner,
    those observables for which a factorization-based approximation is accurate,
    and only apply the approximation to such observables?
\end{enumerate}

The 2+1D Yang-Mills theory results presented in section \ref{sec:ym2h}
show that limitations in the accuracy of the state representation
--- i.e., the loop-list truncation ---
are a critical issue restricting both how far any given calculation can be
pushed into weak coupling,
and how many variational parameters can be usefully included.
Are there superior alternatives to a loop-list truncation?
Specifically:

\begin{enumerate}
\setcounter{enumi}{6}
\item
    Will switching 
    to a finite dimensional master field representation
    produce better results in the weak-coupling regime?
    In the old work \cite{CSVAI}, the Hamiltonian
    one-plaquette model was studied using both master field
    and loop-list state representations.  As shown in Fig.~4
    of that work, the master field results were found to
    exhibit curious non-monotonic behavior in the weak
    coupling regime,
    at couplings where the finite dimensional master field approximation 
    could no longer well represent the tails of the increasingly compact
    eigenvalue distribution,
    leading to a view that loop-list state representations,
    despite their limitations, seemed more promising.
    On the other hand, a unitary finite dimensional master field
    approximation, despite being less accurate
    in the strong coupling regime
    (for comparable computational sizes), has the clearly
    desirable feature of never violating Wilson loop
    positivity constraints.
    Will master field truncation sizes
    achievable with modern computational resources
    allow one to reach weaker couplings than feasible
    with loop-list truncations in 2+1 and 3+1 dimensional
    Yang-Mills?
    This is clearly a key question to answer.

%\item
    %Closely related on the computational side,
    %is the feasiblity of efficient implementation on a high performance
    %cluster significantly different when using a very large master field
    %state representation?

\item
    Are there fundamentally different choices for a coherence group,
    acting irreducible on the gauge invariant Hilbert space \cite{YaffeRMP},
    which might allow completely different types of state representations?
    One possibility to explore could involve traces of products of equal-time
    Dirac propagators plus their time derivatives.

\end{enumerate}

As noted in section \ref{sec:gordion}, the \textit{Gordion} program code,
along with extensive design and implementation notes,
are freely available on 
\href{https://github.com/lgyaffe/Gordion}{Github}.
Interested readers are encouraged to use this code base to explore
some of the above questions.
It is hoped that future work
will shed light on their answers.

\begin{acknowledgments}
Support from U.S. Department of Energy
grant DE-SC\-0011637 is gratefully acknowledged.
\end{acknowledgments}


\begin{thebibliography}{99}

%\cite{YaffeRMP}
\bibitem{YaffeRMP}
L.~G.~Yaffe,
``Large $N$ limits as classical mechanics,''
\href{https://doi.org/10.1103/RevModPhys.54.407}
{Rev.~Mod.~Phys.~\textbf{54}, 407 (1982)}.
%doi:10.1103/RevModPhys.54.407
%305 citations counted in INSPIRE as of 01 Dec 2024

%\cite{CSVAI}
\bibitem{CSVAI}
F.~R.~Brown and L.~G.~Yaffe,
``The coherent state variational algorithm:
A numerical method for solving large-$N$ gauge theories,''
\href{https://doi.org/10.1016/0550-3213(86)90318-4}
{Nucl.~Phys.~B \textbf{271}, 267--332 (1986)}.
%doi:10.1016/S0550-3213(86)80012-8
%24 citations counted in INSPIRE as of 01 Dec 2024

%\cite{CSVAII}
\bibitem{CSVAII}
T.~A.~Dickens, U.~J.~Lindqwister, W.~R.~Somsky and L.~G.~Yaffe,
``The coherent state variational algorithm. 2. Implementation and testing,''
\href{https://doi.org/10.1016/0550-3213(88)90234-9}
{Nucl.~Phys.~B \textbf{309}, 1--119 (1988)}.
%doi:10.1016/0550-3213(88)90234-9
%20 citations counted in INSPIRE as of 01 Dec 2024

%\cite{Witten:1979kh}
\bibitem{Witten:1979kh}
E.~Witten,
``Baryons in the $1/N$ expansion,
\href{https://doi:10.1016/0550-3213(79)90232-3}
{Nucl. Phys. B \textbf{160}, 57--115 (1979)}.
%3050 citations counted in INSPIRE as of 01 Sep 2025

%\cite{Wilson:1974sk}
\bibitem{Wilson:1974sk}
K.~G.~Wilson,
``Confinement of quarks,''
\href{https://doi.org/10.1103/PhysRevD.10.2445}
{Phys.~Rev.~D \textbf{10}, 2445--2459 (1974)}.
%6674 citations counted in INSPIRE as of 04 Jun 2025

%\cite{Kogut:1974ag}
\bibitem{Kogut:1974ag}
J.~B.~Kogut and L.~Susskind,
``Hamiltonian formulation of Wilson's lattice gauge theories,''
\href{https://doi.org/10.1103/PhysRevD.11.395}
{Phys.~Rev.~D \textbf{11}, 395--408 (1975)}
%doi:10.1103/PhysRevD.11.395
%2516 citations counted in INSPIRE as of 04 Jun 2025

\bibitem{Meyer:2004hv}
H.~Meyer and M.~Teper,
``Confinement and the effective string theory in SU(N ${}\to \infty$):
a lattice study,''
\href{https://doi.org/10.1088/1126-6708/2004/12/031}
{JHEP \textbf{12}, 031 (2004)}
\href{https://arxiv.org/abs/hep-lat/0411039}
{arXiv:hep-lat/0411039 [hep-lat]}.
%41 citations counted in INSPIRE as of 30 Dec 2025

\bibitem{Teper:2008yi}
M.~Teper,
``Large $N$,''
\href{https://doi.org/10.22323/1.066.0022}
{PoS \textbf{LATTICE2008}, 022 (2008)},
\href{https://arxiv.org/abs/0812.0085}
{arXiv:0812.0085 [hep-lat]}.
%40 citations counted in INSPIRE as of 29 Dec 2025

\bibitem{Bursa:2012ab}
F.~Bursa, R.~Lau and M.~Teper,
``SO($2N$) and SU($N$) gauge theories in 2+1 dimensions,''
\href{https://doi.org/10.1007/JHEP05(2013)025}
{JHEP \textbf{05}, 025 (2013)},
\href{https://arxiv.org/abs/1208.4547}{arXiv:1208.4547 [hep-lat]}.
%15 citations counted in INSPIRE as of 29 Dec 2025

\bibitem{Athenodorou:2015nba}
A.~Athenodorou, R.~Lau and M.~Teper,
``On the weak $N$-dependence of SO($N$) and SU($N$) gauge theories in 2+1 dimensions,''
\href{https://doi.org/10.1016/j.physletb.2015.08.023}
{Phys. Lett. B \textbf{749}, 448--453 (2015)},
\href{https://arxiv.org/abs/1504.08126}
{arXiv:1504.08126 [hep-lat]}.
%18 citations counted in INSPIRE as of 29 Dec 2025

\bibitem{Athenodorou:2016ebg}
A.~Athenodorou and M.~Teper,
``SU($N$) gauge theories in 2+1 dimensions: glueball spectra and $k$-string tensions,''
\href{https://doi.org/10.1007/JHEP02(2017)015}
{JHEP \textbf{02}, 015 (2017)},
\href{https://arxiv.org/abs/1609.03873}{arXiv:1609.03873 [hep-lat]}.
%55 citations counted in INSPIRE as of 29 Dec 2025

\bibitem{Athenodorou:2021qvs}
A.~Athenodorou and M.~Teper,
``SU($N$) gauge theories in 3+1 dimensions: glueball spectrum, string tensions and topology,''
\href{https://doi.org/10.1007/JHEP12(2021)082}
{JHEP \textbf{12} (2021), 082},
\href{https://arxiv.org/abs/2106.00364}{arXiv:2106.00364 [hep-lat]}.
%98 citations counted in INSPIRE as of 29 Dec 2025

%\cite{Gonzalez-Arroyo:1982hyq}
\bibitem{Gonzalez-Arroyo:1982hyq}
A.~Gonzalez-Arroyo and M.~Okawa,
``The twisted Eguchi-Kawai model: A reduced model for large $N$ lattice gauge theory,''
\href{https://doi.org/10.1103/PhysRevD.27.2397}
{Phys. Rev. D \textbf{27}, 2397 (1983)}.
%doi:10.1103/PhysRevD.27.2397
%365 citations counted in INSPIRE as of 02 Feb 2026

%\cite{Kovtun:2007py}
\bibitem{Kovtun:2007py}
P.~Kovtun, M.~Unsal and L.~G.~Yaffe,
``Volume independence in large $N_c$ QCD-like gauge theories,''
\href{https://doi.org/10.1088/1126-6708/2007/06/019}
{JHEP \textbf{06}, 019 (2007)},
%doi:10.1088/1126-6708/2007/06/019
\href{https://arxiv.org/abs/hep-th/0702021}{arXiv:hep-th/0702021 [hep-th]}.
%202 citations counted in INSPIRE as of 02 Feb 2026

%\cite{Bonanno:2025hzr}
\bibitem{Bonanno:2025hzr}
C.~Bonanno, M.~Garc{\'\i}a P{\'e}rez, A.~Gonz{\'a}lez-Arroyo, K.~I.~Ishikawa and M.~Okawa,
``Non-perturbative determination of meson masses and low-energy constants in large-$N$ QCD,''
\href{https://doi.org/10.1007/JHEP12(2025)096}
{JHEP \textbf{12}, 096 (2025)},
%doi:10.1007/JHEP12(2025)096
\href{https://arxiv.org/abs/2508.05446}{arXiv:2508.05446 [hep-lat]}.
%4 citations counted in INSPIRE as of 02 Feb 2026

%\cite{Bonanno:2023ypf}
\bibitem{Bonanno:2023ypf}
C.~Bonanno, P.~Butti, M.~Garc{\'\i}a Per{\'e}z, A.~Gonz{\'a}lez-Arroyo, K.~I.~Ishikawa and M.~Okawa,
``The large-$N$ limit of the chiral condensate from twisted reduced models,''
\href{https://doi.org/10.1007/JHEP12(2023)034}
{JHEP \textbf{12}, 034 (2023)},
%doi:10.1007/JHEP12(2023)034
\href{https://arxiv.org/abs/2309.15540}{arXiv:2309.15540 [hep-lat]}.
%20 citations counted in INSPIRE as of 02 Feb 2026

\bibitem{Anderson:2016rcw}
P.~D.~Anderson and M.~Kruczenski,
``Loop equations and bootstrap methods in the lattice,''
\href{https://doi.org/10.1016/j.nuclphysb.2017.06.009}
{Nucl. Phys. B \textbf{921}, 702-726 (2017)},
\href{https://arxiv.org/abs/1612.08140}
{arXiv:1612.08140 [hep-th]}.
%71 citations counted in INSPIRE as of 30 Dec 2025

\bibitem{Kazakov:2022xuh}
V.~Kazakov and Z.~Zheng,
``Bootstrap for lattice Yang-Mills theory,''
\href{https://doi.org/10.1103/PhysRevD.107.L051501}
{Phys. Rev. D \textbf{107}, no.5, L051501 (2023)}
\href{https://arxiv.org/abs/2203.11360}
{arXiv:2203.11360 [hep-th]}.
%51 citations counted in INSPIRE as of 30 Dec 2025

%\cite{Makeenko:1979pb}
\bibitem{Makeenko:1979pb}
Y.~M.~Makeenko and A.~A.~Migdal,
``Exact equation for the loop average in multicolor QCD,''
\href{https://doi.org/10.1016/0370-2693(79)90131-X}
{Phys. Lett. B \textbf{88}, 135 (1979)
[erratum: Phys. Lett. B \textbf{89}, 437 (1980)]}.
%496 citations counted in INSPIRE as of 01 Sep 2025

%\cite{Wadia:1980rb}
\bibitem{Wadia:1980rb}
S.~R.~Wadia,
``On the Dyson-schwinger equations approach to the large $N$ limit:
Model systems and string representation of {Yang-Mills} theory,''
\href{https://doi.org/10.1103/PhysRevD.24.970}
{Phys. Rev. D \textbf{24}, 970 (1981)}
%88 citations counted in INSPIRE as of 30 Jan 2026

\bibitem{Kazakov:2024ool}
V.~Kazakov and Z.~Zheng,
``Bootstrap for finite $N$ lattice Yang-Mills theory,''
\href{https://doi.org/10.1007/JHEP03(2025)099}
{JHEP \textbf{03}, 099 (2025)},
\href{https://arxiv.org/abs/2404.16925}
{arXiv:2404.16925 [hep-th]}.
%27 citations counted in INSPIRE as of 30 Dec 2025

\bibitem{Guo:2025fii}
Y.~Guo, Z.~Li, G.~Yang and G.~Zhu,
``Bootstrapping SU(3) lattice Yang-Mills theory,''
\href{https://doi.org/10.1007/JHEP12(2025)033}
{JHEP \textbf{12}, 033 (2025)},
\href{https://arxiv.org/abs/2502.14421}
{arXiv:2502.14421 [hep-th]}.
%9 citations counted in INSPIRE as of 30 Dec 2025

\bibitem{Lin:2020mme}
H.~W.~Lin,
``Bootstraps to strings: solving random matrix models with positivity,''
\href{https://doi.org/10.1007/JHEP06(2020)090}
{JHEP \textbf{06}, 090 (2020)},
\href{https://arxiv.org/abs/2002.08387}
{arXiv:2002.08387 [hep-th]}.
%91 citations counted in INSPIRE as of 30 Dec 2025

\bibitem{Koch:2021yeb}
R.~d.~Koch, A.~Jevicki, X.~Liu, K.~Mathaba and J.~P.~Rodrigues,
``Large $N$ optimization for multi-matrix systems,''
\href{https://doi.org/10.1007/JHEP01(2022)168}
{JHEP \textbf{01}, 168 (2022)}
\href{https://arxiv.org/abs/2108.08803}
{arXiv:2108.08803 [hep-th]}.
%27 citations counted in INSPIRE as of 30 Dec 2025

\bibitem{Mathaba:2023non}
K.~Mathaba, M.~Mulokwe and J.~P.~Rodrigues,
``Large $N$ master field optimization: the quantum mechanics of two Yang-Mills coupled matrices,''
\href{https://doi.org/10.1007/JHEP02(2024)054}
{JHEP \textbf{02}, 054 (2024)},
\href{https://arxiv.org/abs/2306.00935}
{arXiv:2306.00935 [hep-th]}.
%14 citations counted in INSPIRE as of 30 Dec 2025

%\cite{Rodrigues:2025sbu}
\bibitem{Rodrigues:2025sbu}
J.~P.~Rodrigues,
````Glueballs'' in the quantum mechanics of three large massless Yang-Mills coupled matrices,''
\href{https://doi.org/10.1016/j.physletb.2025.139768}
{Phys. Lett. B \textbf{868}, 139768 (2025)},
%doi:10.1016/j.physletb.2025.139768
\href{https://arxiv.org/abs/2503.05304}{arXiv:2503.05304 [hep-th]}.
%1 citations counted in INSPIRE as of 02 Feb 2026

\bibitem{Dawid:2025zxc}
S.~M.~Dawid, Z.~T.~Draper, A.~D.~Hanlon, B.~H{\"o}rz, C.~Morningstar, F.~Romero-L{\'o}pez, S.~R.~Sharpe and S.~Skinner,
``QCD predictions for physical multimeson scattering amplitudes,''
\href{https://doi.org/10.1103/6nql-yrhw}
{Phys. Rev. Lett. \textbf{135}, no.2, 021903 (2025)},
\href{https://arxiv.org/abs/2502.14348}
{arXiv:2502.14348 [hep-lat]}.
%14 citations counted in INSPIRE as of 30 Dec 2025

\bibitem{Dawid:2025doq}
S.~M.~Dawid, Z.~T.~Draper, A.~D.~Hanlon, B.~H{\"o}rz, C.~Morningstar, F.~Romero-L{\'o}pez, S.~R.~Sharpe and S.~Skinner,
``Two- and three-meson scattering amplitudes with physical quark masses from lattice QCD,''
\href{https://doi.org/10.1103/bx16-lp3r}
{Phys. Rev. D \textbf{112}, no.1, 014505 (2025)},
\href{https://arxiv.org/abs/2502.17976}
{arXiv:2502.17976 [hep-lat]}.
%16 citations counted in INSPIRE as of 30 Dec 2025

\bibitem{Fitzpatrick:2022dwq}
A.~L.~Fitzpatrick and E.~Katz,
``Snowmass white paper: Hamiltonian truncation,''
\href{https://arxiv.org/abs/2201.11696}
{arXiv:2201. 11696 [hep-th]}.
%23 citations counted in INSPIRE as of 30 Dec 2025

\bibitem{Witten:1979pi}
E.~Witten,
``\href{https://lib-extopc.kek.jp/preprints/PDF/1980/8002/8002242.pdf}
{The $1 / N$ expansion in atomic and particle physics},''
\href{https://doi.org/10.1007/978-1-4684-7571-5_21}
{NATO Sci. Ser. B \textbf{59}, 403-419 (1980)}
%26 citations counted in INSPIRE as of 31 Dec 2025

\bibitem{Coleman:1980nk}
S.~R.~Coleman,
``1/$N$,''
\href{https://lib-extopc.kek.jp/preprints/PDF/1980/8005/8005163.pdf}
{SLAC-PUB-2484 (1980)}.
%21 citations counted in INSPIRE as of 31 Dec 2025

\bibitem{Coleman:1985rnk}
S.~Coleman,
``Aspects of Symmetry: selected Erice lectures,''
Cambridge University Press, 1985,
\href{https://doi.org/10.1017/CBO9780511565045}
{ISBN 978-0-521-31827-3}.
%227 citations counted in INSPIRE as of 31 Dec 2025

%\cite{Lindqwister:1988xc}
\bibitem{Lindqwister:1988xc}
U.~J.~Lindqwister,
``Numerical studies of large $N$ lattice gauge theories,''
\href{https://catalog.princeton.edu/catalog?search_field=author&q=lindqwister}{Princeton Univ. Ph.D.~thesis, 1988,
UMI-88-09316}.

%\cite{Dickens:1987ih}
\bibitem{Dickens:1987ih}
T.~A.~Dickens,
``Numerical studies of fermionic field theories at large $N$,''
\href{https://catalog.princeton.edu/catalog/993726213506421}{Princeton Univ. Ph.D.~thesis, 1987
UMI-87-16885}.

%\cite{Somsky:1989}
\bibitem{Somsky:1989}
W.~R.~Somsky,
``The coherent state variational algorithm and the QCD deconfinement phase transition,''
\href{https://catalog.princeton.edu/catalog/994859983506421}{Princeton Univ.~Ph.D.~thesis, 1989}.

\bibitem{Gordion}
L.~G.~Yaffe,
\href{https://github.com/lgyaffe/Gordion/blob/main/doc/gordion.pdf}
{Gordion: Design and Implementation}.

%\cite{Gross:1980he}
\bibitem{Gross:1980he}
D.~J.~Gross and E.~Witten,
``Possible third order phase transition in the large $N$ lattice gauge theory,''
\href{https://journals.aps.org/prd/abstract/10.1103/PhysRevD.21.446}
{Phys.~Rev.~D \textbf{21}, 446--453 (1980)}.
%955 citations counted in INSPIRE as of 01 Apr 2025

%\cite{Wadia:2012fr}
\bibitem{Wadia:2012fr}
S.~R.~Wadia,
``A study of U($N$) lattice gauge theory in 2-dimensions,''
\href{https://arxiv.org/pdf/1212.2906}
{arXiv:1212.2906 [hep-th]}.
%{118 citations counted in INSPIRE as of 30 Jan 2026

\bibitem{Friedan:1980tu}
D.~Friedan,
``Some Nonabelian toy models in the large $N$ limit,''
\href{https://doi.org/10.1007/BF01942328}
{Commun. Math. Phys. \textbf{78}, 353 (1981)}.
%46 citations counted in INSPIRE as of 30 Dec 2025

%\cite{Wadia:1980cp}
\bibitem{Wadia:1980cp}
S.~R.~Wadia,
``$N = \infty$ phase transition in a class of exactly soluble model lattice gauge theories,''
\href{https://doi.org/10.1016/0370-2693(80)90353-6}
{Phys.~Lett.~B \textbf{93}, 403--410 (1980)}.
%290 citations counted in INSPIRE as of 01 Apr 2025

%\cite{Jevicki:1980zq}
\bibitem{Jevicki:1980zq}
A.~Jevicki and B.~Sakita,
``Loop space representation and the large $N$ behavior of the one plaquette {Kogut-Susskind} Hamiltonian,''
\href{https://doi.org/10.1103/PhysRevD.22.467}
{Phys.~Rev.~D \textbf{22}, 467 (1980)}.
%100 citations counted in INSPIRE as of 04 Jun 2025

%\cite{Neuberger:1980qh}
\bibitem{Neuberger:1980qh}
H.~Neuberger,
``Nonperturbative contributions in models with a nonanalytic behavior at infinite $N$,''
\href{https://doi.org/10.1016/0550-3213(81)90238-8}
{Nucl.~Phys.~B \textbf{179}, 253-282 (1981)}.
%41 citations counted in INSPIRE as of 04 Jun 2025

\bibitem{Kazakov:1980zj}
V.~A.~Kazakov,
``Wilson loop average for an arbitrary contour in two-dimensional U($N$) gauge theory,''
\href{https://doi.org/10.1016/0550-3213(81)90239-X}
{Nucl. Phys. B \textbf{179}, 283-293 (1981)}.
%115 citations counted in INSPIRE as of 19 Jan 2026

%\cite{Symanzik:1983dc}
\bibitem{Symanzik:1983dc}
K.~Symanzik,
``Continuum limit and improved action in lattice theories. 1. Principles and $\varphi^4$ theory,''
\href{https://doi.org/10.1016/0550-3213(83)90468-6}
{Nucl. Phys. B \textbf{226}, 187-204 (1983)}.
%1121 citations counted in INSPIRE as of 16 Jan 2026

%\cite{Weisz:1982zw}
\bibitem{Weisz:1982zw}
P.~Weisz,
``Continuum limit improved lattice action for pure Yang-Mills theory. 1.,''
\href{https://doi.org/10.1016/0550-3213(83)90595-3}
{Nucl. Phys. B \textbf{212}, 1-17 (1983)}.
%480 citations counted in INSPIRE as of 16 Jan 2026

%\cite{Weisz:1983bn}
\bibitem{Weisz:1983bn}
P.~Weisz and R.~Wohlert,
``Continuum limit improved lattice action for pure Yang-Mills theory. 2.,''
\href{https://doi.org/10.1016/0550-3213(84)90563-7}
{Nucl. Phys. B \textbf{236}, 397 (1984)}
\href{https://doi.org/10.1016/0550-3213(84)90543-1}
{[erratum: Nucl. Phys. B \textbf{247}, 544 (1984)]}.
%223 citations counted in INSPIRE as of 16 Jan 2026

\end{thebibliography}
\end {document}